\documentclass[12pt,a4paper,final]{iopart}
\usepackage{iopams}
\usepackage{setstack}
\usepackage{lmodern} %scalable font
\usepackage{graphicx}
\usepackage{epigraph}
\usepackage{amssymb}
\usepackage{paralist}
\usepackage{color}
\usepackage{xfrac} %reqires scalable font
\usepackage{subfigure}
\usepackage[square,numbers]{natbib}
\usepackage{amscd}
\usepackage{dsfont}
\usepackage{bm}

\usepackage[pdfborder={0 0 0}]{hyperref}

\DeclareGraphicsExtensions{.png, .pdf, .jpg, .bmp}
\graphicspath{{./pics_png/},{./pics/},{./borispics/},{./}}

\newcommand{\eg}{\textit{e.g. }}
\newcommand{\ie}{\textit{i.e., }}

\newcommand{\teq}{\! = \!}
\newcommand{\eqref}[1]{(\ref{#1})}

\newcommand{\iu}{\mathrm{i}}
\newcommand{\eu}{\mathrm{e}}
\newcommand{\dd}{\mathrm{d}}
\newcommand{\Tprim}{T_\gamma^{\scriptscriptstyle(P)}}
\newcommand{\Nprim}{N_\gamma^{\scriptscriptstyle(P)}}
\newcommand{\jcut}{j_\text{cut}}

\newcommand{\mod}{{\rm mod}}
\newcommand{\Sman}{\mathcal{S}_{\rm man}}
\newcommand{\ktm}{\text{(KT)}}
\newcommand{\Sga}{\mathcal{S}_\gamma}
\newcommand{\Aga}{A_\gamma}
\newcommand{\Sact}{\mathcal{S}}
\newcommand{\Sprim}{\mathcal{S}_\gamma^{\scriptscriptstyle(P)}}
\newcommand{\ldmax}{{\tilde{\lambda}_\text{max}}}
\newcommand{\heff}{\hbar_\text{eff}}
\newcommand{\Utk}{\hat{W}_K}
\newcommand{\Uti}{\hat{W}_I}
\newcommand{\Uto}{\hat{W}_0}
\newcommand{\Utp}{\hat{P}}
\newcommand{\Utg}{\hat{G}}
\newcommand{\Ut}{\hat{W}}
\newcommand{\Nrep}{r_\gamma^{\scriptscriptstyle(N)}}
\newcommand{\Trep}{r_\gamma^{\scriptscriptstyle(T)}}
\newcommand{\Mga}{\underline{M}_\gamma}

\newcommand{\Uh}{\hat{U}}
\newcommand{\HI}{\hat{H}_I}
\newcommand{\HK}{\hat{H}_K}
\newcommand{\eig}{\Lambda}

\newcommand{\operatorname}[1]{\text{#1}}

\newcommand{\Zmat}{\underline{\mathcal{Z}}}
\newcommand{\dosc}{d_\text{osc}}
\renewcommand{\mat}{\underline}

\newcommand{\conline}{(\textit{color online}) }

\newcommand{\K}{V}

\newif\ifdraft

\draftfalse %controls whether annotations are shown

\begin{document}

%\title[KIC semi-classics]{Many-Body Semiclassics for the Kicked Spin Chain//Alternative: Large spectral oscillations in the  Kicked Spin Chain }
\title[KIC semiclassical fluctuations]{Semiclassical Prediction of Large Spectral Fluctuations in Interacting Kicked Spin Chains}

\author{Maram Akila$^{*}$, Boris Gutkin$^{\dagger *}$, Peter Braun$^{*}$, Daniel Waltner$^{*}$ and Thomas Guhr$^{*}$}

\address{
$*$: Faculty of Physics, University of Duisburg-Essen, Lotharstr. 1, 47048 Duisburg, Germany\\
$\dagger$: Max Planck Institute for the Physics of Complex Systems, N\"othnitzer Str. 38, 01187 Dresden, Germany}

\ead{maram.akila@uni-due.de}
\ead{boris.gutkin@uni-due.de}

\begin{abstract}
While plenty of results have been obtained  for single-particle quantum systems with chaotic dynamics through a semiclassical theory, much less is known about quantum chaos in the many-body setting. We contribute to recent efforts to make a semiclassical analysis of many-body  systems feasible. This is nontrivial due to both the enormous density of states and the exponential proliferation of periodic orbits with the number of particles.
% We focus on large scale oscillations of the spectral density indicative of the nature of the underlying classical dynamics.
As a model system we study kicked interacting spin chains employing semiclassical methods supplemented by a newly developed duality approach. We show that for this model the line between integrability and chaos becomes blurred. Due to the interaction structure the system features (non-isolated) manifolds of periodic orbits possessing highly correlated, collective dynamics.
 %"As with" should be correct, see https://ell.stackexchange.com/questions/48158/as-with-grammar-and-meaning-in-a-sentence
As with the invariant tori in integrable systems, their presence lead to significantly enhanced spectral fluctuations, which by order of magnitude lie in-between integrable and chaotic cases.
%   \Onote{old:}
%   While plenty of results have been obtained  for 
%   single-particle quantum systems with chaotic dynamics by  semiclassical theories, much less is known about quantum chaos in
%   the many-body setting. We contribute to recent efforts to make a
%   semiclassical analysis of many-body  systems feasible, which is
%   nontrivial due to both the enormous density of states and the exponential proliferation of periodic
%   orbits with the number of  particles. Employing a recently proposed  duality relation, we focus on large scale oscillations of the spectral density. Within a model of a Kicked Spin Chain we show that such density oscillations are dominated by a specific class of  periodic orbits which feature a highly correlated spatial structure.

\end{abstract}

%\pacs{abc}
\vspace{2pc}

%Working draft as of \today

\ifdraft
    \tableofcontents
\else
    \newpage
\fi

%%%%%%%%%%%%%%%%%%%%%%%%%%%%%%%%%%%%%%%%%%%%%%%%
% Introduction
%%%%%%%%%%%%%%%%%%%%%%%%%%%%%%%%%%%%%%%%%%%%%%%%
\section{Introduction}
\label{sec:intro}

 For many years the field of quantum chaos has revolved mostly around questions regarding the spectral statistics of single to few body Hamiltonian systems \cite{stoeckmann,haake}. As has been realized, already in the 80's, the type of spectral statistics  crucially depends on the system's underlying classical dynamics. In particular, for fully chaotic systems the spectral statistics turn out to be universal, see \cite{BGS1984,CVGG1980},  and are well described by Random Matrix Theory (RMT) \cite{guhr}.
 The desire to understand this connection led to the development of  methods based on the semiclassical Gutzwiller trace formula  which allows  one to treat  spectral correlations in systems with fully chaotic dynamics \cite{berry2,sieberRichter,haakePRL1}.
 In  recent years the focus of research has shifted from single to  many-body systems and there is a natural inclination to explore their spectral properties  using semiclassical   tools \cite{regen1,regen2,regen4,dubert}. These systems possess  several  distinct   dynamical  features which set them apart from the single particle case. One of them  is the possibility of  collective dynamics which is   found in countless physical systems, ranging from Bose Einstein condensation \cite{TSimula,h17} over superparamagnetism \cite{superParaRev} to nuclei \cite{h10,VanDerWoude1987217,f12,f14}.
 This phenomenon can appear only when the system possesses  a significant number of particles $N$. So, as opposed to the single particle semiclassical limit, where only the Planck's constant   tends  to zero,   the many-body   problem  should take into account  the  existence of another  (large) parameter $N$. This interplay has a profound impact on the resulting spectrum of the system.
 In this paper we continue our  recent effort \cite{akilaPRL} to explore spectral properties  of the  kicked spin chains  through their classical dynamics when the number of spins $N$ becomes large.   

While  eigenenergies $\{E_1,E_2, \dots\}$ of a  generic integrable Hamiltonian  behave like a bunch of statistically  independent numbers,  spectra of chaotic systems exhibit a  much more rigid structure. A simple  way to validate  this statement is to look at the spectral density \(d(E)\),
where the smooth    and the oscillating parts are separated: 
\begin{equation}
    d(E)=\sum_{n=1}^\infty \delta (E-E_n) =\bar{d}(E)+ \dosc(E)\,.
\end{equation}
Recall that $d(E)$  should be understood as a distribution which requires smoothing with an appropriate test function to remove its singular nature \cite{haake}.
A natural expectation is  that due to the rigidity of the  spectra the smoothed oscillating part $\dosc$ for  Hamiltonians with chaotic dynamics   should be significantly smaller than for integrable ones.

The impact  of classical dynamics  on the  magnitude of $\dosc$ can easily be understood   through a semiclassical theory.  
To this end we consider the Fourier transform, $
    \Tr \Uh{\!(t)} \teq \int_{-\infty}^{\infty}\! d(E) \eu^{-\iu t E/\hbar}\, \dd E
$,
of the  density of states which is given by traces of the time evolution propagator  $ \hat{U}(t)$. In the semiclassical  limit   $\hbar \to 0$ those are related to the periodic orbits (POs) of the corresponding  classical system,
\begin{equation}
    \Tr{\hat{U}(t)}=\sum_n \eu^{-\iu t E_n/\hbar}
    \sim \sum_\gamma \Aga \eu^{\iu \Sga/\hbar}\,,
    \label{eq:intro:traceEq1}
\end{equation}
where the sum on the right-hand side runs over all POs \(\gamma\)  possessing a period $t$. Their associated action is denoted by \(\Sga\) and the prefactor $\Aga$ depends on the stability of the PO. In the case of integrable systems with $D$ degrees of freedom POs are not isolated but rather reside on $D$ dimensional tori. For these systems the prefactor scales as \(\Aga\sim \hbar^{(-D+1)/2}\). If, on the contrary, the orbit is fully isolated one observes no scaling,  $\Aga\sim \hbar ^0 $.
Furthermore, if the orbit has (non-trivial) marginal directions the corresponding scaling is of an intermediate type, $\Aga \sim \hbar^{-\alpha}$, where the fractional exponent  $\alpha $ depends on the type of bifurcation \cite{cat_berry2,cat_schomerus,cat_bestPaper}. 
For purely chaotic systems all periodic orbits are isolated, therefore the oscillating term in the semiclassical limit scales as  $\dosc\sim \hbar^{-1}$ independently of the system dimensions. On the other hand for fully integrable $N$-particle systems  the resulting spectral oscillations have significantly larger scales of the order  $\dosc\sim \hbar^{-(D+1)/2}$ which clearly grow with the number of particles, as  $D\propto N$.

The above line of reasoning  holds  only for the pure semiclassical limit, where $\hbar \to 0$ while the number of particles $N$ is held fixed. Clearly, the purely thermodynamic limit, where $N\to\infty$ at fixed $\hbar$, corresponds to a very different physical picture. For instance, in the case of interacting bosons the first and the  second limit  give rise   to two very different classical systems -- they correspond to the first and second quantization treatment of the system Hamiltonian, respectively \cite{regen2,dubert}.  In the present paper we address the  problem of spectral fluctuations in  the case when both   $\hbar^{-1}$ and $N$ are large.  Within a rigorous mathematical framework  this corresponds to a double limit, where simultaneously $\hbar \to 0$ and $N\to\infty$ is taken while some relation between both parameters is kept.
Specifically, we ask whether  an analog of  \eqref{eq:intro:traceEq1} holds in this case and if yes,  what the magnitude of the prefactors $A_\gamma$ is. As we show, the answer is in general affirmative, but with a twist -- for the  model under consideration   the sum on the right hand side of \eqref{eq:intro:traceEq1}  turns out to be  strongly dominated by a  very particular   class of POs with a low spatial period. 

Going from few to many-particle systems is a hard challenge, especially so in the semiclassical limit \cite{primack}. The primary  reason  for this is pretty obvious: for a \(N\)-body system the density of states  grows exponentially  with $N$, \ie \( \bar d\sim\hbar^{-D+1}\). 
Already for moderate \(N\) a full resolution of the quantum spectrum is beyond the scope of any numerical or experimental approach. As a result, the direct calculation of the left hand side in \eqref{eq:intro:traceEq1} seems intractable. To the best of our knowledge an explicit comparison of periodic orbits and quantum spectra was never attempted before  beyond few-body systems like the helium atom (\(N\teq 2\)) \cite{wintgen2,richterHE}. Similar constraints hold for the overall field of many-body quantum mechanics. Many of the currently employed standard methods try to circumvent this limitation with approximate approaches which reduce the complexity of the system's Hilbert space. 
For instance, mean-field methods treat effective single particle systems, see \eg \cite{MF1}, and  matrix product states restrict the possible maximal entanglement between different particles \cite{MPS1,MPS2}. These methods  are typically  well suited to treat systems in vicinity to the ground state, but fail in the bulk of the spectrum,  where  the eigenstates of the   system are fully entangled.
%\Onote{I fail to see here  connection with anything what goes before and after: 
%For these areas semiclassical approaches seem more suited. Most of the current literature in this direction focuses on the indistinguishability of the particles, discussing back-scattering for Bosons\cite{regen1} or the density of states\cite{regen2} with application to two particle systems\cite{regen3}. Only few results towards fermionic systems are available, for instance \cite{regen4}.
%Still, for a validation of semiclassical results the left hand side of \eqref{eq:intro:traceEq1} has to be evaluated.}
Since we are interested in the semiclassical limit of the Hamiltonian, a different idea is needed to reduce the complexity of the Hilbert space. 
To this end we employ a recently developed approach, see \cite{gutkin,akila2,akilaPRL}, to evaluate traces of the time evolution operator for  kicked chain-like systems  with local interactions. This method, for discrete maps and integer times $T$, is based on the exact  duality relation  
\begin{equation}
\Tr{\hat{U}^T}= \Tr{\Ut}^N\,,
\label{eq:intro:duality}
\end{equation}
which connects traces of the unitary  Floquet evolution $\hat{U}$ in time to those of a dual,  non-unitary   evolution $\Ut$ in space.  Crucially, the dimension of $\Ut$ depends only on $T$ and is  therefore small if times are short. This allows an effective numerical calculation of the left hand side of \eqref{eq:intro:traceEq1} provided that the considered time $T$ encompasses only several periods.  On a more fundamental level, by virtue of its low dimension the operator $\Ut$ is much better suited for studies of large scale many-body spectral fluctuations  in comparison to  the original time evolution $\hat{U}$.

Originally,  the duality approach has been developed for the Kicked Ising Chain model with a fixed spin quantum number of $j=1/2$  \cite{akila2}. 
In the present paper we focus  on a natural extension  of this setting to an arbitrary  $j$. 
This model can be seen as a chain of $N$ Kicked Tops coupled through nearest neighbour interaction, as detailed in section \ref{sec:model}.
The  semiclassical limit is  attained by sending  the effective Planck's constant  \(\heff=j^{-1}\) to zero, \ie $j\to\infty$. 
While the classical dynamics of a single Kicked Top ranges from integrable to almost fully chaotic, its $N$ body extension never exhibits full hyperbolicity. This means, in particular, that a typical PO possesses both elliptic and hyperbolic directions.   As we show in the body of the  paper POs  of the system, in general, can be separated into two classes -- isolated POs  and non-isolated PO manifolds of dimensions $2$ and $4$.
%The later exist only  if  $\mu$ belongs to a special set of values $\mu({p,N_0})=-\cos{\frac{2\pi p}{N_0}}$, $p\in\{ 1,2,\ldots,N_0-1\}$, $N_0\in \mathbb{N}$, where the integer $N_0$ has a meaning of prime spatial period of POs.
Depending on the choice of parameters these PO  manifolds may have a short spatial period \(\Nprim\)  which   can be seen  as  a signature of highly correlated, collective dynamics.
Furthermore, we explore \eqref{eq:intro:traceEq1} for $1$ and $2$ time steps by investigating  the  spectrum of the dual operators   $\Ut$. Our main result is that  $\Tr{\Uh^2}$ is strongly dominated by  a small number of these   PO manifolds, while all isolated POs are suppressed in the large $N$ limit. The  factors $A_\gamma \sim \hbar^{-\alpha(N)}$  associated with each   PO manifold
exhibit a very large scaling exponent $\alpha(N)=\alpha_0 N$,  growing linearly  with $N$. The maximum  scaling $\alpha_0$ appears to be $1/4$ which is exactly half of the corresponding value  for  the integrable case.  Moreover, the model discriminates between different chain lengths. It shows  particularly  strong  spectral oscillations  for chains of length $N=\Nprim k$ with $k\in{\mathds{N}}$, where the prime length $\Nprim$  is solely defined by the PO manifolds.

The paper is structured as follows.  In the next section  we introduce the Kicked Spin Chain model, which in comparison to \cite{akilaPRL} is extended by a non-zero torsion $\K$. Its periodic orbits are then studied in section \ref{sec:po}. The formulation  and proof of the spectral duality relation \eqref{eq:intro:duality} is given in section \ref{sec:dual}. In section \ref{sec:action} the semiclassical trace formula  is  studied numerically for the aforementioned one and two time steps $T$ of dynamical evolution in  the case of $\K=0$. We demonstrate that for $T=2$ the right hand side of \eqref{eq:intro:traceEq1} is indeed strongly dominated  by collective PO manifolds in the large $N$ limit. In the subsequent  section \ref{sec:spec} we explain these empirical findings via a semiclassical theory of the dual operator  $\Ut$.  We extend the above results to the case of $\K\neq 0$ in section \ref{sec:nlkic}.  Finally, the conclusions are  presented in section \ref{sec:conclusion}. Technical details are relegated to the appendices.

\section{Model}
\label{sec:model}

Throughout this section we introduce the Kicked Spin Chain model \cite{prosen2,prosenJt-2,prosen2007b,prosenB3-d} for general spin quantum numbers \(j\). The Hamiltonian
\begin{equation}
    \hat{H}(t)=\hat{H}_I+\hat{H}_K \sum_{T=-\infty}^\infty\delta(t-T)
    \label{eq:model:Habst}
\end{equation}
describes the general dynamics of \(N\) spins where \(\hat{H}_I\) describes the (Ising) coupling between spins and \(\hat{H}_K\) provides local kicks acting on each spin separately. Time (in between kicks) is measured in terms of integer unit steps. The corresponding time  (Floquet) evolution operator for a single time step is thus given by
\begin{equation}
    \hat{U}=\hat{U}_I\hat{U}_K
    \qquad\text{with}\qquad
    \hat{U}_{I,K}=\eu^{-\iu (j+1/2) \hat{H}_{I,K}}\,.
    \label{eq:model:u}
\end{equation}
Therein \((j+1/2)\), as detailed later, takes on the role of the inverse Planck constant \(\hbar^{-1}\). As a side remark, an exchange of \(\hat{H}_I\) and \(\hat{H}_K\) in \eqref{eq:model:Habst} leads to the same evolution \(\hat{U}\).

\subsection{Kicked Top}
\label{sec:model:kickedTop}

To begin with we recall the Kicked Top \cite{haake} as the \(N\teq 1\) limit of the model.
The corresponding Hamiltonians are
\begin{eqnarray}
    \hat{H}_K^\ktm &=& \frac{2\,\vec{b}\cdot\hat{\vec{S}}}{j+1/2}\,,
    \label{eq:hkt:k}
    \\
    \hat{H}_I^\ktm &=& \frac{4 J^\ktm}{(j+1/2)^2}\,(\hat{S}^z)^2\,
    \label{eq:hkt:i}
\end{eqnarray}
with the spin operator \(\hat{\vec{S}}=(\hat{S}^x,\ \hat{S}^y,\ \hat{S}^z)^\mathrm{T}\) for spin quantum number \(j\), \ie \((\hat{\vec{S}})^2\teq j(j+1)\). The Ising part contains a non-trivial quadratic term which can be thought of as a shear or torsion. This term singles out the \(z\)-direction and therefore the magnetic field \(\vec{b}\) in \(\hat{H}_K^\ktm\) can be restricted, without loss of generality, to the \(xz\)-plane, \(\vec{b}= (b^x,\,0,\,b^z)^T=b\,(\sin{\varphi},\,0,\,\cos{\varphi})^T\), where \(\varphi\) is the angle between the magnetic field and the \(z\)-axis, \(\tan{\varphi}\teq b^x/b^z\).
To avoid a dependence of the coupling parameter strengths on \(j\) in the classical limit
%, the maximal eigenvalues of \(\hat{S}^z\) are \(\pm j\), \Onote{write more to keep this}
we rescale both Hamiltonians by their respective powers in \(\hat{S}\).
Minimal uncertainty is given in terms of \(j\) for spin coherent states, which therefore replaces \(\hbar^{-1}\) as a measure of Planck cell size.
% thesis: would be good if you could refer back to spin coherent states somewhere

The corresponding classical model can be found replacing \(\hat{\vec{S}}\) by a spin vector \(\sqrt{j(j+1)}\vec{n}\), \(|\vec{n}|\teq 1\), precessing on the Bloch sphere.
The relation to the canonical coordinates \((q,p)\) is given by \cite{haake,keppeler}
\begin{equation}
 \vec{n}=\left(\sqrt{1-p^2}\cos q,\sqrt{1-p^2}\sin q, p \right)^T
 \label{eq:nInQP}
\end{equation}
with the corresponding Hamiltonian
\begin{equation}
H(q,p)=4 J p^2+
2\left(b^z p+b^x\sqrt{1-p^2}\cos q\right)
\sum_{T=-\infty}^\infty\delta(t-T)\,
\label{eq:kt:hamiltonian}
\end{equation}
having only a single degree of freedom.

\begin{figure}
\includegraphics[width=0.3\textwidth]{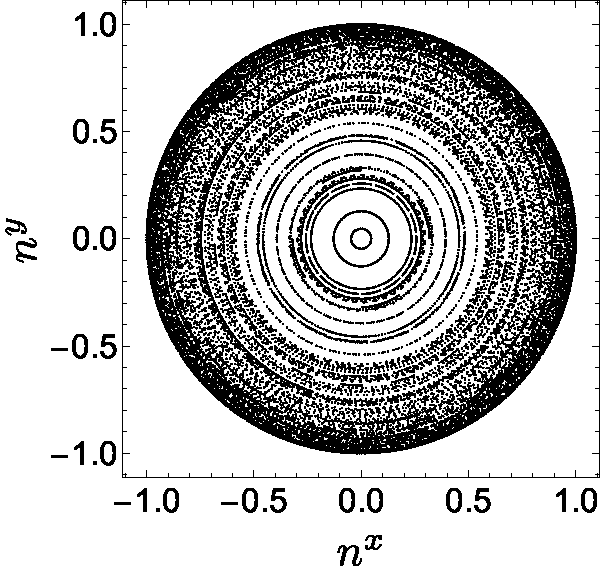}
\hfill
\includegraphics[width=0.3\textwidth]{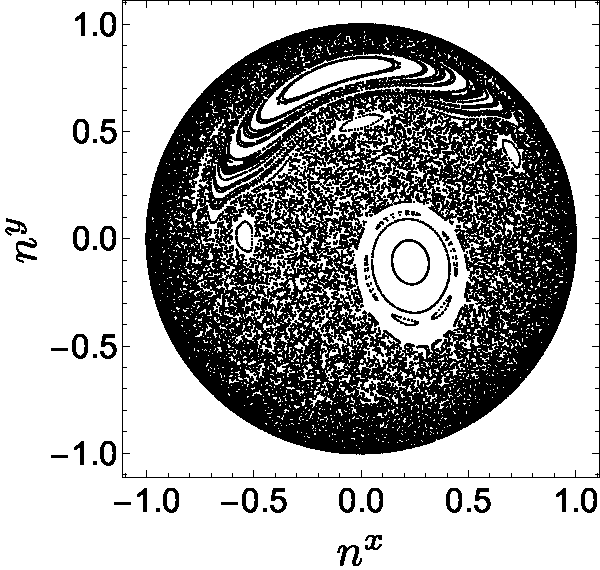}
\hfill
\includegraphics[width=0.3\textwidth]{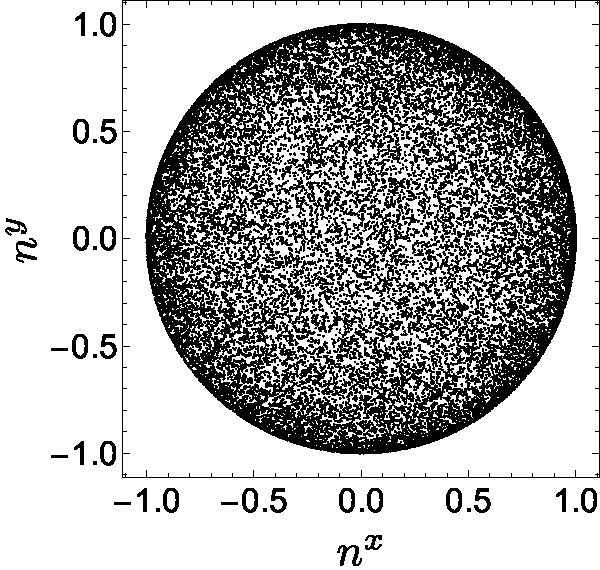}
\caption[Kicked top phase-space for integrable, chaotic and intermediate parameters.]{Upper hemisphere (\(n^z>0\)) of the classical phase-space for the kicked top, see eq. \eqref{eq:kt:classRot}, after 200 iterations for several hundred randomly chosen initial points. Parameters are chosen as \(J^\ktm \teq 0.7\), \(b \teq 0.9\sqrt{2}\approx 1.27\). The angle varies with \(\varphi= 0,\, 0.2,\, \pi/4\) (from left to right).}
\label{fig:KickedTopPoincare1}
\end{figure}

The classical action of the kick onto \(\vec{n}\) is a rotation around the \(\vec{b}\)-axis by the angle \(2b\), denoted by \(\underline{R}_{\vec{b}} (2b)\). The Ising part also acts as a rotation around the \(z\)-axis, however, its angle depends on the value of \(n^z\) creating the torsion. Combining both rotations one finds the new position of \(\vec{n}\) after a single time step as
\begin{equation}
\vec{n}(T\!+\!1)=
\underline{R}_{z} \big( 4J^\ktm n^z \big)\,
\underline{R}_{\vec{b}} (2b)\,
\vec{n}(T)\,.
\label{eq:kt:classRot}
\end{equation}
Figure \ref{fig:KickedTopPoincare1} provides Poincare sections of this dynamics for different values of \(\varphi\). In the case of \(\varphi \teq 0\), left panel, the system is integrable and the phase-space is filled by tori corresponding to fixed actions. Changing the angle slightly leads to a breakup of those tori, leaving the system in a mixed state including remaining regular islands, while for \(\varphi\teq \pi/4\) the system is (almost) fully chaotic.

In several cases it turns out to be useful to decompose the \(\vec{b}\)-rotation into three rotations around the coordinate axes,
\begin{equation}
   \underline{R}_{\vec{b}} (2b)=
   \underline{R}_z(\alpha) \underline{R}_x(\beta)\underline{R}_z(\gamma) 
   =
   \underline{R}_z(\alpha-\pi/2) \underline{R}_y(\beta)\underline{R}_z(\gamma+\pi/2)
   %Signs of the \pm\pi/2 in second decomposition changed, should be correct now - MA
    \,.
    \label{eq:spec:eulerDecomp}
\end{equation}
The angles \(\alpha,\beta,\gamma\) are the corresponding Euler angles of this decomposition and for our choice of \(\vec{b}\), namely \(b^y\teq 0\), they are given by
\begin{equation}
    \alpha = \gamma, \quad
    b^z \tan (\pi/2-\alpha) =b \cot{b}, \quad
    \cos{\beta} =\left(\frac{b^z}{b}\right)^2+ \left(\frac{b^x}{b}\right)^2 \cos{2b}\,.
    %changed sign in tan, seems to be correct now - MA
    \label{eq:spec:eulerAnglesDef}
\end{equation}
Due to \(\underline{R}_z(x)\underline{R}_z(y)\teq \underline{R}_z(x+y)\) this allows us, classically,  to express the whole dynamics in terms of alternating \(x,z\) or \(y,z\) rotations, respectively.

\subsection{Kicked Spin Chain}
\label{sec:model:kic}
For the \(N\)-body extension of the Kicked Top into a one dimensional chain we introduce a homogeneous bilinear coupling between neighbouring spins,
%by a slight modification of the interaction part,
\begin{equation}
    \hat{H}_I=\sum_{n=1}^N \left( \frac{4 J\,\hat{S}_{n+1}^z\hat{S}_n^z}{(j+1/2)^2}
    + \frac{4 \K \,(\hat{S}_{n}^z)^2}{(j+1/2)^2}\right)\,,
    \label{eq:kic:hi}
\end{equation}
with the inter-spin Ising coupling \(J\) and an additional local non-linearity governed by \(\K\). Throughout the paper we mostly restrict ourselves to the special case \(\K\teq 0\) and assume this condition if not stated otherwise. Discussion of the \(\K\neq 0\) extension is relegated to section \ref{sec:nlkic}. The kicked part is kept local and identical to \eqref{eq:hkt:k},
\begin{equation}
    \hat{H}_K = \sum_{n=1}^N \frac{2\,\vec{b}\cdot\hat{\vec{S}}_n}{j+1/2}\,.
    \label{eq:kic:hk}
\end{equation}
Boundary conditions for the interaction are chosen periodic, \ie\ \(\hat{S}_{N+1}^z\teq \hat{S}_{1}^z\), making the system translation invariant. In consequence, the special case of \(N\teq 1\) of the Kicked Spin Chain corresponds to the Kicked Top above with \(J^\ktm\teq J+\K\).
%The relevance of this observation for the \(N\)-spin system will be discussed in the next section. The translational symmetry leads to a decomposition of the spectrum of \(\hat{U}\) into \(N\) blocks corresponding to the different eigenvalues \(\eu^{2\pi\iu k/N}\) of the translation operator. The spectra of conjugated sectors \(k\leftrightarrow -k\) coincide (double degeneracy) while the cases of \(k\teq  0\) and \(k\teq N/2\) (existing for \(N\in\text{even}\)) possess a further internal symmetry \cite{prosen2007b}.
%This is given by an exchange of \(q_n\to -q_n\), equivalently \(n_m^y\to -n_m^y\), which leads to time reversed dynamics. This symmetry relates the \(k\) and \(-k\) sectors, mapping the two special cases onto themselves.

The modifications on the classical side compared to \eqref{eq:kt:classRot} and \eqref{eq:kt:hamiltonian} are straightforward. The Hamiltonian \eqref{eq:kt:hamiltonian} is adjusted to
\begin{equation}
\fl
H(\vec{q},\vec{p})=\sum_{n=1}^N\left[4J p_{n+1}p_n+4\K p_n^2+\
2\left(b^zp_n+b^x\sqrt{1-p_n^2}\cos q_n\right)
\sum_{T=-\infty}^\infty\delta(t-T)\right]\,,
\label{eq:kic:hamiltonian}
\end{equation}
which includes additional interaction between neighbouring momenta.
The Hamiltonian equations of motions give rise to the rotation of \(N\) classical spin vectors \(\vec{n}_m\),
\begin{equation}
\vec{n}_m(T\!+\!1)=
\underline{R}_{z} \big( 4J (\chi_m+2 \K /J\, n_m^z) \big)\,
\underline{R}_{\vec{b}} (2b)\,
\vec{n}_m(T)\,.
\label{eq:kic:classRot}
\end{equation}
In this case the angle of rotation, \(\chi_m \teq n_{m-1}^z + n_{m+1}^z\), encodes the bilinear interaction between the spins.

The system remains integrable for \(b^x\teq 0\). In the special case of \(j \teq 1/2\) the Kicked Spin Chain (for \(\K\teq 0\)) possesses another (non-trivial) integrable regime for \(b^z\teq 0\) \cite{lieb,atas,akila2} which ceases to exist for higher spin quantum numbers.
%thesis: K should act as local magnetic field, ie non zero h, connect to s1/2 case

%%%%%%%%%%%%%%%%%%%%%%%%%%%%%%%%%%%%%%%%%%%%%%%
% PO
%%%%%%%%%%%%%%%%%%%%%%%%%%%%%%%%%%%%%%%%%%%%%%%
\section{Periodic Orbits}
\label{sec:po}

For the semiclassical analysis knowledge of the PO  actions and their stabilities is essential. In the limit of large $j$  the trace of the propagator  can be expressed by  a Gutzwiller-type of sum over POs of  period $T$:
\begin{equation}
    \Tr\,\hat{U}^T\sim\sum_{\gamma(T)}
    A_\gamma \eu^{\iu (j + 1/2) \Sga}\,.
    \label{eq:action:TrGutzwill}
\end{equation}
 This relation was explicitly derived for spin systems in~\cite{waltner3}. Here \(\Sga\) is the classical action as given in \ref{sec:apx:classAct} 
and the prefactor \(A_\gamma\) is determined by the stability of the orbit. If the orbit is sufficiently isolated in phase-space,  it is given by
\begin{equation}
    A_\gamma
    =\frac{\Tprim \eu^{\iu G_\gamma}}{\sqrt{\left|\det\left(\underline{M}_\gamma-\mathds{1}\right)\right|}}
    \label{eq:action:stability}
\end{equation}
where  \(G_\gamma\) is  the Maslov phase and \(\underline{M}_\gamma\) is the monodromy matrix  determining the stability of the orbit under small perturbations.

After establishing  basic properties of the POs due to the system's  chain like structure we look,  in more detail,  at the case of integrable dynamics in section \ref{sec:po:int}. The general case, with $\K =0$, is covered by  section \ref{sec:po:generalCase}, where  the  primary focus is  on manifolds of non-isolated POs which play a crucial role in the subsequent  semiclassical analysis.

\subsection{General Properties}
\label{sec:po:general}

 A periodic orbit \(\gamma\) of duration \(T_\gamma\) for \(N_\gamma\) spins is a set \(\{\vec{n}_m\}_{m=1}^N\) of Bloch vectors satisfying
\begin{eqnarray}
\label{eq:po:abstractDef}
\vec{n}_m(T_\gamma)=\vec{n}_m(0), \qquad\text{where}
\\
\nonumber
\vec{n}_m(0)=\vec{n}_m\,,\quad
\vec{n}_m(T_\gamma)=\left(\underline{R}_m(J,\K,\vec{b})\right)^{T_\gamma}\,\vec{n}_m(0)
\end{eqnarray}
with the classical propagation matrix \(\underline{R}_m(J,\K,\vec{b})=\underline{R}_{z} \big( 4J \chi_m+8 \K n_m^z \big)\,\underline{R}_{\vec{b}}(2b)\), compare section \ref{sec:model:kic}. As \(\gamma\) is a valid orbit for \(T_\gamma\) time steps it will, by further repetition, also be a valid orbit for \(k T_\gamma\) time steps (\(k\in \mathds{N}\)). This is a direct consequence of the system's translation invariance in time.
%thesis: ie the explicit time Independence of H\neq H(t)
The minimal number of time steps required to close the orbit (for the first time) is the primitive time period \(\Tprim\). Such an orbit leads to \(\Tprim\) different fixed point solutions to \eqref{eq:po:abstractDef} corresponding to changed initial starting points along the orbit.

By construction we have a  translational symmetry not only in time but also along the chain direction. Accordingly, a periodic orbit of the \(N\) spin system induces, by repetition, an orbit for a \(kN\) particle system with the same parameters. For instance, every periodic orbit of the Kicked Top is also a periodic orbit of the Kicked Spin Chain for \(J^\ktm\teq J+\K\). We introduce the primary spatial period \(\Nprim\) as the minimal number of spins required to accommodate the orbit \(\gamma\). 
The cyclic permutation of the motion of individual spins along the chain does, due to the translation symmetry, not change the overall dynamics and any given orbit is thus part of a family of \(\Nprim\) identical orbits with identical action and stabilities.

Periodic orbits can be expressed in terms of repetitions of the prime orbits which encompasses the minimal number of particles and time steps necessary to accommodate it.
These types of repetitions imply a linear scaling of the action \(\Sga\) of an orbit,
\begin{equation}
\Sga=\Trep\Nrep\,\Sprim
\,,
\label{eq:po:abstractSScale}
\end{equation}
where \(\Sprim\) is the action of the prime orbit and
\begin{equation}
    \Trep=\frac{T}{\Tprim}\,,\qquad
    \Nrep=\frac{N}{\Nprim}
\end{equation}
are the repetitions in time and space, respectively.
The actions  \(\Sga\) can be  calculated as the sum of  local spherical areas swept by the  \(\vec{n}_i\)'s on the  Bloch's spheres. The specific calculations are relegated to \ref{sec:apx:classAct}. 

 Numerics shows that for a generic choice of parameters most of the orbits comprise both hyperbolic and elliptic directions. In other words,  for a typical $\gamma$ the set of eigenvalues of the corresponding monodromy matrix \(\Mga\) includes ones for which $|\lambda_i|>1$, as well as ones with   $|\lambda_i|=1$. The relation \eqref{eq:action:stability} breaks down when one of the directions becomes marginal, \ie one of the eigenvalues turns into 1 and changes from hyperbolic to elliptic, or vice versa, under infinitesimal change of the system parameters.
If an orbit is marginal this also holds for its repetitions in time and space.  We comment further on non-isolated orbits in section  \ref{sec:action}.

%In such a case a more refined uniform approximation is needed to correctly treat the contribution of such orbits \cite{cat_schomerus,cat_manderfeld}.
%One typically finds an algebraic scaling of \(\Aga\) with the spin quantum number \(j\) which is descriptive of a particular type of bifurcation, see \cite{cat_bestPaper} for the \(N\teq 1\) case.

% For the Lyapunov exponents, which determine the linear stability of \(\gamma\) \gcite, the picture is more complicated. They remain invariant under arbitrary repetitions in time. In particle direction any perturbation direction in phase-space (with a prescribed Lyapunov) will keep this property for all larger spin systems by simple repetition of the direction\Onote{Clear enough?}. This is of special consequence if the direction in question is marginal as all larger repetitions also remain marginal, for a further discussion see section \todo. But with increased \(N\) more directions emerge whose Lyapunov exponent is \textit{a priori} not clear. Further insight into their properties gives the concept of the dual monodromy matrix \todo.

\subsection{Integrable Case}
\label{sec:po:int}

For  \(b^x\teq 0\) all rotations are around the \(z\)-axis and therefore commute with the Hamiltonian making the system integrable. As a result the dynamics of the kicked system for arbitrary times is equivalent to one at fixed time, \eg \(T\teq 1\) with rescaled system parameters \(J\to JT\) and \(b^z\to b^z T\). Moreover, the flow induced by the Hamiltonian $ \HI+ \HK$  for time $T$ is identical to the evolution of the kicked system for \(T\) time steps with the same parameters.

%Therefore, the momenta form the action variables of the integrable system restricting the angles \(\vec{q}\) to a \(N\) dimensional subspace.
For the classical trajectories $p_n=\text{const.}$ holds for each $n$ and  periodic orbits form \(N\) dimensional manifolds.
% All trajectories in this subspace are equivalent with respect to their dynamic properties. If for certain values of \(\vec{p}\) a periodic orbit exists it is therefore a \(N\) dimensional manifold.
To close a trajectory in phase-space after \(T\) iterations it is sufficient that the total change in angles \(\Delta q_n\) is a multiple of \(2\pi\),
\begin{equation}
 \Delta q_n=4 T\left(J (p_{n-1}+p_{n+1})+ 2\K p_n \right)+2b^z T=2\pi m_n\,.
 \label{eq:po:deltaQint}
\end{equation}
The \(m_n\in\mathds{Z}\) is a local winding number for spin \(n\).
Since the  momenta are bounded, \(|p_n|\leq 1\),   \(\chi_n\teq p_{n-1}+p_{n+1}\) resides within the interval $[-2,+2]$. Therefore, this equation  has no solution if, for instance, \(b^z>4(J\!+\!\K )\) and \(4(J+\K )+b^z<\pi/T\).
In such cases the system does not posses any classical periodic orbits of period \(T\) or shorter.
If all parameters (times \(T\)) are sufficiently small, the first accessible winding number is necessarily zero.
With increasing time $T$ the number of possible \(m_n\) grows linearly and with it the number of possible (distinct) periodic orbits grows algebraically.
% In contrast to the other cases time in the integrable case can be seen as a scaling factor of the system parameters, making arbitrary real times accessible.
With respect to \(N\) the number of periodic orbits is determined by all admissible combinations of the winding numbers.
If there is more than one allowed \(m_n\) the growth is thus exponential in \(N\). This exponential growth also holds for non-integrable parameter choices.

\subsection{General Case}
\label{sec:po:generalCase}

A perturbation of the integrable model by a non-zero \(b^x\) breaks up the $N$-dimensional  periodic tori into  isolated periodic orbits and some low dimensional manifolds of non-isolated periodic orbits.  We first comment on the general properties of the isolated ones and later detail on the  manifolds which, as it turns out,  play  a significant role in the semiclassical treatment of the corresponding quantum model.

\subsubsection{Isolated Periodic Orbits.}
\label{sec:po:generic}The observed exponential proliferation of periodic orbits for increasing $N$ within the integrable model carries over to the general case.
 For large $N$ the stabilities of orbits are of a mixed type, \ie  both hyperbolic and elliptic directions are present in the same orbit. The behavior of the prefactors $|\Aga|$, however, substantially   depends on the time $T$. For a generic set of parameters and $T=1$ a typical orbit  is well isolated, so that $|\Aga|$ is an exponentially small quantity. On the other hand, for 
$T=2$ we found   many  $\gamma$'s  for whom a large number of eigenvalues of $\Mga$ are  close to $1$. In other words, an essential number of periodic orbits is almost marginal implying
quite small determinants $\det(1-\Mga)$. In such cases the approximation \eqref{eq:action:stability} is no longer applicable. 

\subsubsection{Periodic Orbit Manifolds.}
\label{sec:po:manifold}
Besides the  isolated orbits the case \(T\teq 2\) (\(\K =0\)) also  features  four dimensional manifolds of periodic orbits, \ie regions in phase-space where every point constitutes a periodic orbit. 
As we explain below, this phenomenon occurs  when the length  of the  spin chain is equal to \(N\teq 4k\),  \(k\in\mathds{N}\). This peculiar condition  can be traced back to a special feature of the four-spin system whose periodic orbits, by repetition, also induce periodic orbits of  larger systems with \(N\teq 4k\). According to \eqref{eq:po:abstractDef} for $\K =0$ and $N=4$  the time evolution of the first and the  third spin vectors \(\vec{n}_1, \vec{n}_3\)  are provided by one and the same rotation matrix $\underline{R}_{z} \big( 4J \chi_1 \big)\underline{R}_{\vec{b}} (2b)$. This immediately implies that the scalar product  \((\vec{n}_1\cdot\vec{n}_3)\) is a conserved quantity. Similarly, \((\vec{n}_2\cdot\vec{n}_4)\) is preserved, as well.  In other words, the $N\teq 4$ spin chain possess two integrals of motion. Particularly, in the case of $b^x\teq 0$ the system is over integrable
having $6$ integrals of motion rather than $4$: In addition to the four momenta $ p_i, i=1,\dots, 4$  the differences between  coordinates  \(q_1-q_{3}\), \(q_2-q_{4}\) are conserved under time evolution.

\begin{figure}
\includegraphics[width=0.95\columnwidth]{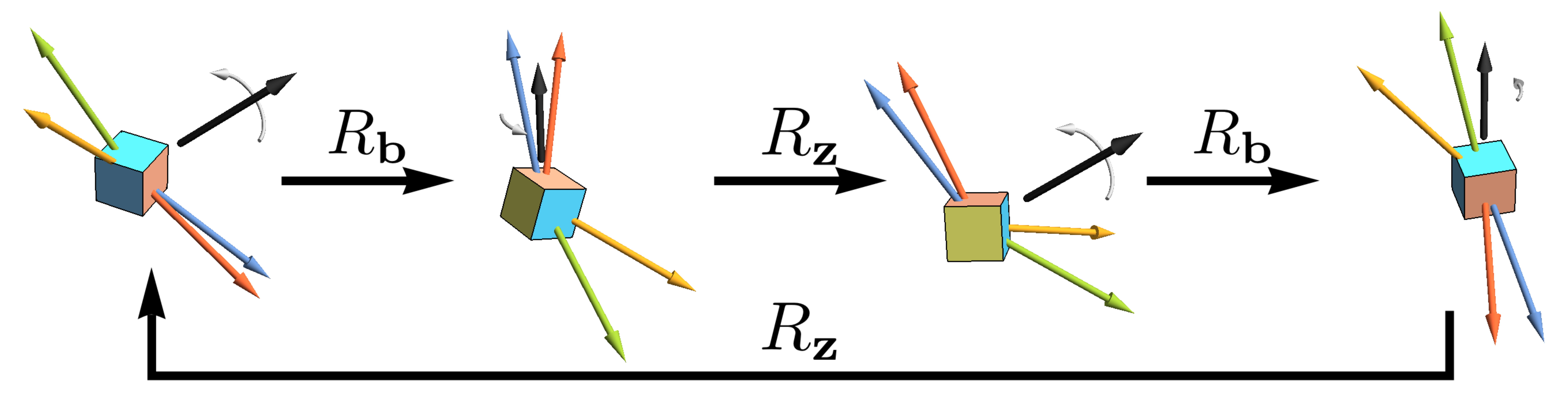}
\caption[Example of a PO trajectory on the 4D manifold]{\conline Trajectory of a PO on the manifold, depicted after each rotation step for \(J\teq 0.7\) and
  \(b^x\teq b^z\teq 0.9\). Spins are ordered along the chain according to their colors: blue,yellow, green and red. Visible is a solid-body rotation leaving all angles between the spins constant. The co-rotating cube serves as reference.}
\label{fig:po:manifoldSchema}
\end{figure}

\textit{4D manifolds} --- In the general case we provide an explicit construction of periodic orbit manifolds. Since the dynamics  of spin $i$  depends exclusively on the  time evolution of the variable  $\chi_i\teq n_{i-1}^z+n_{i+1}^z$, any trajectory satisfying the condition
\begin{equation}
\underline{R}_{z} \big( 4J \chi_i^{(1)} \big)\,
\underline{R}_{\vec{b}} (2b)\underline{R}_{z} \big( 4J \chi_i^{(2)} \big)\,
\underline{R}_{\vec{b}} (2b)=\mathds{1}, \qquad i=1,\dots, N\,,
\label{eq:po:rSquared}
\end{equation}
where  $\chi_i^{(1)}$, $\chi_i^{(2)}$ are the values at the time-steps $t\teq 1,2$, respectively, is automatically periodic. The most simple way to satisfy this condition is to assume that  $4J\chi_i^{(1)}\mod\; 2\pi\teq 4J \chi_i^{(2)}\mod\; 2\pi\teq \chi$  is  constant for all spins.  This implies that $(\underline{R}_{z} \big( 4J \chi_i^{(t)} \big)\allowbreak\underline{R}_{\vec{b}} (2b))^2=\mathds{1}$ such that $\underline{R}_{z} \big( 4J \chi_i^{(t)} \big)\underline{R}_{\vec{b}} (2b) = \underline{R}_{\vec{\zeta}}(\pi) $ is a rotation about $\pi$ around some axis $\vec{\zeta}$. This forces  the  value of \(\chi_i^{(t)}\) to satisfy the following equation:
\begin{equation}
b^z\,\tan{\left(2J \chi_i^{(t)}\right)}=b\,\cot{b}, \qquad i=1\dots, N, \qquad t=1,2. 
\label{eq:po:chiManifold}
\end{equation}
Fixing the values of \(\chi\) by eq.~\eqref{eq:po:chiManifold} imposes  restrictions  onto the positions of each spin at each time-step $t\teq 1,2$,
\begin{eqnarray}
\label{eq:po:manConditions}
\chi_i^{(1)}&=&n_{i-1}^z+n_{i+1}^z\,,\label{eq:po:rSquared1}\\
\label{eq:po:rSquared2}
\chi_i^{(2)}&=&-2 \sin^2{b} \sin{\varphi} \cos{\varphi} \left(n_{i-1}^x+n_{i+1}^x\right)
    +\sin{\varphi} \sin{2 b} \left(n_{i-1}^y+n_{i+1}^y\right) \\
   && +\left(\cos{2 b} \sin^2{\varphi}+\cos^2{\varphi}\right)\left(n_{i-1}^z+n_{i+1}^z\right)
   \nonumber
\,,
%for the derivation of this non-sense, see WB/1079 where R_b is presented. Then require that chi as sum of the nz components remains constant in terms of previous n(t=1) components (note: R_I leaves chi naturally invariant and can be ignored)
\end{eqnarray}
where  the  constants \(\chi_i^{(t)}\) satisfy   \eqref{eq:po:chiManifold} for all \(i\) and  \( t\). 
The second equation results from the fact that the second time step \(\chi_i^{(2)}\) is obtained from the original spin vectors via a rotation,   $\chi_i^{(2)}=\vec{e}_z \cdot \underline{R}_{\vec{b}} (2b) \left(\vec{n}_{i-1}+\vec{n}_{i+1} \right)$ (the $z$-component is not changed by $\underline{R}_{z} \big( 4J \chi_i^{(t)} \big)$ and it thus does not need to be considered).
For any sequence of $2N$ solutions of \eqref{eq:po:chiManifold}  obeying the conditions  \(-2\leq \chi_i^{(t)} \leq +2\),   the equations \eqref{eq:po:rSquared1}, \eqref{eq:po:rSquared2} fix  a \( 4\)-dimensional manifold of initial conditions for periodic orbits.
An example of  such a periodic orbit   is given  in figure \ref{fig:po:manifoldSchema} which shows that the relative motion between the spins is frozen due to the identical $\underline{R}_{z} \big( 4J \chi_i^{(t)} \big)$.
All periodic orbits  belonging to these manifolds have one and the same action provided by an elegant formula,
 \begin{equation}
    \Sman=   J \sum_{i=1}^N \sum_{t=1}^{2} \chi_i^{(t)}\chi_{i+1}^{(t)}\,,
    \label{eq:po:smanBase}
\end{equation}
whose explanation is relegated to \ref{sec:apx:t2action}.

We can distinguish three different regimes, where \eqref{eq:po:chiManifold} has none, one or several solutions in the interval \(-2\leq \chi_i^{(t)} \leq +2\), each having unique consequences for the system behavior.
The first case occurs, when    \(J\) or \(b\) are  sufficiently small, bringing the model  close to the integrable/non-interacting regime. Most of the paper is devoted to the single manifold regime, where \eqref{eq:po:chiManifold} admits one unique solution such that $\chi_i^{(t)}\teq \chi$ for all $i,t$. In this case the action of the manifold orbits is given by
\begin{equation}
    \Sact= N \Sman
    \qquad\text{with}\quad
    \Sman= 2 J \chi^2
    \quad\text{and}\quad
    N=4k
    \,.
    \label{eq:po:smanBase:single}
\end{equation}
In this case equation \eqref{eq:po:rSquared2} reduces, using $\chi_i^{(2)}\teq\chi_i^{(1)}\teq\left(n_{i-1}^z+n_{i+1}^z\right) $, to the simpler form
\begin{equation}
 \chi=\left(n_{i-1}^x+n_{i+1}^x\right) \cot{\varphi}
+\left(n_{i-1}^y+n_{i+1}^y\right) \frac{\cot{b}}{\sin{\varphi}}\,.
\end{equation}
For the case of several possible solutions  \(\chi_i^{(t)}=\chi+ m_i^{(t)}(\pi/2J),\)   \(m_i^{(t)}\in \mathbb{Z}\) of eq.~\eqref{eq:po:chiManifold} the number of different manifolds of periodic orbits starts to grow exponentially with $N$. This can be understood if we  compare the role  of the  $m_i^{(t)}$  to the spin winding numbers in the integrable case. There  the number of orbits with respect to \(N\) was determined by the exponentially growing amount of different possible combinations of winding numbers. In a similar way we can exchange the possible values of \(m_i^{(t)}\) along the spin chain leading to the  exponential growth  of different  periodic orbit  manifolds.

%A different class of manifolds, which appears to play only a minor role for semiclassical treatment of the model, appears under similar conditions.
\textit{2D manifolds} --- 
%calculations in WB/1086
So far, we considered the cases where \eqref{eq:po:rSquared} holds for all spins.
For the existence of manifolds it is sufficient to demand this condition for only half of the spins, \eg the odd indexed ones. In this case the manifold will be only two dimensional as we are sparing out half of the chain. For the even indexed spins this implies that they still fulfill \eqref{eq:po:rSquared1}. While this ensures that the trajectories of the odd spins are closed regardless of their initial conditions, we need further restrictions to ensure that also the trajectories of the even ones are periodic.
An exemplary way to realize this, already present for 4 spins, is by aligning the even spins along the rotational axis, \ie \(\underline{R}_{z} \big( 4J \chi_i^{(t)} \big)\underline{R}_{\vec{b}} (2b)\vec{n}_i\teq \vec{n}_i\), where \(i\) is even. From the components of this equation we derive several constraints, one on the angles of each of the two even spin,
\begin{equation}
    \sin{(q_i+\gamma)}=
    \frac{p_i\tan{\beta/2}}{\sqrt{1-p_i^2}}\,,
    \label{eq:po:2DmanFixedSpins:q}
\end{equation}
which is surprisingly independent of the Ising interaction. Therein \(\beta,\gamma\) are the Euler angles of the kick rotation as defined in \eqref{eq:spec:eulerAnglesDef}. Another constraint fixes the value of $\chi_i$ and therefore the axis along which the spins are aligned. As before for the 4D manifolds this value has to be maintained by the other set of spins, in this case the odd indexed ones, at both time steps. The resulting equations,
\begin{eqnarray}
    p_{i-1}+p_{i+1}&=&\frac{q_{i}}{2J} \qquad\mod\, \pi/(2J)\,,
    \label{eq:po:2DmanMovingSpins:T1}
    \\
    \label{eq:po:2DmanMovingSpins:T2}
    \left(p_{i-1}+p_{i+1}\right) \tan{\beta/2}& = &
    \sqrt{1-p_{i-1}^2}\sin{(q_{i-1}+\gamma)}
    \\&+&\sqrt{1-p_{i+1}^2}\sin{(q_{i+1}+\gamma)}
    \qquad\mod\, \pi/(2 J \sin{\beta})\,,
    \nonumber
\end{eqnarray}
are, up to the differing value of $\chi_i$, similar in nature to the ones used in \eqref{eq:po:rSquared1} and \eqref{eq:po:rSquared2}, including the possible multiplicity in the values of $\chi$. The last constraints concern the values of the odd $\chi_{i\pm1}$, due to demanding the original manifold condition \eqref{eq:po:rSquared} they are given by \eqref{eq:po:chiManifold} and have to be fixed via the even indexed spins. They are aligned along the same axis and we thus find
\begin{equation}
    p_i=\frac{\chi_{i\pm1}}{2}\,.
    \label{eq:po:2DmanFixedSpins:p}
\end{equation}
These six constraints fully fix the even spins, \eqref{eq:po:2DmanFixedSpins:q} and \eqref{eq:po:2DmanFixedSpins:p} hold for each of the spins separately, while \eqref{eq:po:2DmanMovingSpins:T1} and \eqref{eq:po:2DmanMovingSpins:T2} impose only two further conditions.

\subsection{Weak coupling regime.}
\label{sec:po:weak}
In the special case of the non-interacting regime $J\teq 0$, all periodic orbits are given by compositions of solutions for the single spin case. As a result, a non-interacting spin chain  becomes  fully chaotic if the corresponding $N\teq 1$ Kicked Top possesses chaotic dynamics, as  happens for certain choices of the parameters $\vec{b}$ and $\K $. In such a case each periodic orbit of the non-interacting spin chain is isolated and fully hyperbolic. This situation still persists after introducing  a weak coupling $J$ between the spins, at least,  if  $T$ is sufficiently short. In this regime periodic orbits are fully hyperbolic and  can be related to their non-interacting counterparts  making their identification  an easy achievable goal.  Accordingly, a leading order semiclassical approximation    \eqref{eq:action:TrGutzwill}  works significantly better for weakly interacting spin chains in comparison to the general case, where the dynamics is plagued by bifurcations.    

%%%%%%%%%%%%%%%%%%%%%%%%%%%%%%%%%%%%%%%%%%%%%%%%%%%%%%%%%%%%%%%%%%%%%
% Duality
%%%%%%%%%%%%%%%%%%%%%%%%%%%%%%%%%%%%%%%%%%%%%%%%%%%%%%%%%%%%%%%%%%%%%
\section{Duality Relation}
\label{sec:dual}

The dynamics  of chain-like  models with nearest neighbour interactions  is governed by Hamiltonian  equations  which are local both in time and particle indices \(n,t\), respectively. This suggests that in certain situations it might be useful to reverse the roles of $n$ and $t$ looking at  $t$ as particle and $n$ as time index, see \cite{gutkin}.
We first illustrate this on the classical level  and later extend these ideas to the quantum setting.

\subsection{Classical Duality}
\label{sec:dual:classical}

%For instance,  taking $n$ in equation \todo  as time,  converts the problem of finding  $T \teq 1$ periodic orbits in a chain of length $N$  into one of  finding  $N$-periodic  orbits in a single particle system.

It  is a simple observation that,  in general, one and the same set of Newtonian equations  with nearest neighbor interactions,
\begin{equation}
    q_{n,t+1}= \phi(q_{n,t},q_{n,t-1};q_{n-1,t},q_{n+1,t})
    \label{eq:dual:phi}
\end{equation}
leads to two possible dynamical systems. The first one  is provided by the conventional symplectic map $\Phi$: $(\vec{ q}_{t}, \vec{p}_{t}) \to (\vec{q}_{t+1}, \vec{p}_{t+1})$, $\vec{q}_{t}=(q_{1,t}, \dots q_{n,t})$, $\vec{p}_{t}=(p_{1,t}, \dots p_{n,t})$ describing the  propagation of the system in time. On the other hand, 
the same set of  equations  \eqref{eq:dual:phi} can be used   to connect  the  
``future" coordinate $q_{n+1,t}$ in space through  its spacial predecessors:
\begin{equation}
    q_{n+1,t}=\tilde{\phi}(q_{n,t},q_{n-1,t};q_{n,t-1},q_{n,t+1})\,. 
    \label{eq:dual:dualClassMap}
 \end{equation}
 Under the condition that  such an inversion is unique  this defines the second map  $\tilde{\Phi}$:
$(\vec{ q}_{n}, \vec{p}_{n}) \to (\vec{q}_{n+1}, \vec{p}_{n+1})$,  $\vec{q}_{n}=(q_{n,1}, \dots q_{n,T})$, $\vec{p}_{n}=(p_{n,1}, \dots p_{n,T})$  which we call  dual.
 It corresponds to  the propagation in ``space``, \ie in particle index, rather than in time.  For the case of the considered spin chain such a dual map can be defined   if $b$ and $J$ are sufficiently small as \eqref{eq:dual:dualClassMap} possesses a unique solution only in this case. Both maps are, except for some special cases (see \cite{gutkin}), quite different.  In particular $\tilde{\Phi}$ is typically  not even symplectic. Nevertheless, the  two  maps posses  one and  the same set of periodic orbits. 
Indeed,  $\tilde{\Phi}$ for a chain of length $T$  and  $\Phi$ for a  chain of length  $N$  have  the same set of fixed points for \(N\)  (respectively \(T\)) steps of dynamical evolution. In other words, both $\tilde{\Phi}$ and   ${\Phi}$ can, in principle, be used to find periodic orbits of the system. In the next section we show  how the above classical  duality  reappears   in the quantum setting.

\subsection{Quantum Duality}
\label{sec:dual:quantum}

A central object of our calculations are the traces of \(\Uh^T\) which encode  information on the quantum spectrum. 
Straightforward calculations of this quantity are not possible for long chains  due to exponentially growing matrix  dimension, \(\operatorname{dim}{\Uh}\teq (2j+1)^N\times (2j+1)^N\). 
 Even for the smallest spin quantum number \(j\teq1/2\) only spectra  of chains with  around 20 spins are easily accessible.
In \cite{gutkin} it was observed that  this problem can be, in fact, circumvented due to the exact relation
\begin{equation}
\Tr \hat{U}^T=\Tr {\Ut}^N\,
\label{eq:dual:dual}
\end{equation}
which identifies the traces of the quantum time  evolution operator  with those of a dual one \(\Ut\) of dimension \((2j+1)^T\times (2j+1)^T\). Informally speaking, the evolution operators $\hat U$  and \(\Ut\)  can be regarded as quantizations of    \(\Phi\) and  \(\tilde{\Phi}\), respectively. In contrast  to   $\hat U$  the dual operator  \(\Ut\) is in general non-unitary due to the non-symplectic  nature of its classical counterpart $\tilde{\Phi}$. Most significantly,   \(\Ut\) has a rather small  ($N$-independent)  dimension, as long as the considered time \(T\) is short. This 
   allows the  calculation of $\Tr \Uh^T$ for  small $T$ and arbitrary $N$, even if $j$ is relatively large.

In \cite{akila2} such a duality was shown for the \(j\teq 1/2\) Kicked Ising Chain.
Here, we extend it to a broad class of kicked systems with nearest neighbor interactions and arbitrary $j$. To this end  we consider the \((2j+1)^N\) dimensional product basis,
\begin{equation}
|\vec{\sigma}\rangle= |\sigma_1\rangle \otimes|\sigma_2\rangle\otimes\dots \otimes|\sigma_N\rangle\,,\label{eq:dual:basis}
\end{equation}
with discrete single particle states \(|\sigma_n\rangle \in \{|-j\rangle,\,\allowbreak|-j+1\rangle\,\allowbreak,\ldots \allowbreak|j\rangle \}\).  It is  assumed that the time  evolution of the system can be split into  two parts, \(\hat{U}\teq\hat{U}_I\hat{U}_K\), where \(\hat{U}_I\) and \(\hat{U}_K\) correspond to interaction and kick, respectively. 
Further on, we assume that the interaction part \(\hat{U}_I\) couples only nearest-neighbours, is diagonal  and in addition translation invariant with respect to the particle number in the basis \eqref{eq:dual:basis}. Its matrix elements are thus given by
\begin{equation}
\langle {\vec{\sigma}}| \hat{U}_I | {\vec{\sigma}'}\rangle
= \exp{\left( \sum_{n=1}^N f_I(\sigma_n, {\sigma}_{n+1})\right)}\delta_{\vec{ \sigma}, \vec{\sigma}' }\,,
\label{eq:dual:ui}
\end{equation}
where we introduced the  function \(f_I\) which represents the interaction between neighboring spins. For our choice of Hamiltonian, eq.~\eqref{eq:kic:hi}, it is given by
\begin{equation}
f_I(\sigma_n, {\sigma}_{n+1})
=\frac{-4\iu}{j+1/2} \left(J\sigma_n \sigma_{n+1}+\K \sigma_n^2\right)\,.
\label{eq:dual:fint}
\end{equation}
The kicking part \(\hat{U}_K\) is subjected to only one constraint that it has to be local:
\begin{equation}
\hat{U}_K = 
\bigotimes_{n=1}^N \hat{u}_K
\qquad\text{with}\quad
\langle \sigma | \hat{u}_K | \sigma' \rangle
=\eu^{f_K(\sigma,\sigma')}\,.
\label{eq:dual:uk}
\end{equation}
Here the  function \(f_K\) may be arbitrary, as long as $\hat{u}_K$ is a   unitary matrix. For the case considered in the  paper, see  \eqref{eq:kic:hk}, we obviously find:
\begin{equation}
f_K(\sigma,\sigma')=\mathrm{Ln}\ \langle \sigma| 
\exp{\left(-2\iu \, \vec{b}\cdot \hat{\vec{S}}\right)}
| \sigma' \rangle.
\label{eq:dual:fk}
\end{equation}
%It is, however, crucial that both \(f_I\) and \(f_K\) are defined on the same support, \(\{\sigma_i\}\times \{\sigma_i\}\to \mathds{C}\). The requirement of translation invariance in both cases may be relaxed at the price of ``time'' dependence in \(\Ut\).
We now introduce said dual matrix \(\Ut\teq \Uti \Utk\) via
\begin{eqnarray}
\langle \vec{\sigma}| \Uti &| \vec{\sigma}'\rangle
=& \exp{\left(\sum_{t=1}^T f_K(\sigma_t, {\sigma}_{t+1})\right)}\delta_{\vec{\sigma}, \vec{ \sigma}' }\,,
\label{eq:dual:uidual}
\\
&\Utk = &
\bigotimes_{t=1}^T \hat{w}_K
\quad\text{with}\quad
\langle \sigma | \hat{w}_K | \sigma' \rangle
=\eu^{f_I(\sigma,\sigma')}\,,
\label{eq:dual:ukdual}
\end{eqnarray}
where we exchanged the position of \(f_I\) and \(f_K\) and consider a chain  of \(T\) spins.
In contrast to the  work of \cite{gutkin}, this new operator is non-unitary, this also holds for $\Uti$ and $\Utk$  separately.
For the model at hand, it   can be given in a more explicit form. The new interaction part retains a diagonal structure,
\begin{equation}
\langle{\vec{\sigma}}| \Uti | {\vec{\sigma}'}\rangle
=\delta_{\vec{\sigma}, \vec{ \sigma}' }
\prod_{t=1}^T
\langle \sigma_{t}| \exp\big(-2\iu \, \vec{b}\cdot \hat{\vec{S}}\big)\, | \sigma_{t+1} \rangle
\,,
\label{eq:dual:uti}
\end{equation}
which is fully determined  by the  kick part of the original model. Contrary, the original interaction provides the form of the dual kick:
\begin{equation}
\Utk = 
\bigotimes_{t=1}^T w_K\,,
\quad
\langle \sigma' | w_K | \sigma \rangle
=\exp{\left(\frac{-4\iu \left(J \sigma \sigma'+\K \sigma^2\right)}{j+1/2}\right)}\,.
\label{eq:dual:utk}
\end{equation}
We discuss the  spectrum of $\Uti \Utk$ in section \ref{sec:spec}.

To recover the trace duality we  rewrite  the traces on the left side of eq.~\eqref{eq:dual:dual} as 2D partition function  by inserting identities for the different times \(t\),
\begin{equation}
\Tr \hat{U}^T
=\sum_{\{\vec{\sigma}(t)\}}
\langle \vec{\sigma}(1)| \hat{U} |\vec{\sigma}(T) \rangle\;
\langle \vec{\sigma}(T) | \hat{U} |\vec{\sigma}(T-1) \rangle\;
\langle \vec{\sigma}(T-1) | \ldots | \vec{\sigma}(1) \rangle\,,
\label{eq:dual:traceDecomp}
\end{equation}
and expressing this further as a sum over all possible combinations of \(\sigma_{n,t}\in \{-j,\dots,j\}\) per time--step and spin index,
\begin{equation}
\Tr \hat{U}^T =\sum_{\sigma_{n,t}\in \{-j,\dots,j\}}
\exp\bigg(\iu \sum_{n=1}^N \sum_{t=1}^T
 f_I(\sigma_{n,t},\sigma_{n+1,t} )+ f_K (\sigma_{n,t},\sigma_{n,t+1} \bigg)\,.
\label{eq:dual:trace2Dpart}
\end{equation} 
Since the  result  is symmetric under the exchange $n\leftrightarrow t$, $N\leftrightarrow T$, $f_I\leftrightarrow f_K$, an  analogous procedure leads to the same  expression    for \(\Tr\Ut^N\). In the context of 2D classical Ising models the operators $\hat U$ and $\Ut$ are nothing more than transport  operators along  the ``temporary"  and ``spatial" directions  which express the partition function in two different ways.

Finally, let us comment on a certain peculiarity of  the integrable case (\(b^x=0\)). Due to the identity $ \hat{U}(J, \vec{b})^T=\hat{U}(TJ, T\vec{b})$ the evolution for $T$ time steps can be equivalently thought of as one for a single time step with rescaled parameters. Therefore, the dual operator always takes on the form of a $(2j+1)\times (2j+1)$ matrix (rather than $(2j+1)^T\times (2j+1)^T$) for a single spin system:  
\begin{equation}
\Ut_{nm}=
\exp \left( -\iu \frac{4JT}{j+1/2}nm 
-\iu \frac{4J\K }{j+1/2}m^2
 - 2\iu T b^z n \right)\,,
 \label{eq:dual:utint}
\end{equation}
where the indices \(n,m\) run from \(-j\) to \(+j\).

%%%%%%%%%%%%%%%%%%%%%%%%%%%%%%%%%%%%%%%%%%%%%%%%%%%%%%%%%%%
% Action Spectrum
%%%%%%%%%%%%%%%%%%%%%%%%%%%%%%%%%%%%%%%%%%%%%%%%%%%%%%%%%%%
\section{Action Spectrum}
\label{sec:action}

For quantum Hamiltonian systems the underlying classical POs can be  revealed by taking an appropriate Fourier transform of the  spectral  density with respect to  an energy like   parameter \cite{stoeckmann,wintgen3,stockStein,welge}. For  quantum maps, however, energy is not defined. Still,  it is possible to   extract classical POs   out of traces of the quantum evolution taking a Fourier transform over the inverse of the effective Planck's constant, see e.g.,  \cite{haake,haakePRL1,kus}. The linear scaling with the spin quantum number $j$ in the exponent of \eqref{eq:action:TrGutzwill} makes it a good quantity for such a Fourier transformation. Employing this procedure we obtain
\begin{eqnarray}
    \rho(\Sact)&=&\frac{1}{\jcut}\sum_{j=1}^{\jcut}\eu^{-\iu (j+1/2)\Sact}\Tr\,\hat{U}^T
\nonumber
    \\
    & \sim&\,\frac{1}{\jcut}\sum_{\gamma(T)} A_\gamma\,\delta_{\jcut}(\Sact-\Sga)\,,
     \label{eq:action:spec}
\end{eqnarray}
where $ \delta_{\jcut}$ stands for a periodized approximation of th \(\delta\)-distribution with width \(\sim \pi/\jcut\) and height \(\jcut\). The cut-off \(\jcut\) is introduced  in order to keep the dimension of \(\hat{U}\), or more precisely \(\Ut\), numerically accessible. As \(\jcut\to\infty\)  the function  $\rho(\Sact)$   resolves the classical orbit actions \(\Sga\)  for  orbits of period \(T\),  up to a modulus of \(2\pi\). 
It is worth noting that for technical reasons  the sum in \eqref{eq:action:spec} is restricted to  integer values of \(j\). The inclusion of half-integer  \(j\)  allows, in principle, the resolution of \(\Sga\) up to a modulus of  \(4\pi\). We do not pursue this issue further.

In this  section  we numerically   study the  action spectrum $\rho(\Sact)$  for one and two time steps.  To this end we first  evaluate the spectrum of the dual operator $\Ut$ for  $T=1,2$  and then calculate  $\Tr \hat {U}^T$ using  the duality relation  \eqref{eq:dual:dual}. As a result,  we are able to obtain  $\rho(\Sact)$ for an arbitrarily large spin chain and some finite $\jcut$.
%The cut-off  parameter $\jcut$ determines  the maximum   dimensions \((2\jcut+1)^{T}\times (2\jcut+1)^{T}\)  of $\Ut$  and its value  is restricted by the condition that   spectrum of $\Ut$   remains  numerically accessible. 

\subsection{The effect of bifurcations}
\label{sec:action:sc}

%eventually thesis, but I don't really like it:Consider \eqref{eq:po:abstractDef} which describes periodic orbits as roots of some equation \Onote{assuming the equation is of a polynomial type}. Under change of the system parameters the roots may collide and drift into the complex plane. An over-simplified example would be \(x^2-a\teq 0\) whose two solutions are either real \(a>0\) or complex \(a<0\).
%At the bifurcation point \(a\teq 0\) both solutions coincide. For the classical system this implies a break down of linear stability as both orbits come arbitrary close. Necessarily, some of the orbits Lyapunov exponents are \(0\) creating a divergence in \(A_\gamma\) as given by \eqref{eq:action:stability}.
%This is amended taking higher order corrections into account, \Onote{for few degree of freedom systems} they are given by uniform approximations \gcite. The adjusted \(A_\gamma\) contains a \(j^\alpha\)\Onote{check exponent parameter for consistent labeling, later} dependent scaling correction which depends on the type of bifurcation, \ie the number of roots involved.
%\Onote{It is, however, unclear how this scaling law depends on the degrees of freedom \(N\).}

\begin{figure}[tbh]
    \includegraphics[width=0.45\textwidth]{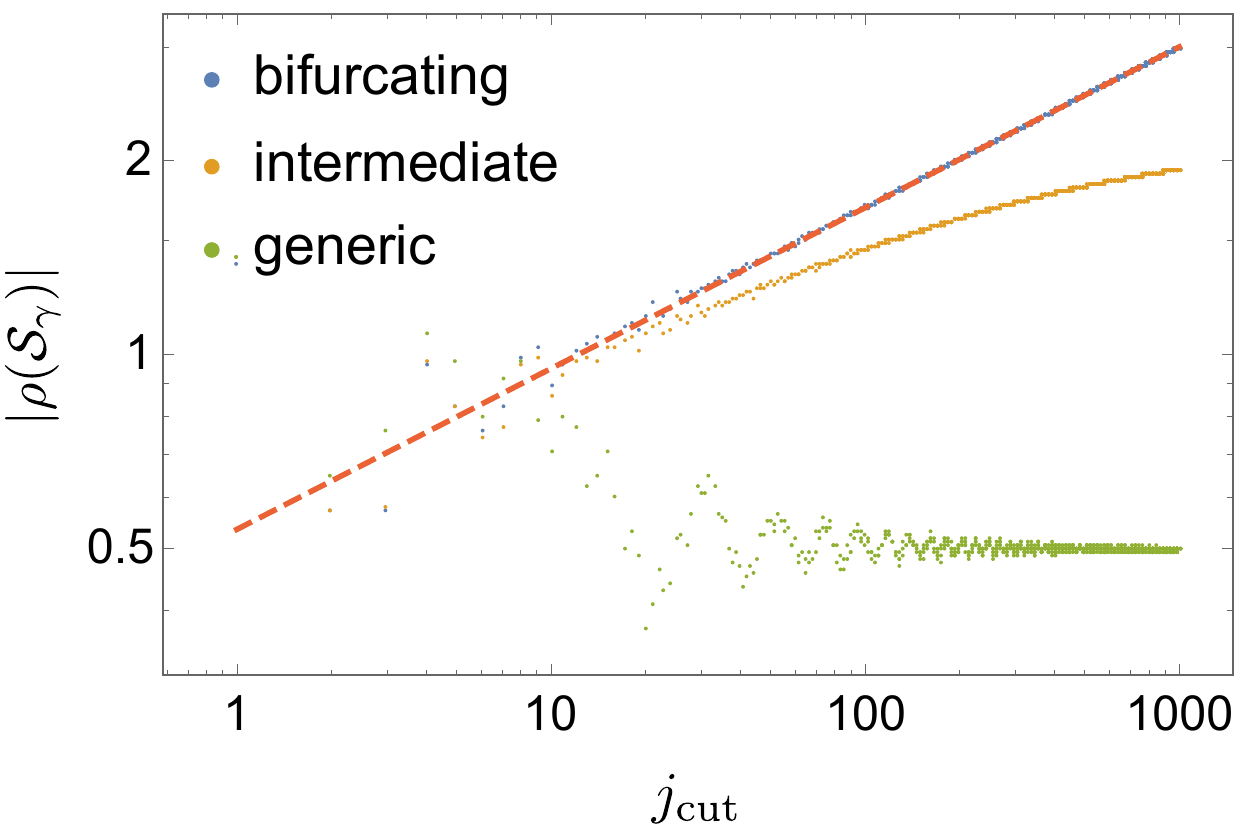}
    \hfill
    \includegraphics[width=0.45\textwidth]{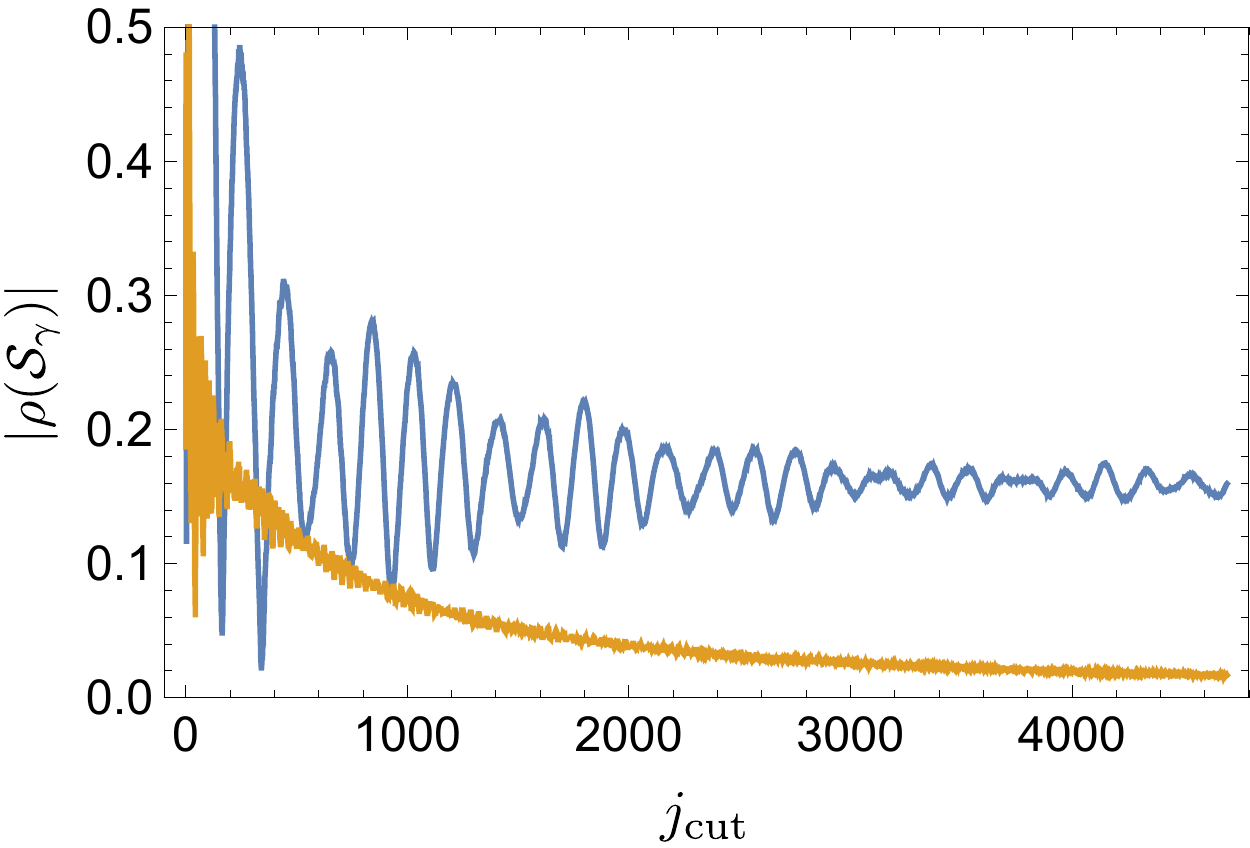}
    \caption[Scalings of the action spectrum at specific orbit positions, with and without bifurcations.]{
        \conline\textit{Left hand side:} Scaling of the action spectrum peak height \(|\rho(\Sga)|\) for \(N\teq 1, T\teq 2\) over the cut-off parameter \(\jcut\). Shown are three cases, for a position which features an (isochronous) pitchfork bifurcating orbit (\(J\teq 0.7\), \(b^x\approx 0.94\) and \(b^z\approx 0.90\)), the same orbit with slightly detuned couplings (\(J\teq 0.68\)) and a generic (\ie isolated) orbit for the detuned parameters.\\
        \textit{Right hand side:} Dependence of the height of \(|\rho(\Sga)|\) on the cut-off parameter \(\jcut\) for two selected orbits of the action spectrum for \(N\teq 7\) particles given in figure \ref{fig:action:sftT1}. The selected orbit shown in blue is the largest one in figure \ref{fig:action:sftT1} at \(\Sga\approx 5.77\), which also shows the strongest deviations from the semiclassical prediction. The other one (orange) is a small ghost orbit at \(\Sga\approx 2.75\) for the same parameters. While the first one saturates, for large $\jcut$, to a limiting value, the ghost decays exponentially.
    }
    \label{fig:action:pitchForkScaling}
    \label{fig:action:T1jcutDep}
\end{figure}

Recall that for  isolated POs    the prefactors $\Aga$ are given by \eqref{eq:action:stability}. Accordingly, if $\det (\Mga -\mathds{1})\neq 0$ for all POs of period $T$ the function  $\rho(\Sact)$  does not  scale with $\jcut$. On the other hand, when a periodic orbit changes its stability type, one of the corresponding eigenvalues turns into \(\eig_i\teq 1\). This immediately implies a divergence of \eqref{eq:action:stability}. In this case the linearized dynamics in terms of $\Mga$ is insufficient to describe the weight of an orbit to the sum in \eqref{eq:action:TrGutzwill}. Instead, higher orders have to be taken into account in the form of uniform approximations \cite{cat_schomerus,cat_bestPaper}. The adjusted \(A_\gamma\)  has the  scaling   \(j^\alpha\),  where the exponent $\alpha >0$  depends on the type of bifurcation. 
To demonstrate how bifurcations affect the action spectrum \(\rho(\Sact)\) we show, as an example,   a (isochronous) pitchfork bifurcation for single particle systems on the left hand side of figure  \ref{fig:action:pitchForkScaling}, where  the algebraic scaling  \(\alpha\teq 1/4\)  is clearly observed. In addition, we provide a slightly detuned system which shows at first algebraic growth and for larger \(\jcut\) tends towards saturation.

So far,  studies of the bifurcation effects  on the quantum  spectrum   have been mostly restricted to systems with a single degree of freedom \cite{cat_bestPaper,cat_manderfeld,cat_schomerus,cat_biff}.
While an exact bifurcation is a singular  event nearly bifurcating orbits with \(\eig_i\approx 1\) are generic  in many-body systems with mixed dynamics. In general, for $N$-body  systems the number of elliptic directions increases with the number of degrees of freedom $N$. Assuming that phases of the corresponding (elliptic) eigenvalues of $\Mga$ are distributed uniformly, the probability to come close to one should grow with $N$.       While one might argue that in the limit \(j\to\infty\)    
equation \eqref{eq:action:stability} must be  recovered for such  nearly bifurcating orbits,  this is only true for the pure semiclassical limit with  fixed \(N\). In practice 
this is never the case as \(j\) is necessarily finite.
In other words,  for a limit where both  $N$ and $j$ tend to infinity the prefactors $\Aga$ (resp. $\rho(\Sact)$) might still possess  a non trivial scaling $j^{\alpha(N)}$ due to the presence of quasi-marginal directions.

%Instead, one needs to take into account that orbits can not be approximated by isolated Gaussians but instead by multi-modal distributions over nearby extrema, giving rise to similar \(A_\gamma\) scaling corrections in the vicinity of bifurcations \cite{cat_manderfeld,cat_schomerus}\qm. This statement holds independent of whether the periodic orbit solutions are real or complex, the latter leading to ghost orbits which decay in the limit \(j\to \infty\)~\cite{kus}\qm. 
%For practical purposes we consider orbits not fully isolated if for one of the stabilities holds \(|\ln{\Lambda_i} |<1\).

%%%%%%%%%%%%%%%%%%%%%%%%%%%%%%%%%%%%%%%%%%%%%%%%%%%%%%%%%
% Action:T1
%%%%%%%%%%%%%%%%%%%%%%%%%%%%%%%%%%%%%%%%%%%%%%%%%%%%%%%%%
\subsection{Single Time Step}
\label{sec:action:T1}

\begin{figure}
    \centering
    \includegraphics[height=3cm]{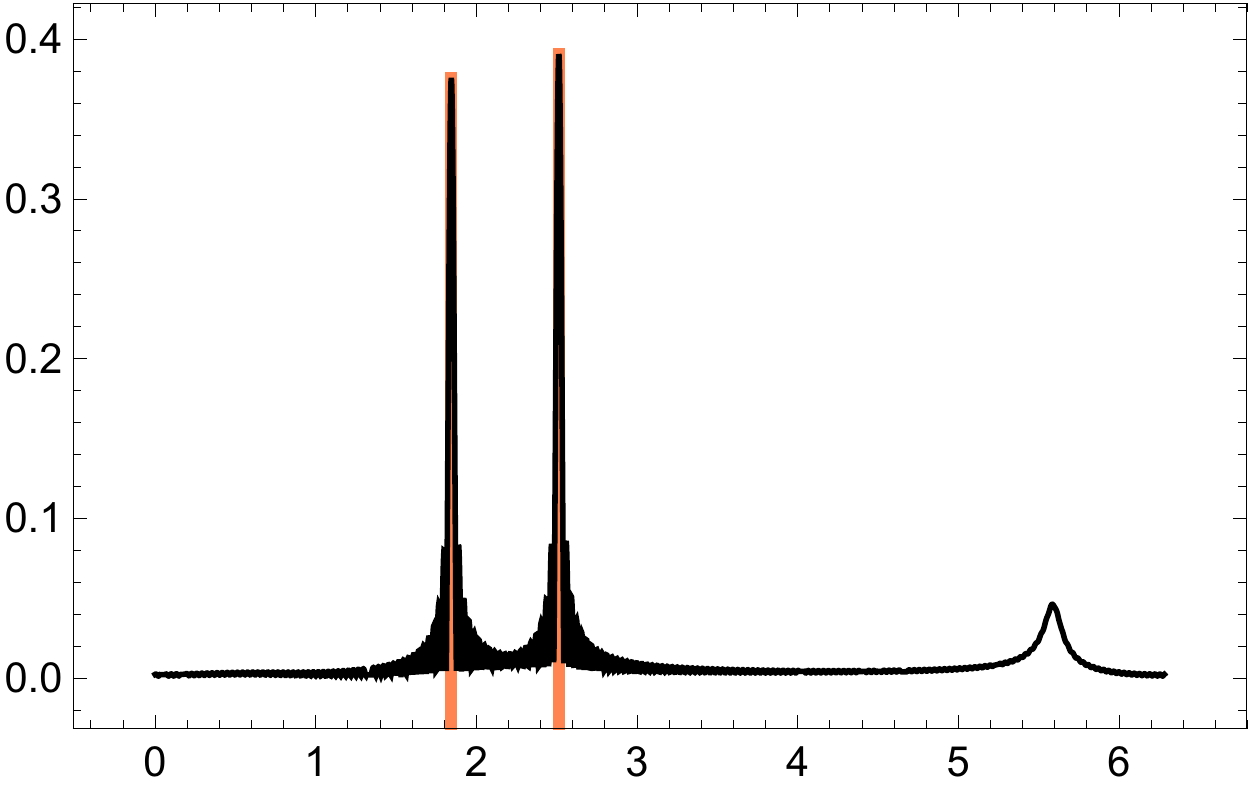}
    \hfill
    \includegraphics[height=3cm]{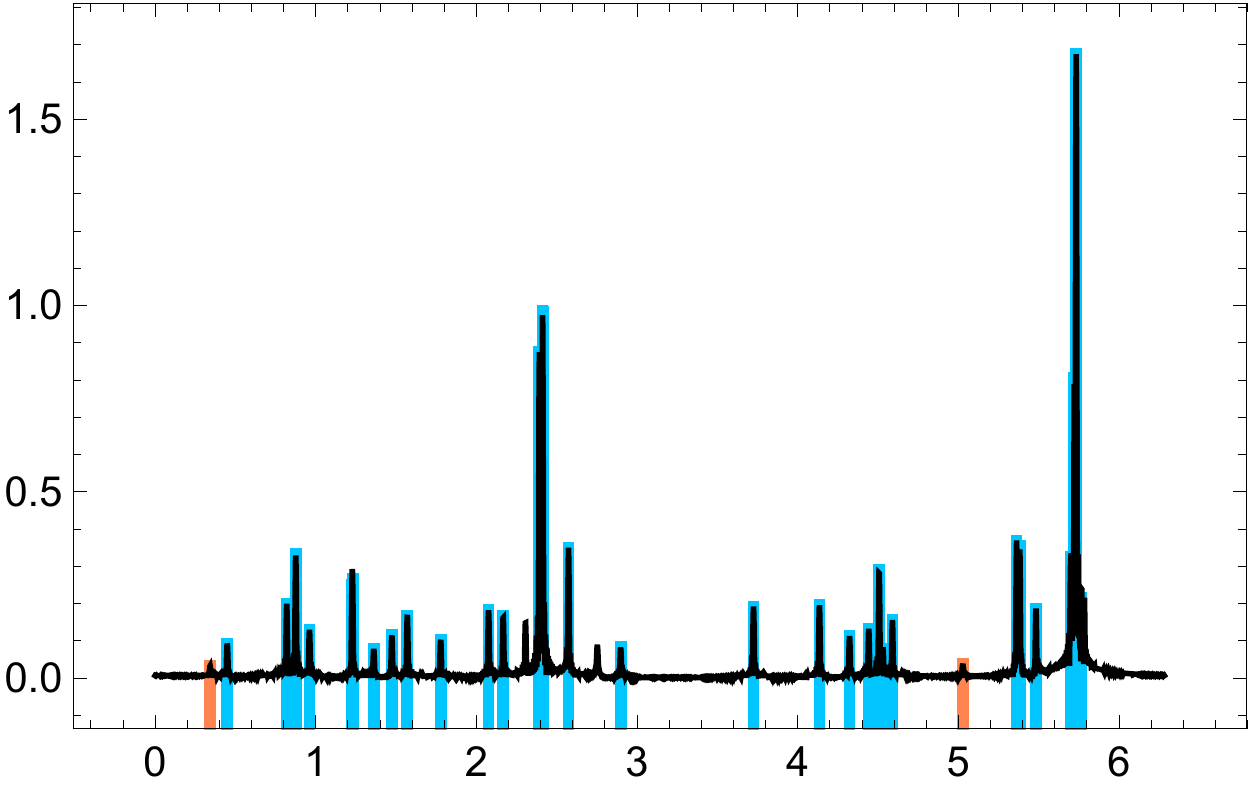}
    \hfill
    \includegraphics[height=3cm]{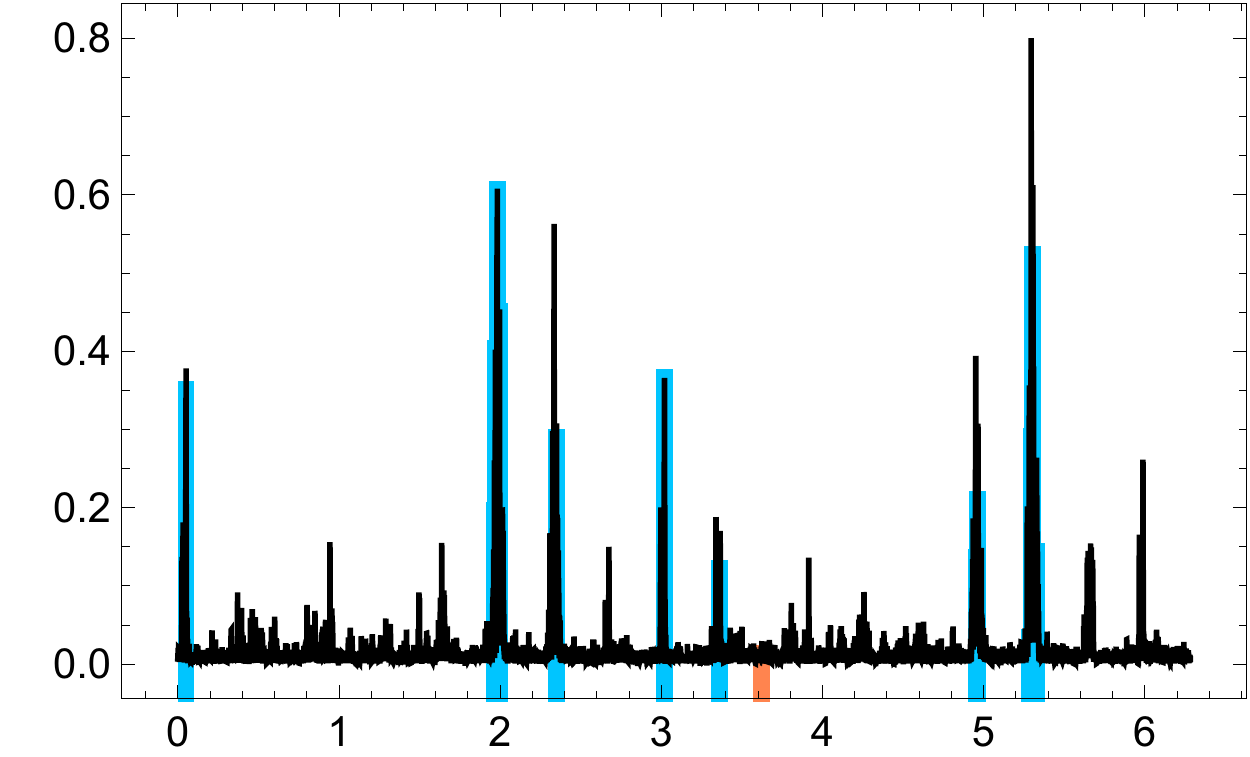}
    \\
    \includegraphics[height=3cm]{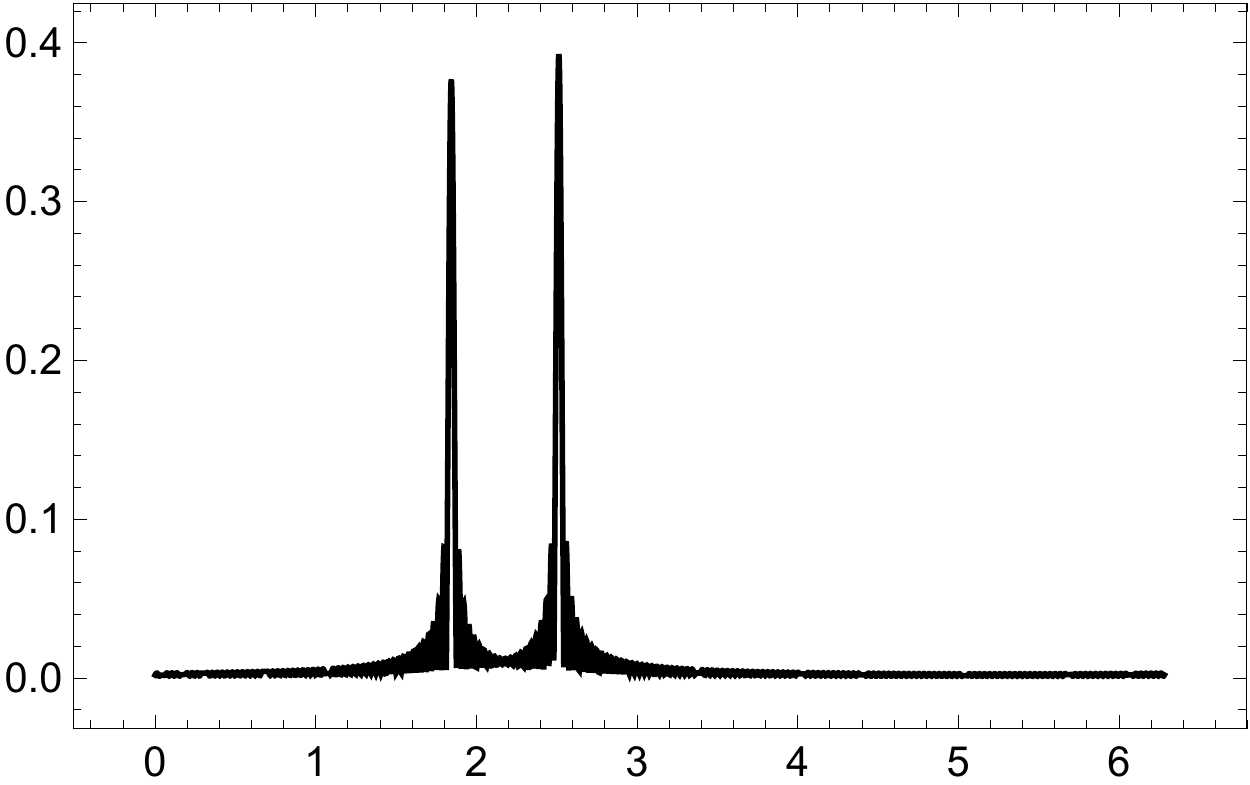}
    \hfill
    \includegraphics[height=3cm]{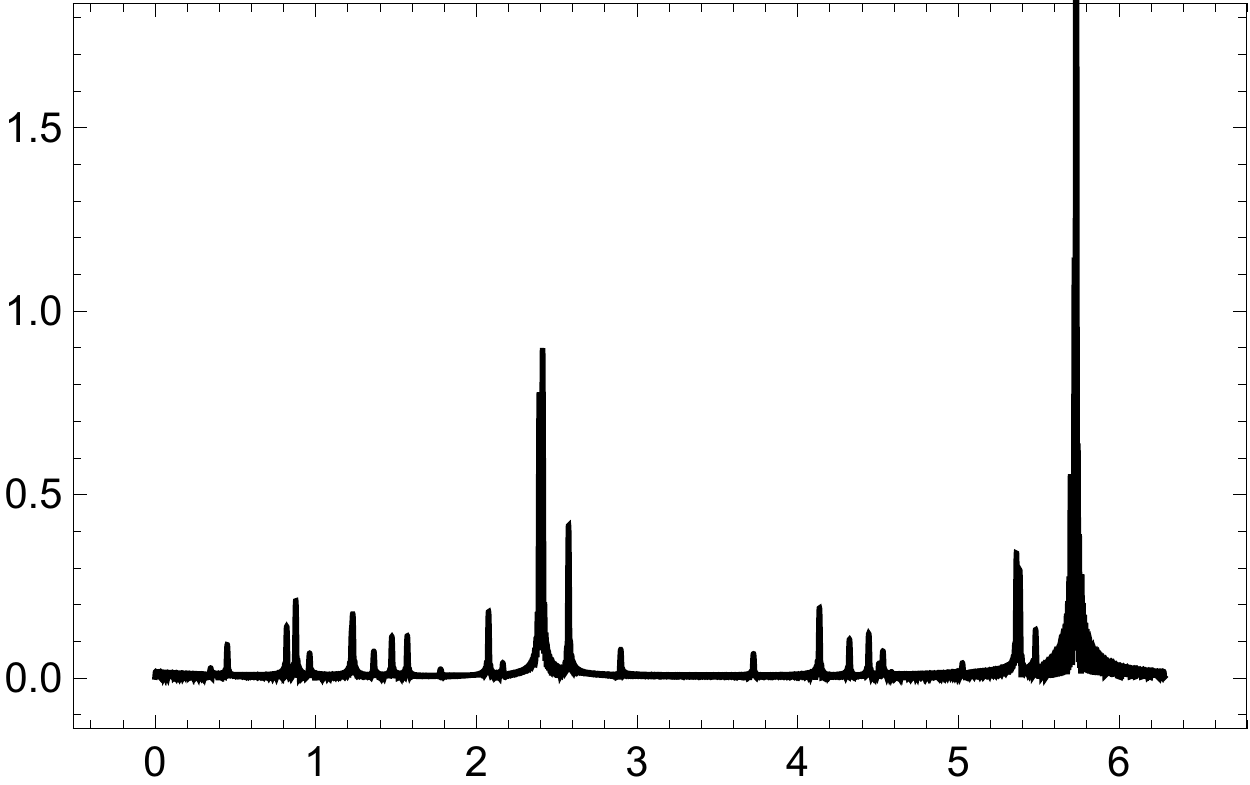}
    \hfill
    \includegraphics[height=3cm]{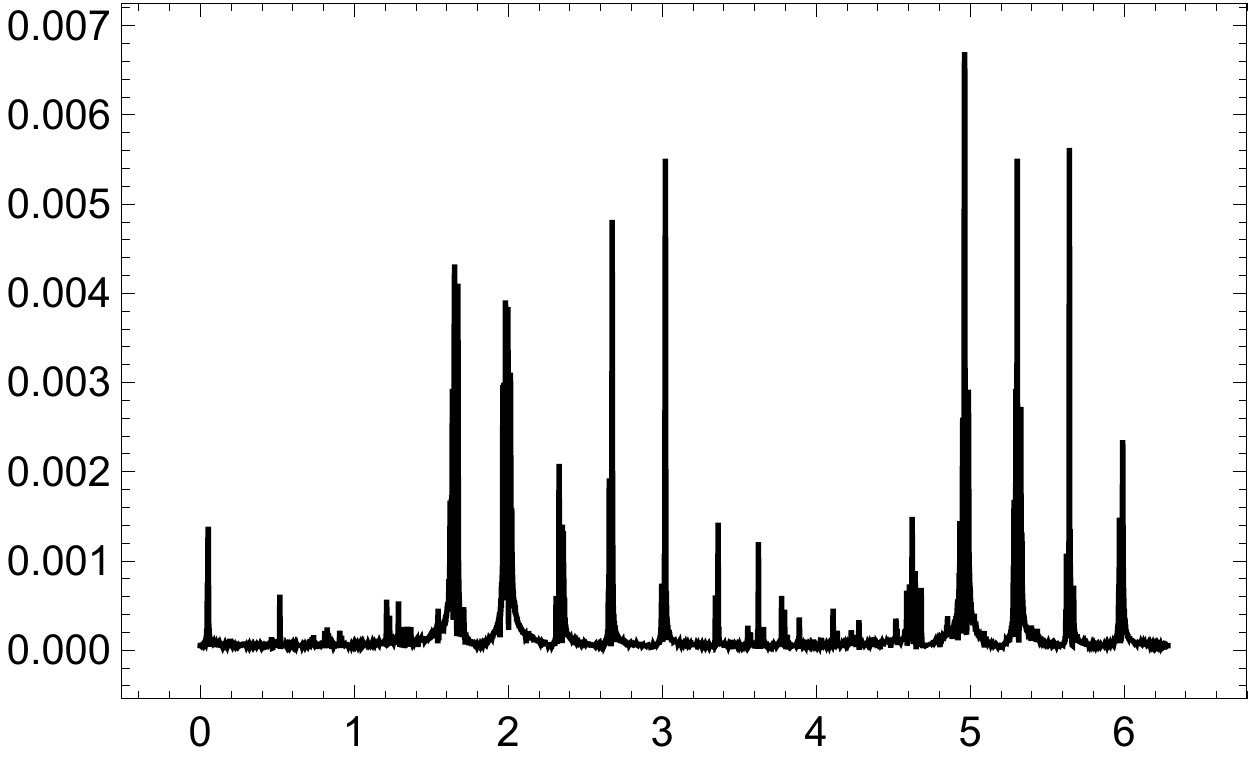}
    \caption[Action Spectrum for $T\teq 1$ and $N\teq 1,7,19$ in comparison to a semiclassical prediction.]{\conline Absolute value \(|\rho(\Sact)|\) of the approximate action spectrum, eq.~\eqref{eq:action:spec}, over \(\Sact\) for different particle numbers, \(N\teq 1,7,19\) (from left to right), and a single time step. The upper row corresponds to the actual quantum data while in the lower row traces are repalced by a semiclassical approximation. System parameters are the same for all panels, \(J\teq 0.7\) and \(b^x \teq b^z \teq 0.9\), but the cut-off parameter is chosen differently to resolve larger particle numbers, \(\jcut\teq 200, 801, 4700\) (left to right). Colored bars in the upper row correspond to the position of classical orbits, the color specifies the spatial period: $\Nprim\teq 1$ (orange), $\Nprim\teq N$ (blue).}
    \label{fig:action:sftT1}
\end{figure}

In the case of \(T\teq 1\) the spectrum  of \(\Ut\) can be easily calculated  for a relatively large cut-off parameter  \(\jcut \sim 10^4\), while  the number of periodic orbits grows weaker with \(N\) in comparison  to longer times. This allows  a good resolution of the action spectrum for moderate spin chain lengths and isolated POs. Figure \ref{fig:action:sftT1} shows the (absolute) action spectrum \(|\rho(S)|\), see \eqref{eq:action:spec}, for identical parameters but different numbers of spins. The upper row depicts numerical calculations based on the spectrum of the  dual quantum operators and  colored bars therein mark the positions  of classical periodic orbits. For comparison the lower row contains a semiclassical approximation for which we use the right hand side of \eqref{eq:action:TrGutzwill} instead of the actual traces in \eqref{eq:action:spec}. In contrast to the upper row this one relies solely on classical information --   actions $\Sga$ of the POs and their  stabilities $\Aga$ provided by eq.~\eqref{eq:action:stability}. 

\subsubsection{$N=1$.} The left panels shows the single particle case of the Kicked Top, which features only two periodic orbits for such short times. The broader peak to the right is a ghost orbit (emerging for larger \(J\)) which is naturally not reproduced  in the lower panel. Otherwise the agreement   is excellent.

\subsubsection{$N=7$.}  The middle panel shows $|\rho(\Sact)|$ for \(N\teq 7\) spins, containing significantly more orbits with very good agreement between the classical positions of their actions and the corresponding   peaks of \(|\rho(S)|\).  As \(N\) is prime these orbits necessarily possess either \(\Nprim\teq 1\)  or \(\Nprim\teq 7\)  marked by different colors in fig.~\ref{fig:action:sftT1}. Naturally, POs with  \(\Nprim\teq 1\)   are just  repetitions  of POs encountered in the $N=1$ case. 
%Still, the former are present in the spectrum but play only an insignificant role.
%Recalling section \ref{sec:dual:linearized} this appears to be a generic feature for \(T\teq 1\) as most dual stabilities \(\eigd_i\) are hyperbolic leading to an exponential decay of \(\Aga\) with \(N\).

Comparison to the semiclassical approximation shows good agreement for approximately half of the POs, but the others exhibit  some  deviations in height due to the proliferating bifurcations. This is most apparent for the highest  peak at \(\Sga\approx 5.77\)  (which height is deliberately cut in the lower panel). The  particular PO contains $6$ elliptic, $4$ mixed and $4$ purely hyperbolic directions,  with \(|\ln \Lambda|\approx 0.86\) bringing it sufficiently  close to a bifurcation. To check it in more details  we take a look at the peak heights  as a function of  \(\jcut\), see the right hand side of fig.~\ref{fig:action:T1jcutDep}. Indeed, the function shows strong  oscillations   due to the existence  of accompanying  orbits with close actions and  saturation is achieved only for considerably high values of \(\jcut\). 

\subsubsection{$N=19$.} For the right hand panels in figure \ref{fig:action:sftT1} the number of spins is increased further to \(N\teq 19\). In this case we are no longer able to resolve individual orbits despite an increased \(\jcut\).
%Figure \ref{fig:action:T1zoom} displays a magnified region of the action spectrum showing some correlation between found classical orbits and the oscillations in the approximate spectrum. 
The semiclassical approximation, based on \(\sim2000\) found orbits, resembles the actual function \(|\rho(S)|\) only for some of the largest peaks. Given the huge amount of underlying POs the clear structure of the action spectrum with only a few dominant peaks is quite remarkable. The positions of  these   peaks indeed correspond to  the POs with the largest  prefactors  \(\Aga\). In fig.~\ref{fig:action:sftT1} we mark  only orbits $\gamma$ which surpass a  fixed threshold,  \(|\det{(  \underline{M}_\gamma-\mathds{1} )} |^{-1/2} > 10^{-3}\).
%\(\Aga>10^{-3}\).
As  one can see, their positions coincide with the largest 
spikes of \(|\rho(S)|\).
%old thresh.: \(\Aga^{-2}\propto|\det{(  \underline{M}_\gamma-\mathds{1} )} |< 10^6\)

%From a semiclassical point of view it is not clear why such orbits should give a dominant contribution as their magnitude appears small. The largest orbit found for the used parameters has a stability of \(\Aga\approx 0.003\), which is still insignificant compared to the observed height of peaks. An alternative explanation might be a bunching of orbits, such that the peaks are not given by a single orbit but instead by several of them with close actions. Figure \ref{fig:action:T1actionHist} shows the distribution of orbits over the action partially backing this assumption. One should point out that for high dimensional systems close actions do not necessarily imply closeness in phase-space. It seems reasonable that a mixture of both, bunching and ``large'' stabilities, causes the peaks. \Onote{Should tell something about system dynamics. Maybe there are more ``stable'' regions that carry the bunching orbits?}

% \begin{figure}
% \centering
% \includegraphics[width=0.45\columnwidth]{WLP_exponentScalingOnly4v2}
% \caption{Scaling exponent \(\alpha\) of $|\rho(S_\gamma)|\sim(j_{\rm cut})^\alpha$
%   at the position \(S_\gamma\) of the largest peak,
% versus the number of spins. For \(T\teq 1,2\), \(J\teq 0.7\), \(b^x\teq b^z \teq 0.9\) and \(N\teq 4k\).}
% \label{fig:alphascale}
% \end{figure}

%%%%%%%%%%%%%%%%%%%%%%%%%%%%%%%%%%%%%%%%%%%%%%%%%%%%%%%
% Action:T2
%%%%%%%%%%%%%%%%%%%%%%%%%%%%%%%%%%%%%%%%%%%%%%%%%%%%%%%
\subsection{Two Time Steps}
\label{sec:action:T2}

\begin{figure}
    \includegraphics[width=0.4\textwidth]{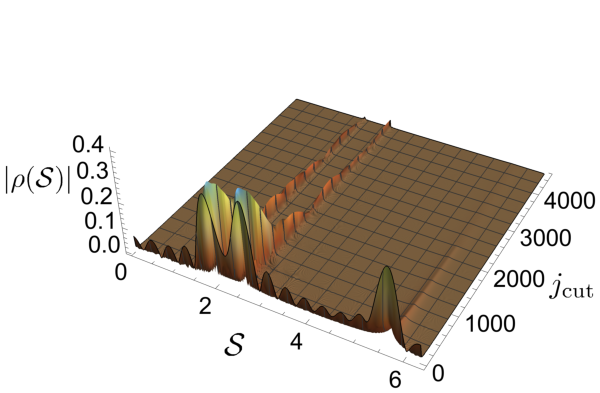}
    \hspace{2cm}
    \includegraphics[width=0.4\textwidth]{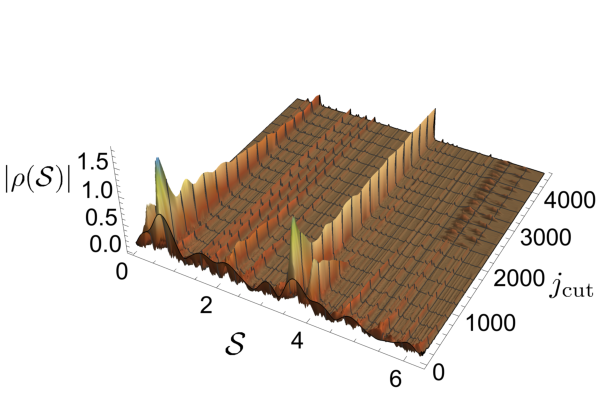}
    \\
    \includegraphics[width=0.4\textwidth]{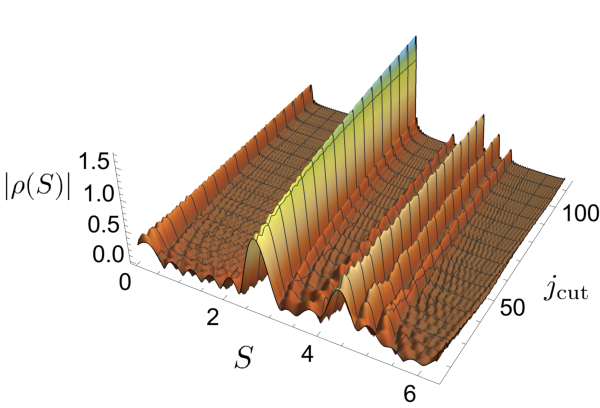}
    \hspace{2cm}
    \includegraphics[width=0.4\textwidth]{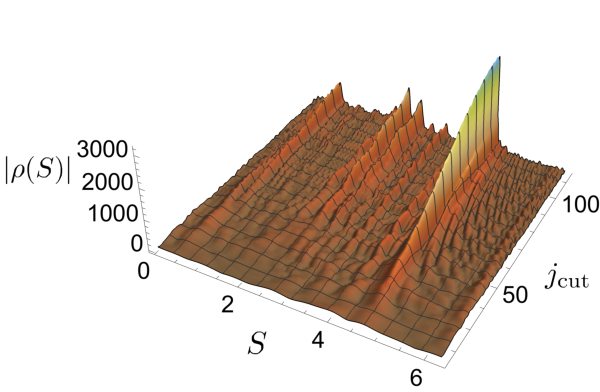}
    \caption[Action Spectrum over both $\Sact$ and $\jcut$ for $N\teq 1, 10$ and $T\teq1,2$]{\conline Approximate action spectrum plotted over the action \(\Sact\) and the chosen cut-off parameter \(\jcut\). Upper row corresponds to \(T\teq 1\) lower to \(T\teq 2\), left column features \(N\teq 1\), right \(N\teq 10\). Parameters are given by \(J\teq 0.7\) and \(b^x\teq b^z \teq 0.9\). \(\jcut\) is necessarily integer, the graphic shows an interpolation. Noise in the upper right panel is caused by computational difficulties of the visualization.}
    \label{fig:action:sFTscaling3D}
\end{figure}

For  two time steps (\(T\teq 2\)) the  number of POs  is substantially larger in  comparison  to the \(T\teq 1\) case with the same parameters. Furthermore, many of them are close  to bifurcations, making a semiclassical reconstruction of the spectrum even harder.  In addition, significantly smaller achievable values of \(\jcut\) limit our resolution.
We  illustrate this with a direct comparison of  the $T=1$ and $T=2$ cases  in figure \ref{fig:action:sFTscaling3D} where we depict the action spectrum using \(\jcut\) as additional variable. For the \(T\teq1\) cases in the upper row we reach sufficiently high values of \(\jcut\) to observe both the initial interference of nearby orbits and  saturation of $|\rho(S_\gamma)|$ for larger  \(\jcut\).  As stated previously the \(N\teq 1\) case (left) also contains a separated ghost orbit whose decay becomes apparent in this visualization. For the  \(T\teq2\) case   it is no longer feasible to resolve such scales and we, instead, only observe  the initial growth associated with  close to bifurcation orbits.
%While this is hindering direct semiclassical analysis with respect to the peak heights section \ref{sec:nlkic} demonstrates that these divergences help single out specific orbits compared to fully hyperbolic systems.

\begin{figure}[tbp]
\includegraphics[width=0.32\textwidth]{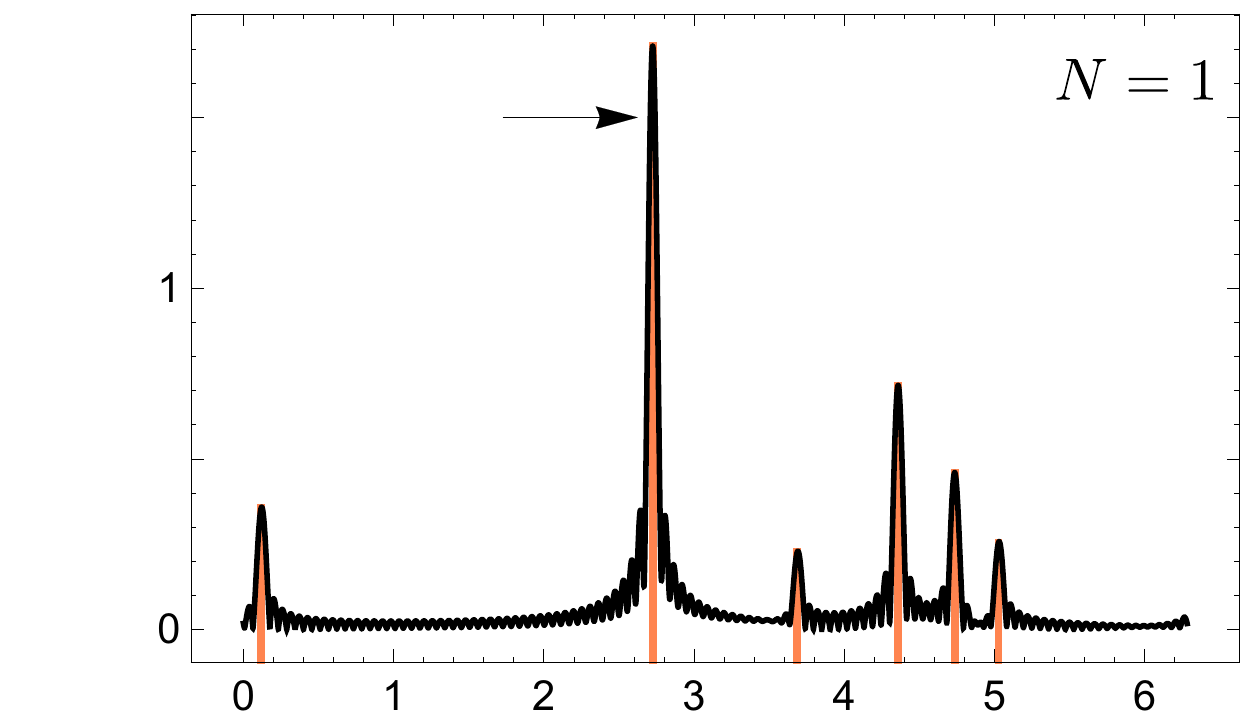}
\hfill
\includegraphics[width=0.32\textwidth]{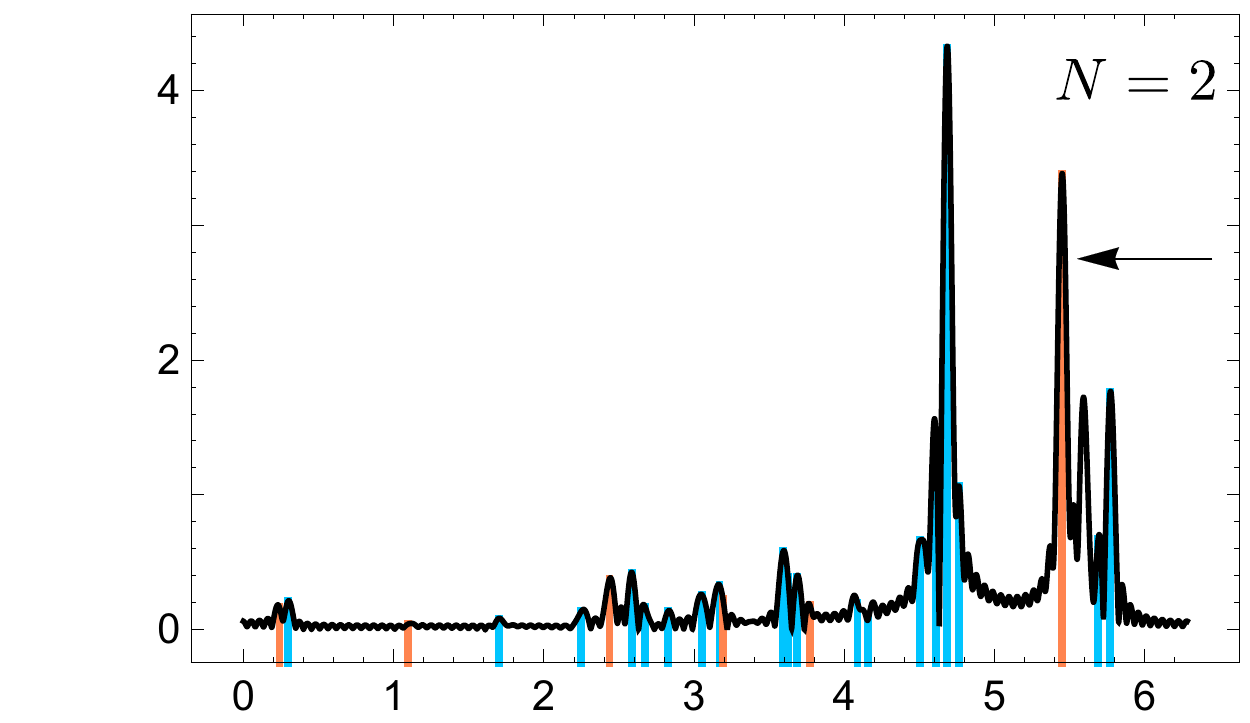}
\hfill
\includegraphics[width=0.32\textwidth]{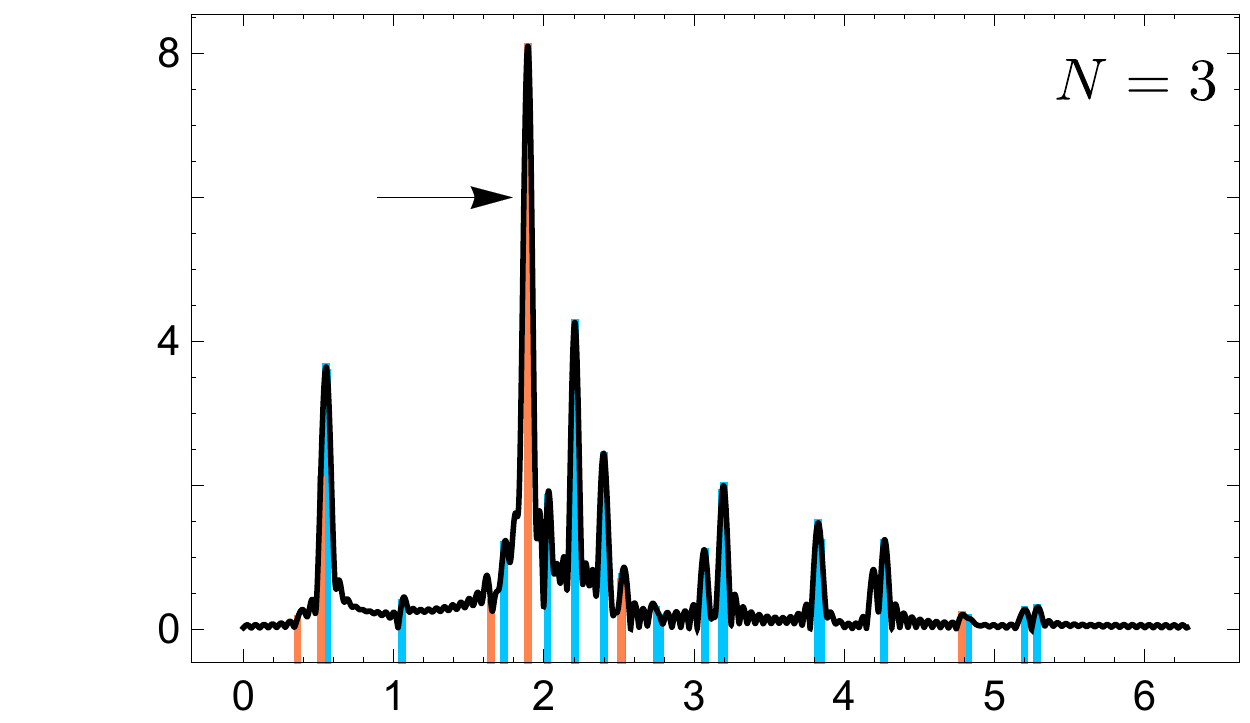}
\\
\includegraphics[width=0.32\textwidth]{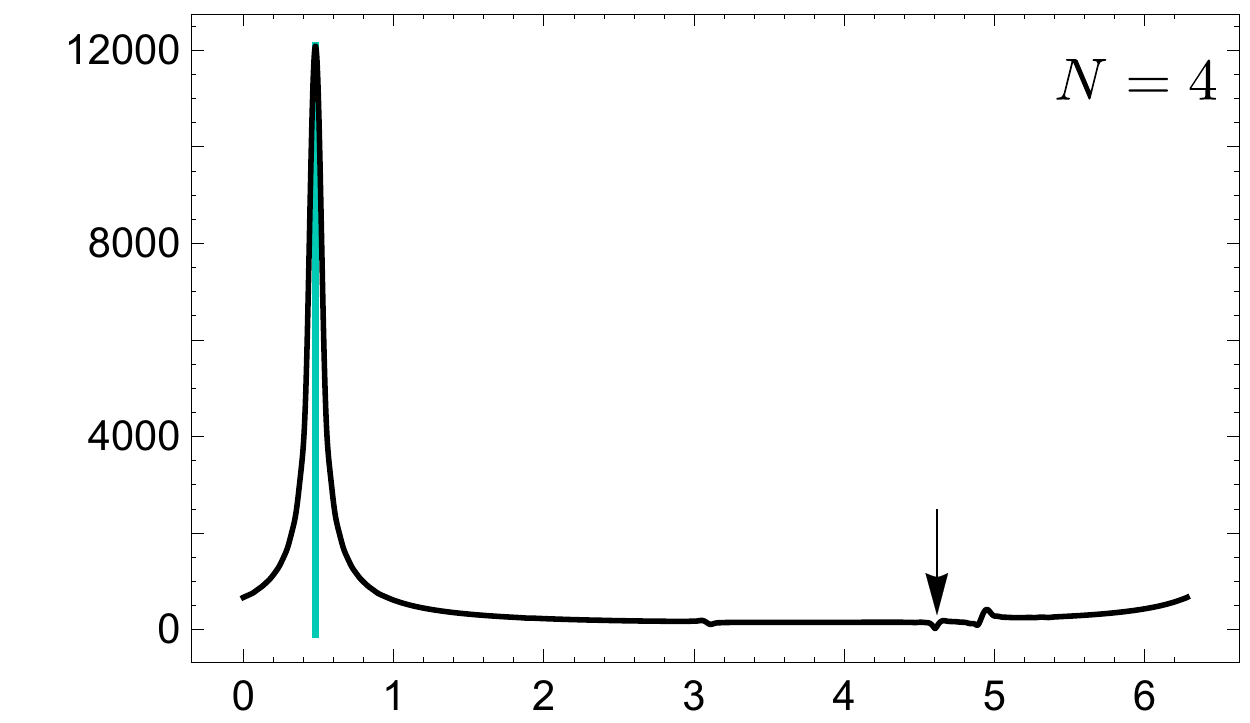}
\hfill
\includegraphics[width=0.32\textwidth]{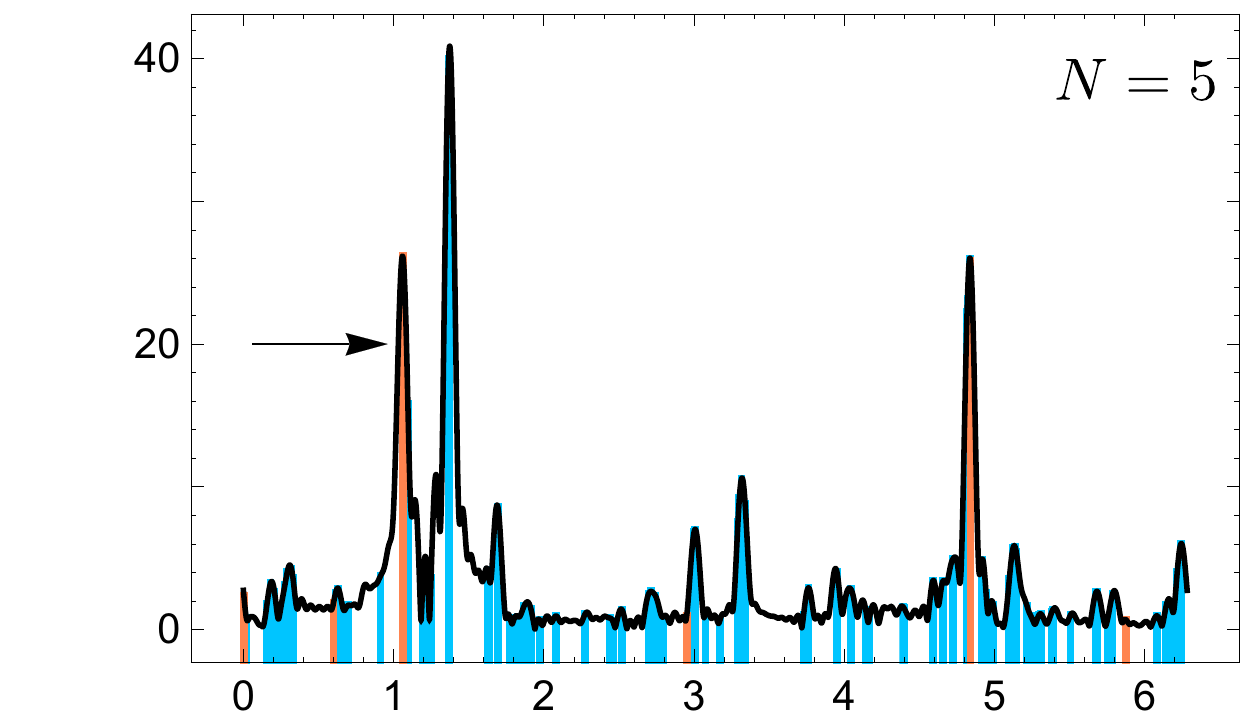}
\hfill
\includegraphics[width=0.32\textwidth]{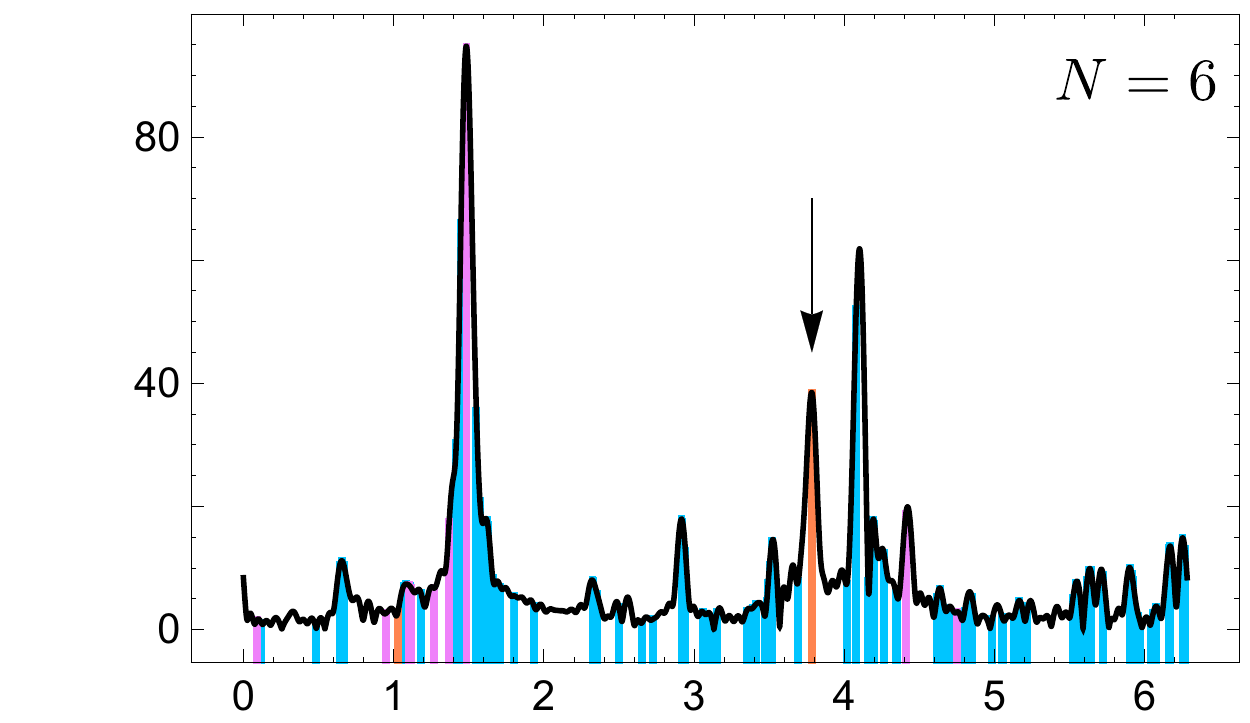}
\\
\includegraphics[width=0.32\textwidth]{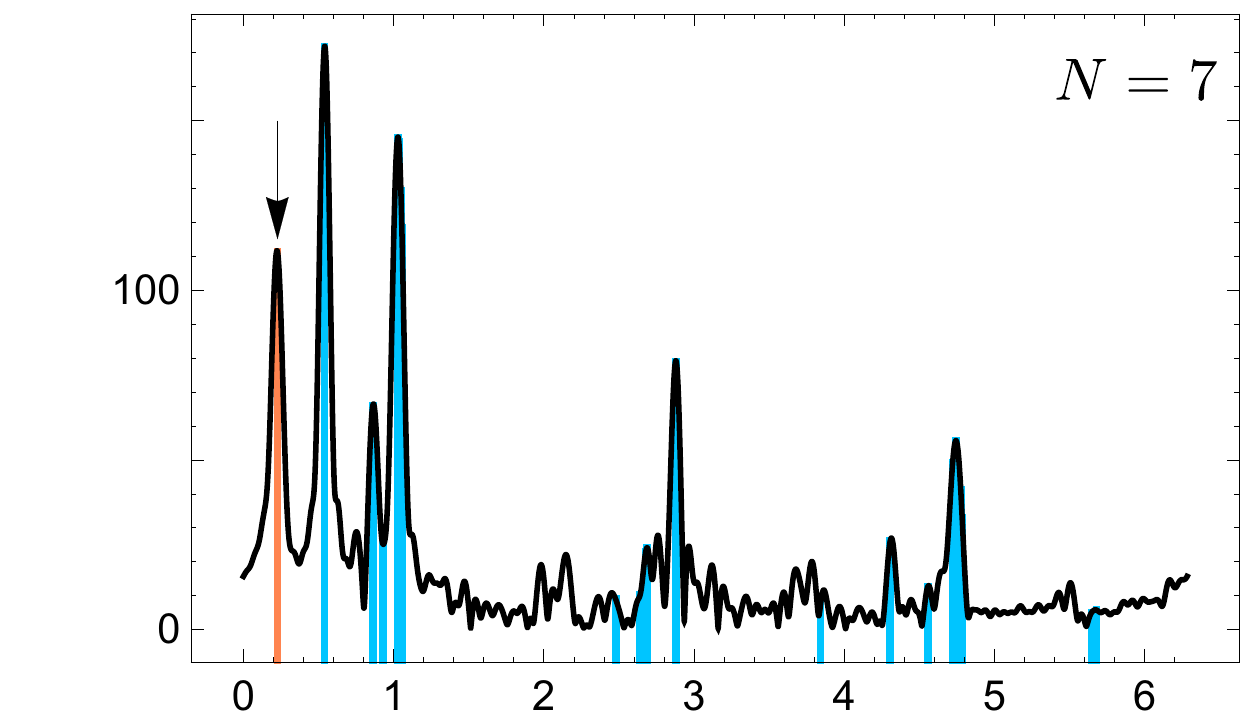}
\hfill
\includegraphics[width=0.32\textwidth]{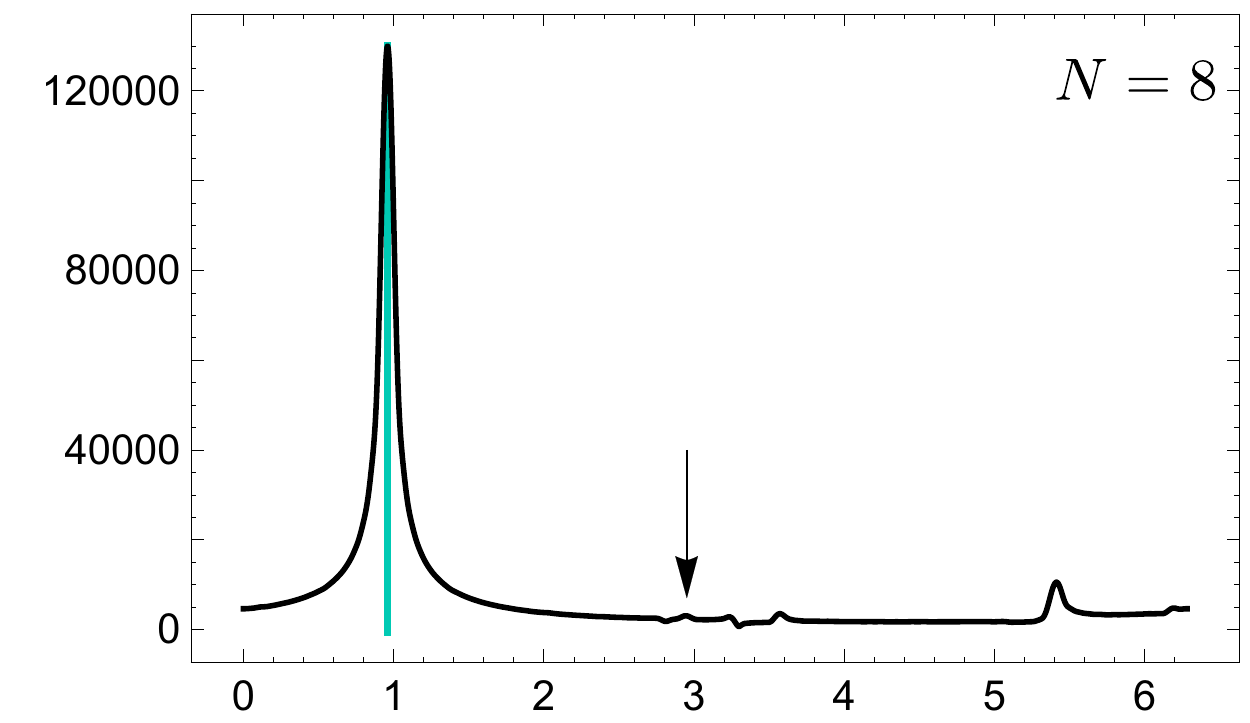}
\hfill
\includegraphics[width=0.32\textwidth]{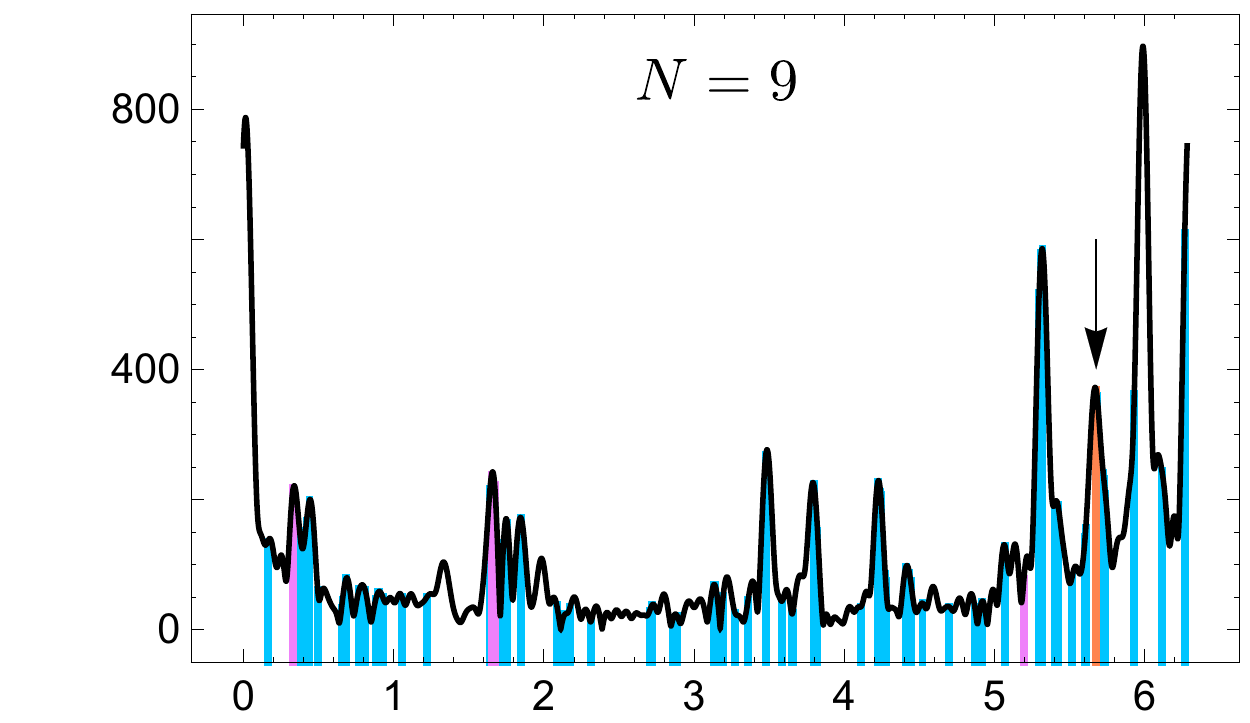}
\\
\includegraphics[width=0.32\textwidth]{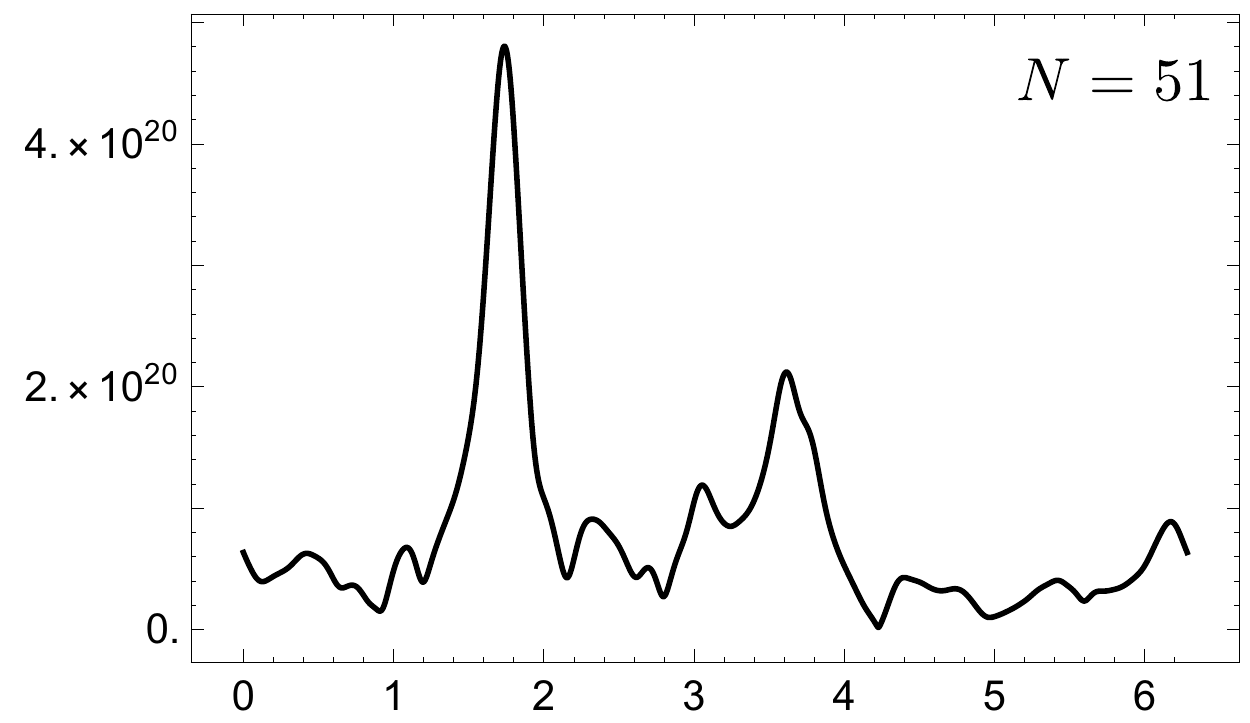}
\hfill
\includegraphics[width=0.32\textwidth]{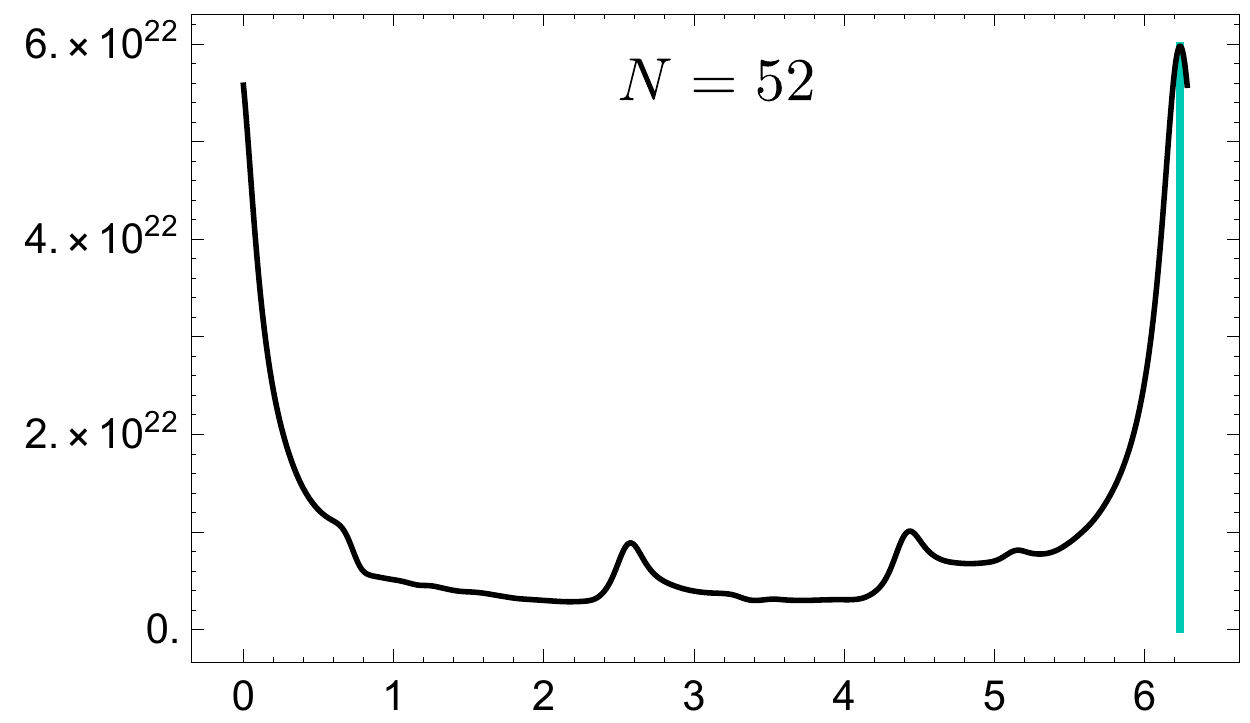}
\hfill
\includegraphics[width=0.32\textwidth]{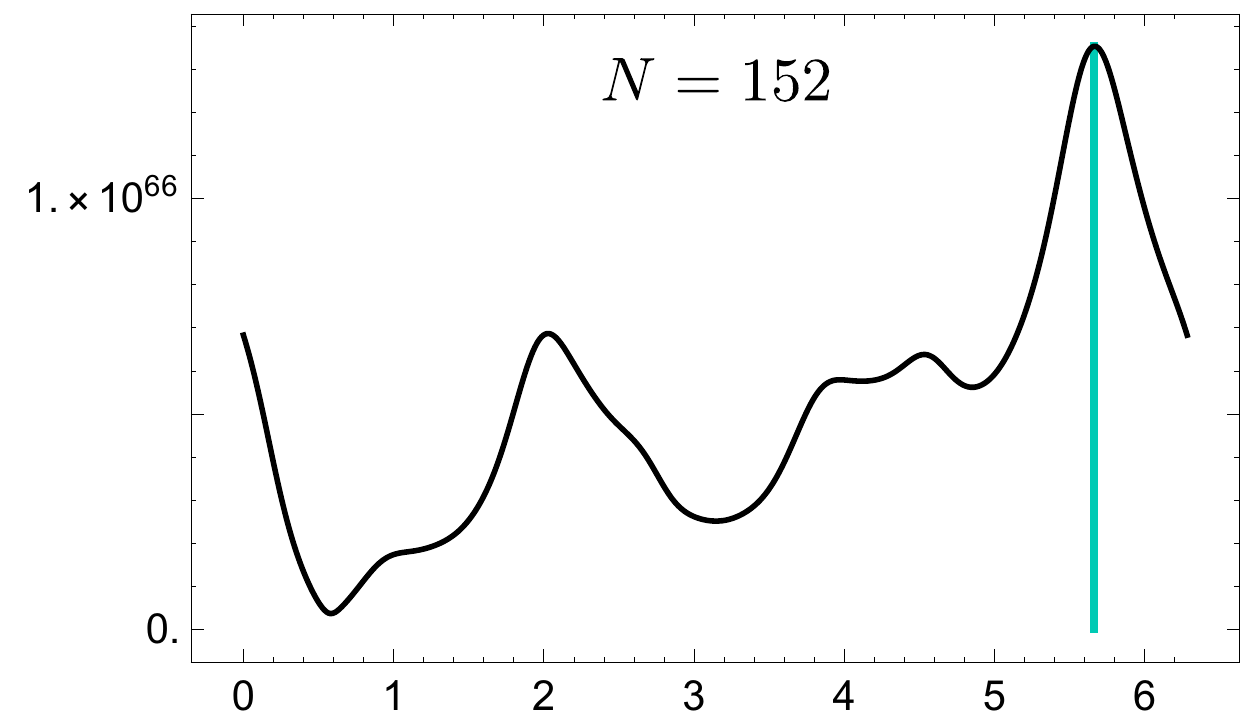}
\caption{\conline
    Absolute value \(|\rho(\Sact)|\) of the approximate action spectrum, eq.~\eqref{eq:action:spec}, over \(\Sact\) for \(T\teq 2\) time steps using \(\jcut\teq 114\). The system parameters are \(J\teq 0.7\) and \(b^x \teq b^z \teq 0.9\), only the number \(N\) of spins is varied. Coloured lines indicate classical orbit positions, the color corresponds to the primitive period: \(\Nprim\teq 1\) (orange), \(\Nprim\teq N\) (blue) and purple otherwise. Green lines correspond to \(N\Sman\), see eq.~\eqref{eq:po:smanBase:single}, indicative of the manifolds position. In the cases of \(N\teq 4\) and \(N\geq 6\) only selected orbits are shown, see text. The arrows indicate the position of an $\Nprim\teq 1$ orbit close to an isochronous pitchfork bifurcation, compare fig.~\ref{fig:action:pitchForkScaling}.  Its impact can be followed up to $N\approx 19$}
\label{fig:action:sftT2}
\end{figure}

In contrast to the previous subsection we therefore omit the semiclassical reconstruction but provide, in figure \ref{fig:action:sftT2}, the numerically calculated action spectra for different particle numbers and identical system parameters. Remarkably, for chain lengths divisible by $4$   the action spectrum \(\rho(\Sact)\) turns out to be  strongly dominated by the PO manifolds. This means that for $N=4k, k\in \mathbb{N}$ one observes only few strong peaks exactly at the positions of the PO manifolds actions, \eqref{eq:po:smanBase:single}, while all other POs are essentially suppressed. Furthermore, for these
length sequences  \(\rho(\Sact)\) exhibits a particularly large magnitude.
%Most of our attention will be devoted towards the effect of the manifolds (see section \ref{sec:po:manifold}) for the \(N\teq 4k \) cases.

%However, the seemingly divergent peak in the \(N\teq 1\) case, see the arrow marking, poses another interesting structure. It is caused by the periodic orbit that undergoes the isochronous pitchfork bifurcation described in section \ref{sec:dual:beaf}, see specifically figure \ref{fig:action:pitchforkBif}. At this choice of parameters it is already very close to the bifurcation point and its impact on \(\rho(\Sact)\) can be followed up to \(N\approx 19\) with decreasing importance/magnitude and is marked by the arrows in the remaining figures. The change in position merely results from the linear scaling of its action as given by \eqref{eq:po:abstractSScale}. We should point out that its continuing importance might well be related to the fact that it spawns \(2^N\) new orbits under bifurcation and this number grows exponentially with \(N\). In addition, it belongs to the special class of orbits for which the vanishing of the dual stability does not stem from \(\eigd_i\to 1\) but instead from \(\mathcal{D}\to 0\).
%While we already demonstrated that the scaling behaviour of \(|\rho(\Sga)|\sim\jcut^\alpha\) is in excellent agreement with the existing theory for \(N\teq 1\) it becomes close to impossible to resolve \(\alpha\) for larger repetitions as the signal is strongly distorted by interferences of neighbouring orbits. 

\begin{figure}[tbh]
    \centering
    \includegraphics[width=0.45\textwidth]{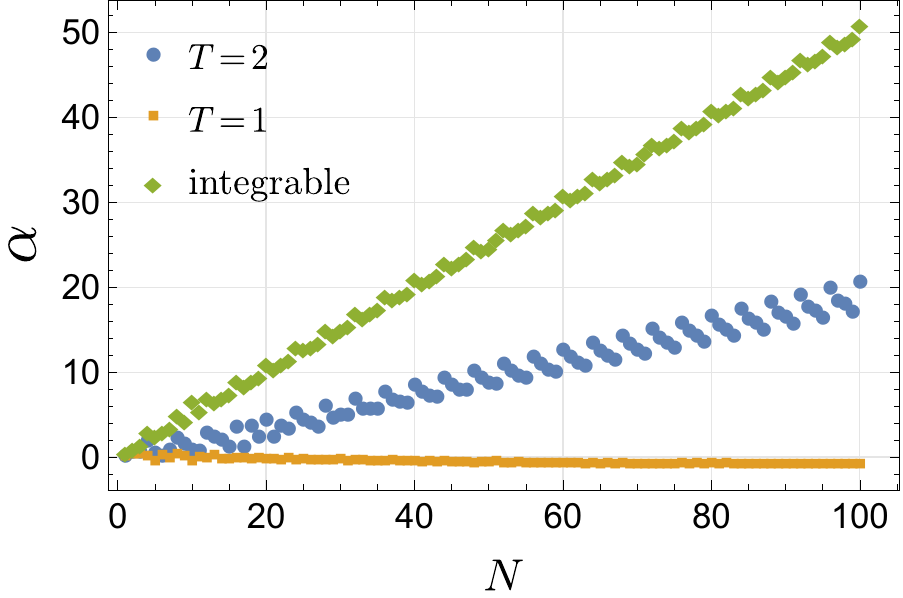}
    \caption[Scaling exponent $\alpha$ over $N$ for $T\teq 1,2$ in the chaotic case and for integrable parameters.]{\conline Estimated scaling exponent \(\alpha\) of the largest peak in the action spectrum for various particle numbers and fixed system parameter \(J\teq 0.7\) and \(b^x\teq b^z \teq 0.9\) for \(T\teq 1,2\). In the integrable case \(b^x\teq 0\) is chosen. For numerical fitting the heights in the range \(\jcut\teq 95\) to \(114\) (for \(T\teq 2\)) and \(\jcut\teq 200\) to \(400\) (for \(T\teq 1\) and integrable) are taken into account. For $T\teq 1$ a close inspection shows that the value of $\alpha$ is not yet fully saturated but instead slightly negative.}
    \label{fig:action:NdepScaling}
\end{figure}

As one can check,  the  height of the peaks at $\Sman$  follows a scaling law,
\begin{equation}
|\rho(\Sman)|\sim (\jcut)^{\alpha(N)}\qquad\alpha(N)\sim\alpha_0 N
\,,
\label{eq:action:rhoSmanNScaling}
\end{equation}
with a constant \(\alpha_0\)  only weakly dependent  on the system parameters
%, as long as, only a single manifold is present in the system,
(for further details on its value see section \ref{sec:spec:ev}).
This scaling is shown in figure \ref{fig:action:NdepScaling} in comparison to the integrable case, where \(\alpha(N)=N/2\). Clearly visible is a strong enhancement whenever the particle number is \(N\teq 4k\) \ie when the PO manifolds appear. However, a linear growth  of  scaling with $N$ is a general trend, independent of whether the  particle number is a multiple of four or not. Compare \eg the general magnitude for the \(N\teq 7\) case in figure \ref{fig:action:sftT2} to the case of \(N\teq 9\), in both cases the manifold is absent. In contrast, \(T\teq 1\) shows no scaling of \(\alpha\) with \(N\). A slight, visible decay for this case in figure \ref{fig:action:NdepScaling} can be attributed to strong interference between neighboring orbits, which influences the actual results.

While the scaling \eqref{eq:action:rhoSmanNScaling}  in  the  integrable system    is easily understood in terms of the classical \(N\) dimensional invariant tori, recall the growth of $\Aga\sim j^{(N-1)/2}$, the increase of \(\alpha\) with $N$ in the case of the four-dimensional PO manifolds  seems to be, at first, a perplexing phenomenon, given that the number of their marginal directions does not grow with $N$. In the strict semiclassical limit $j\to \infty$ with fixed $N$ the existence of four marginal directions would imply only  the  constant scaling $\alpha(N)=2$. The anomalously large 
scaling in the double limit case  can be attributed to the increase of quasi-marginal directions for which the corresponding Lyapunov exponents  are close to one. 
A hand-waving, qualitative explanation of  \eqref{eq:action:rhoSmanNScaling} can be attempted in terms of counting quasi-marginal directions, for whom the Lyapunov exponents are near zero. As numerics shows, their  numbers do indeed grow with \(N\), but  
 correct accounting of such directions is already a challenge for single particle systems, see \cite{cat_schomerus,cat_manderfeld}. Taking into  account contributions  of all nearly bifurcating orbits  for a large $N$  seems to be an extremely difficult problem  and we avoid  this path in what follows. Rather, we will provide an explanation for   \eqref{eq:action:rhoSmanNScaling}  through the  study of  spectral properties  of the  dual operator $\Ut$.

%%%%%%%%%%%%%%%%%%%%%%%%%%%%%%%%%%%%%%%%%%%%%%%%%%%%%%%%%%
% Dual Spectrum
%%%%%%%%%%%%%%%%%%%%%%%%%%%%%%%%%%%%%%%%%%%%%%%%%%%%%%%%%%
\section{Spectrum of the Dual Operator}
\label{sec:spec}

\begin{figure}
    \centering
    %\begin{minipage}{.49\textwidth} %used to build this into a combo-figure
    \includegraphics[width=0.42\textwidth]{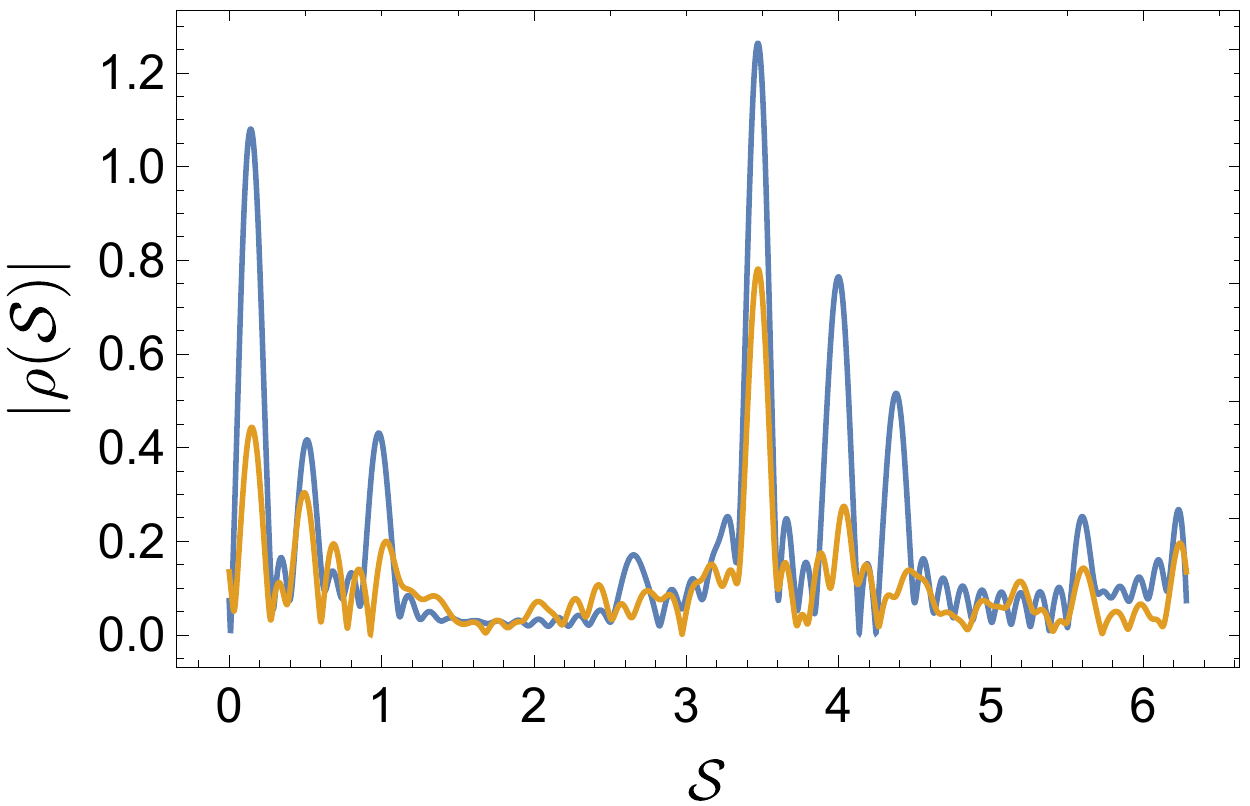}
    \hfil
    \includegraphics[width=0.42\textwidth]{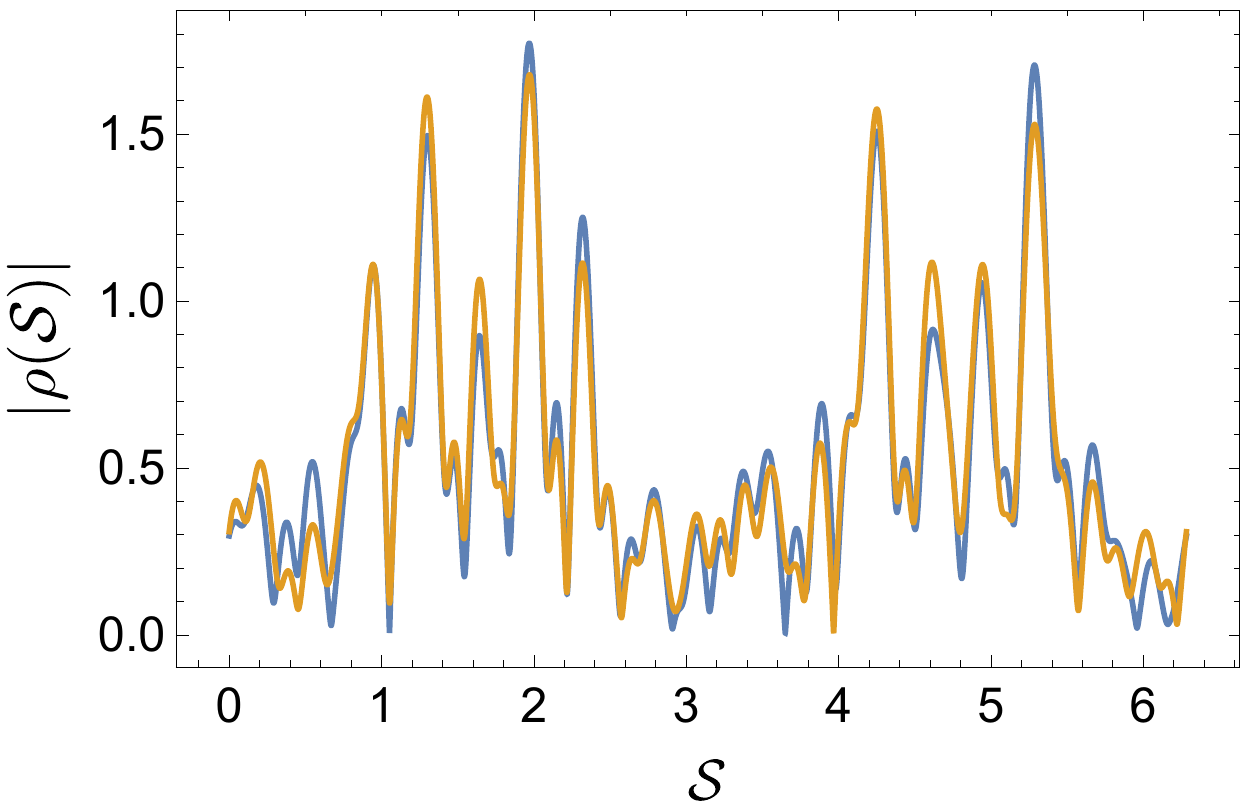}
    \\
    \includegraphics[width=0.42\textwidth]{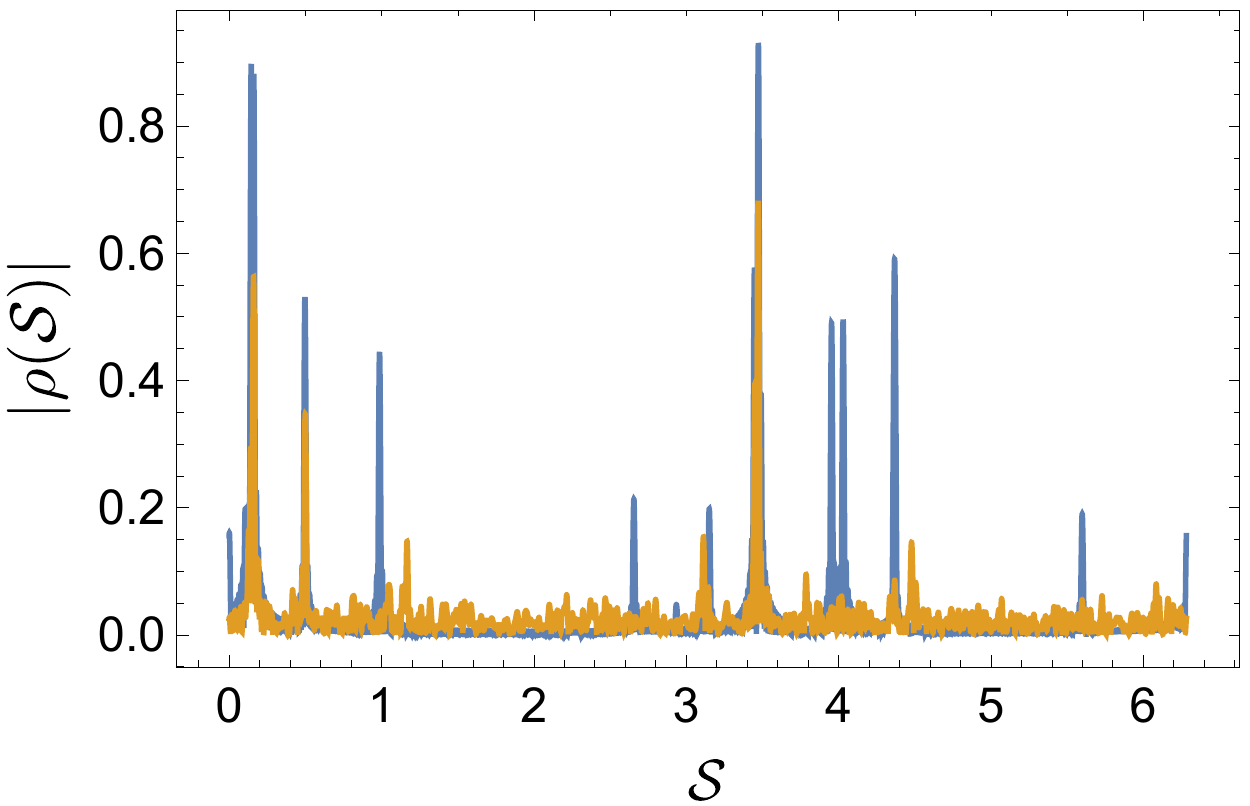}
    \hfil
    \includegraphics[width=0.42\textwidth]{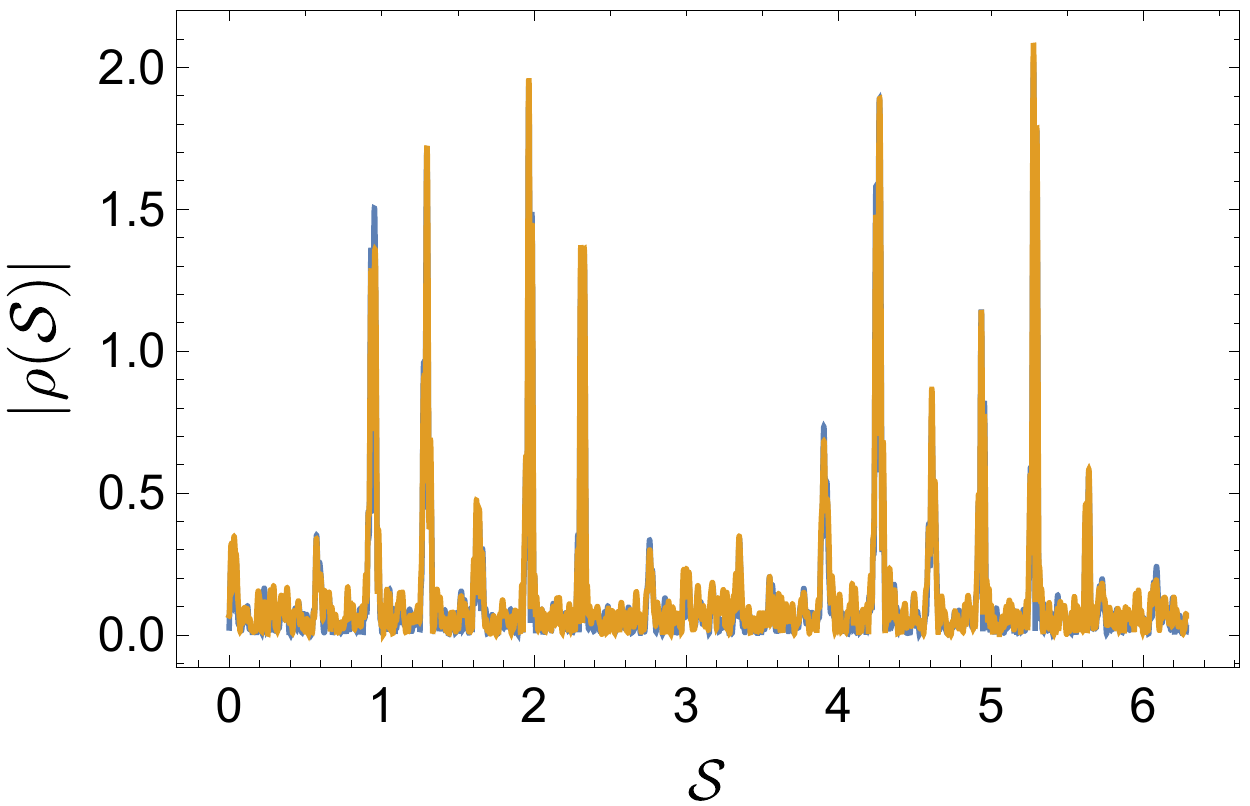}
    %\end{minipage}
    \caption[Approximation of the $T\teq 1$ action spectrum by the largest eigenvalues.]{\conline Comparison between the single time step action spectrum \(|\rho(S)|\) (blue curves) and an approximated variant using only the largest eigenvalue in \eqref{eq:spec.large:ev} instead of the full traces. The four panels correspond to \(N\teq 5\) (left) and \(N\teq 20\) (right), the upper row uses a low cut-off, \(\jcut\teq 50\), the lower one features \(\jcut\teq 500\). Parameters are given by \(J\teq 0.7\) and \(b^x\teq b^z\teq 0.9\).}
    \label{fig:spec:sFTldmax}
\end{figure}

The question of the anomalously large spectral fluctuations associated with the PO manifolds, specifically its scaling with \(N\) as observed in the last section, can be addressed in terms of the largest eigenvalues   of the dual operator \(\Ut\). Indeed,  for large $N$ the traces of  \(\Ut^N\) are  dominated by their  largest eigenvalues,
\begin{equation}
\Tr \hat{U}^T=\Tr\Ut^N =   \sum_l \tilde{\lambda}^N_{l}(1+O(e^{-\delta N})), \qquad \delta >0
\,,
\label{eq:spec.large:ev}
\end{equation}
where the sum can be  restricted to  several   eigenvalues $\tilde{\lambda}_{l}$ with the maximal absolute value.
 The validity of this approximation greatly depends on the magnitude of \(N\). In figure \ref{fig:spec:sFTldmax} we depict both the actual action spectrum (blue curve) and an approximate result (orange), for which we leave in the  sum \eqref{eq:spec.large:ev} only  the  largest eigenvalue. The agreement between the two curves greatly improves with the number of spins  \(N\). Besides \(N\) also \(j\) plays a role as it governs the dimension of \(\Ut\) and therefore the gap $\delta$ between the largest eigenvalues and their successors.
% \begin{figure}
%     \includegraphics[width=0.2\textwidth]{KLM_sFTldmaxComp_N1T1_j0d7bxbz0d9s50}
%     \hfill
%     \includegraphics[width=0.2\textwidth]{KLM_sFTldmaxComp_N5T1_j0d7bxbz0d9s50}
%     \hfill
%     \includegraphics[width=0.2\textwidth]{KLM_sFTldmaxComp_N10T1_j0d7bxbz0d9s50}
%     \hfill
%     \includegraphics[width=0.2\textwidth]{KLM_sFTldmaxComp_N20T1_j0d7bxbz0d9s50}
%     \\
%     \includegraphics[width=0.2\textwidth]{KLM_sFTldmaxComp_N1T1_j0d7bxbz0d9s500}
%     \hfill
%     \includegraphics[width=0.2\textwidth]{KLM_sFTldmaxComp_N5T1_j0d7bxbz0d9s500}
%     \hfill
%     \includegraphics[width=0.2\textwidth]{KLM_sFTldmaxComp_N10T1_j0d7bxbz0d9s500}
%     \hfill
%     \includegraphics[width=0.2\textwidth]{KLM_sFTldmaxComp_N20T1_j0d7bxbz0d9s500}
%     \caption{Comparison between the single time step action spectrum \(|\rho(S)|\) (blue curves) and an approximated variant using the replacement \eqref{eq:spec:ldmaxApprox} for the involved traces. Panels correspond to \(N\teq 1,5,10,20\) from left to right, the upper row uses a low cut-off, \(\jcut\teq 50\), the lower one feature \(\jcut\teq 500\). Parameters are given by \(J\teq 0.7\) and \(b^x\teq b^z\teq 0.9\).}
%     \label{fig:spec:sFTldmax}
% \end{figure}

\begin{figure}
    \centering
    \includegraphics[width=0.42\textwidth]{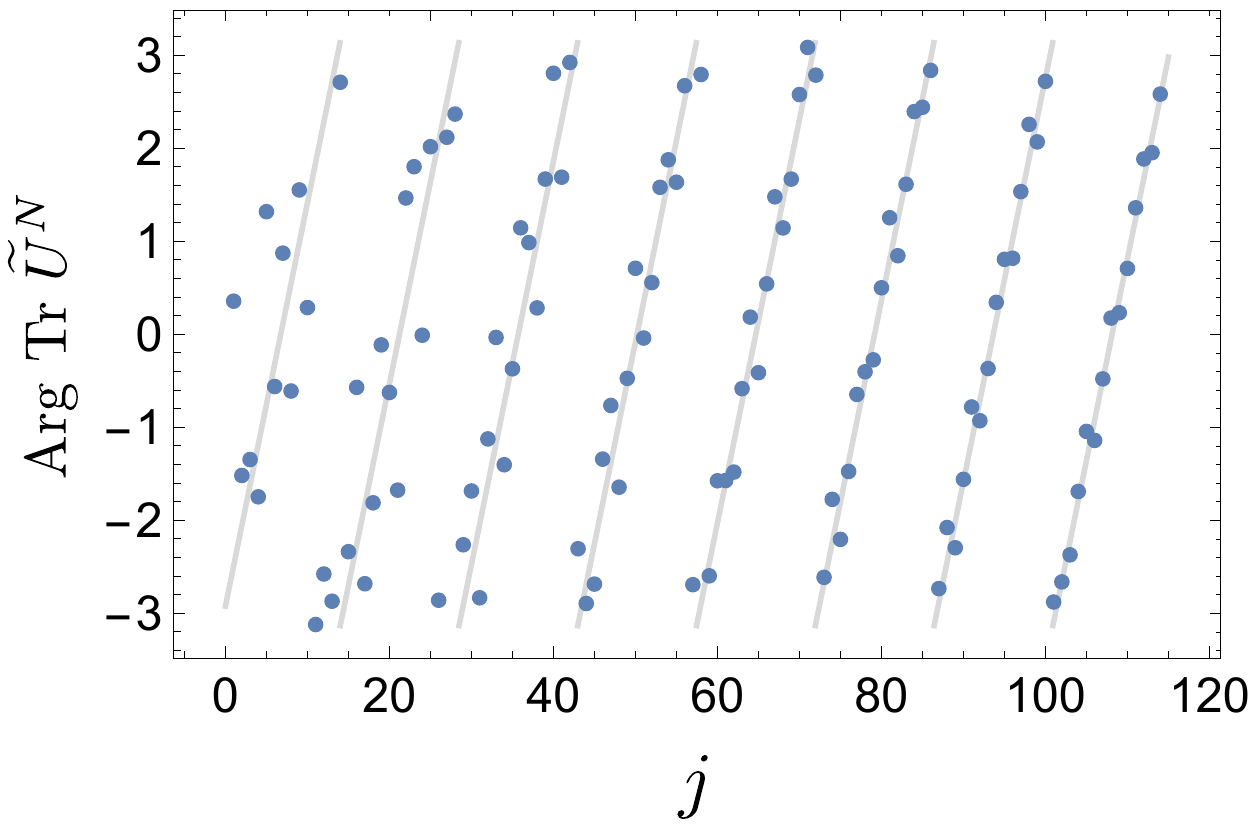}
    \caption[Trace of the $T\teq 2$ dual operator and its domination by $\ldmax$.]{\conline Phase of the trace of the dual operator for differing spin quantum numbers \(j\) and \(J\teq 0.7\), \(b^x\teq b^z \teq 0.9\) where we consider \(T\teq 2 \) time-steps for \(N\teq 56\) particles. The (rescaled) contribution of the manifold's action (gray line proportional to $(j+1/2)\Sman \,\mod\,2\pi $) is clearly visible and works more accurately for larger \(j\).}
    \label{fig:spec:manPhaseOverSmallJ} 
\end{figure}

%%%%%%%%%%%%%%%%%%%%%%%%%%%%%%%%%%%%%%%%%%%%%%%%%%%%%%%%%%%%%%%%
% Spectrum: Numerics
%%%%%%%%%%%%%%%%%%%%%%%%%%%%%%%%%%%%%%%%%%%%%%%%%%%%%%%%%%%%%%%%
\subsection{Numerical Findings} 
\label{sec:spec:ev}

As has been explained above, it    is of crucial importance to understand how the largest eigenvalues of  \(\Ut\)  depend on \(j\) in the  semiclassical limit \(j\to\infty\). Below we provide the results of a numerical study of the dual  operator spectrum  and give their  explanations based on a semiclassical theory in the next section.  

\begin{figure}
    \centering
    \includegraphics[width=0.4\textwidth]{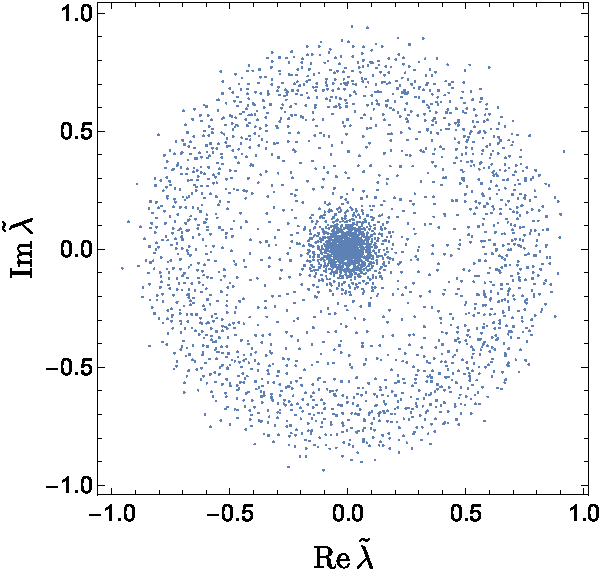}
    \caption[Eigenvalues of $\Ut$ for $T\teq 1$.]{\conline Eigenvalue spectrum $\tilde{\lambda}$ of \(\Ut\) for \(T\teq 1\) in the complex plane. System parameters are chosen as \(J\teq 0.7\) and \(b^x\teq b^z \teq 0.9\) with \(\jcut\teq4700\).}
    \label{fig:spec:T1spec}
\end{figure}

For only a single time step $T=1$ the spectrum \(\{\tilde{\lambda}_i|i=1,\dots, 2j+1\}\) of \(\Ut\) is  uniformly  distributed  in the angular direction, see figure \ref{fig:spec:T1spec} for a generic example. As the operator is non unitary, the eigenvalues are not restricted to the unit circle and, in fact, many of them reside close to the origin indicating the non-unitary nature of   the dual evolution.
\begin{figure}
    \includegraphics[height=0.25\textwidth]{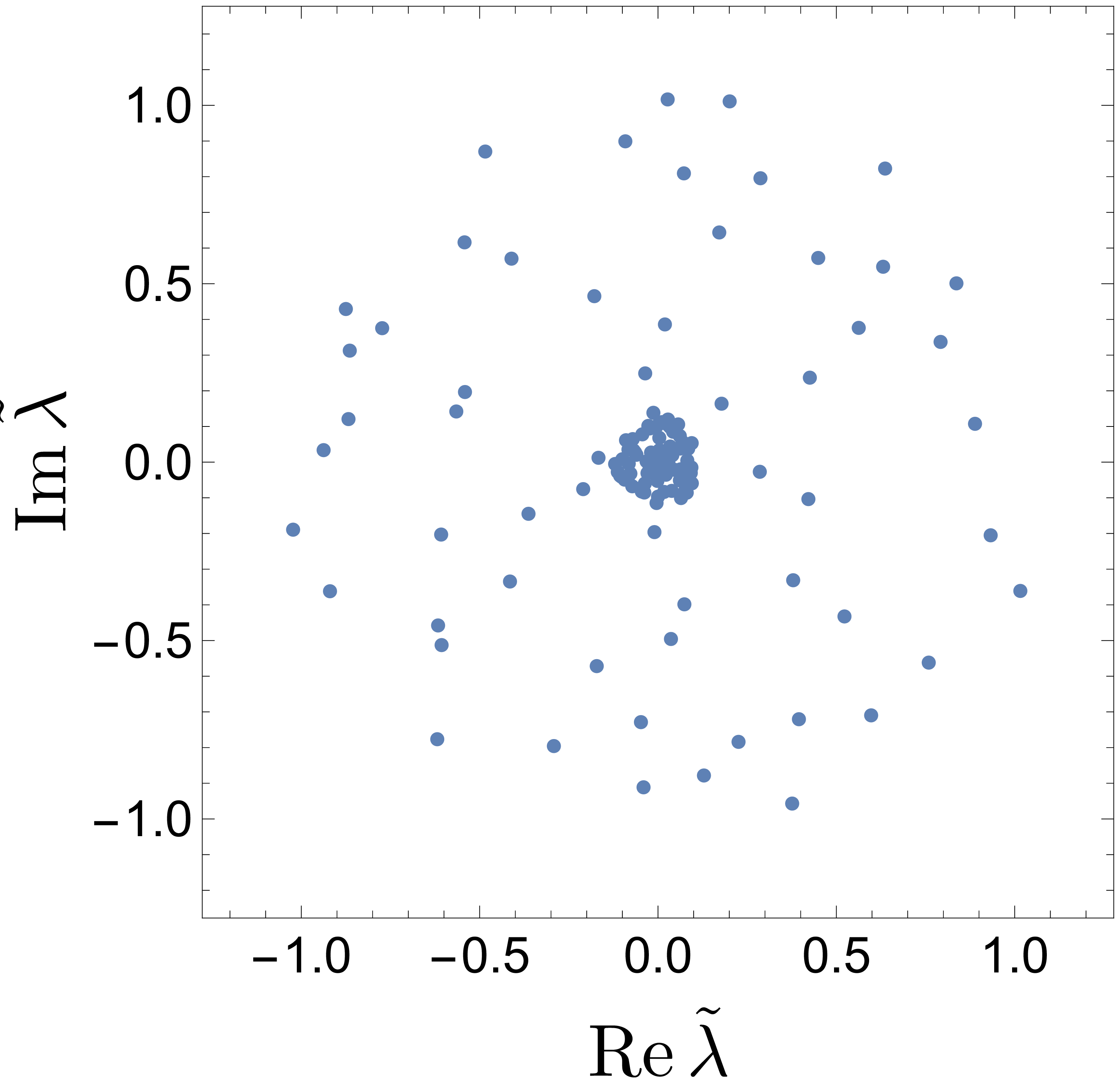}
    \hfill
    \raisebox{0.15\height}{\includegraphics[height=0.23\textwidth]{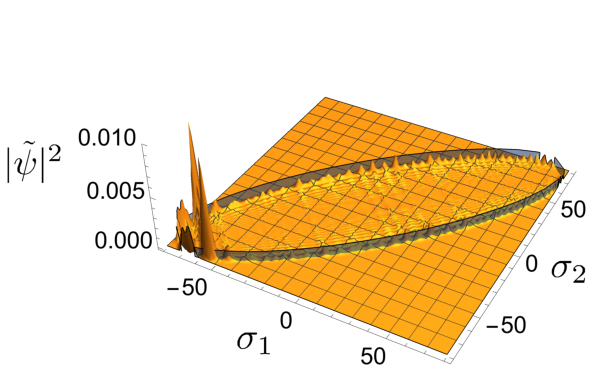}}
    \hfill
    \includegraphics[height=0.25\textwidth]{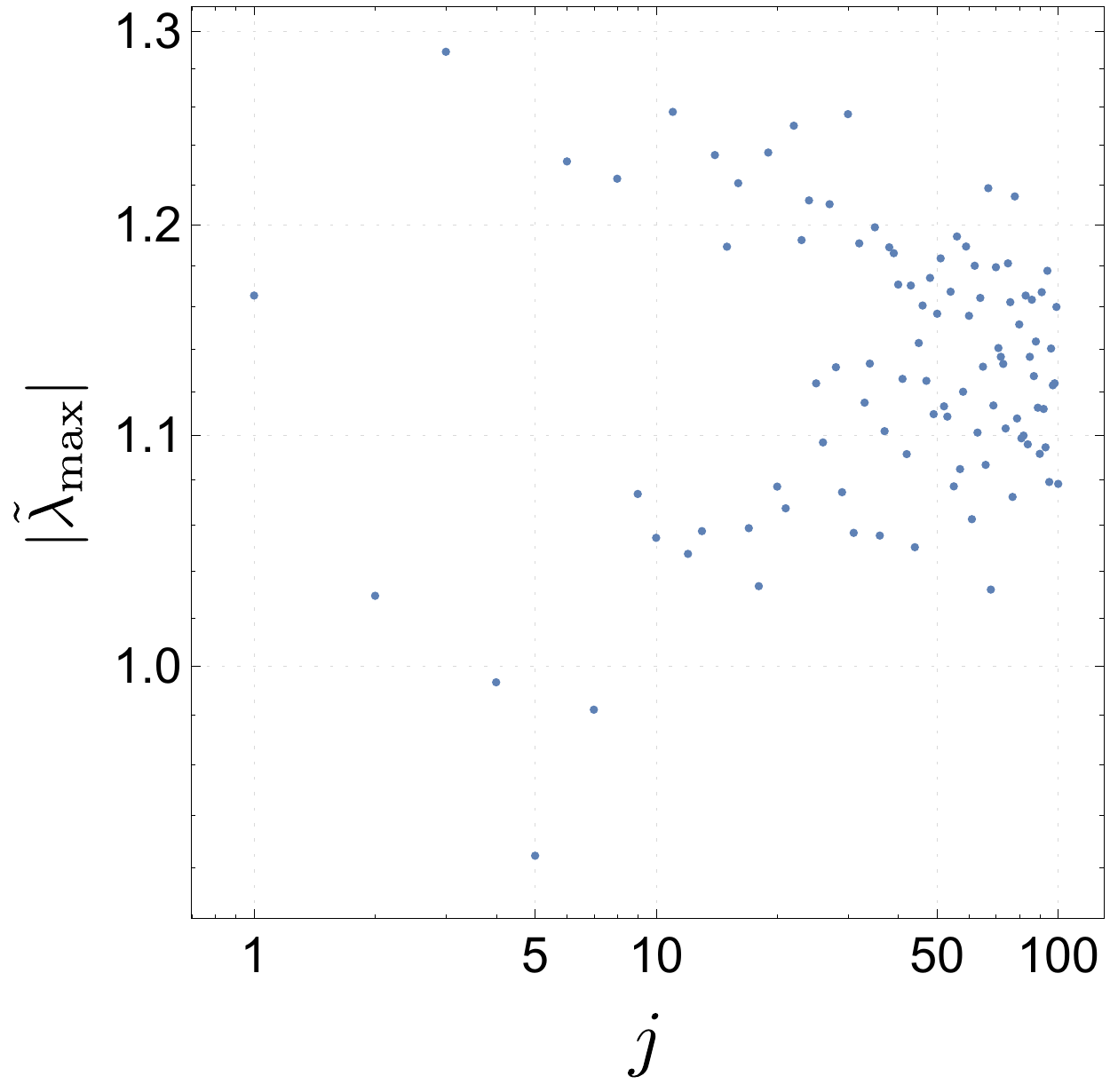}
    \\
    \includegraphics[height=0.25\textwidth]{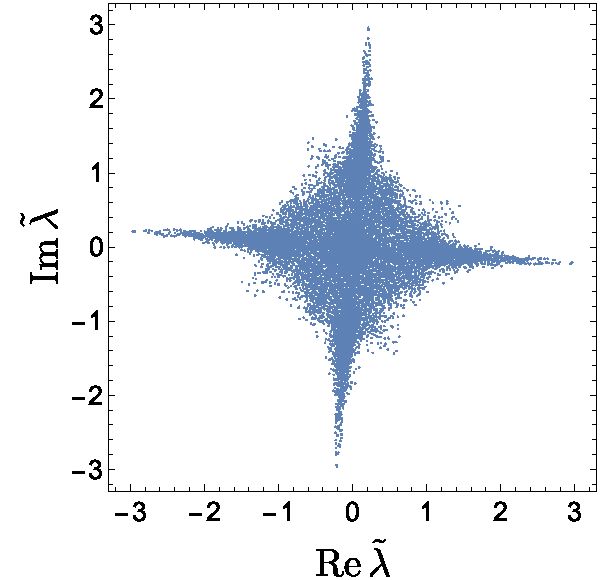}
    \hfill
    \raisebox{0.15\height}{\includegraphics[height=0.23\textwidth]{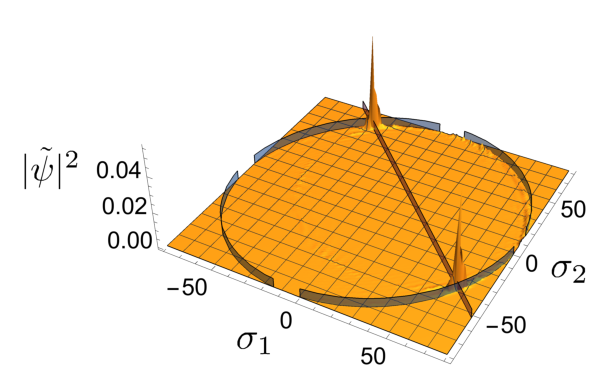}}
    \hfill
    \includegraphics[height=0.25\textwidth]{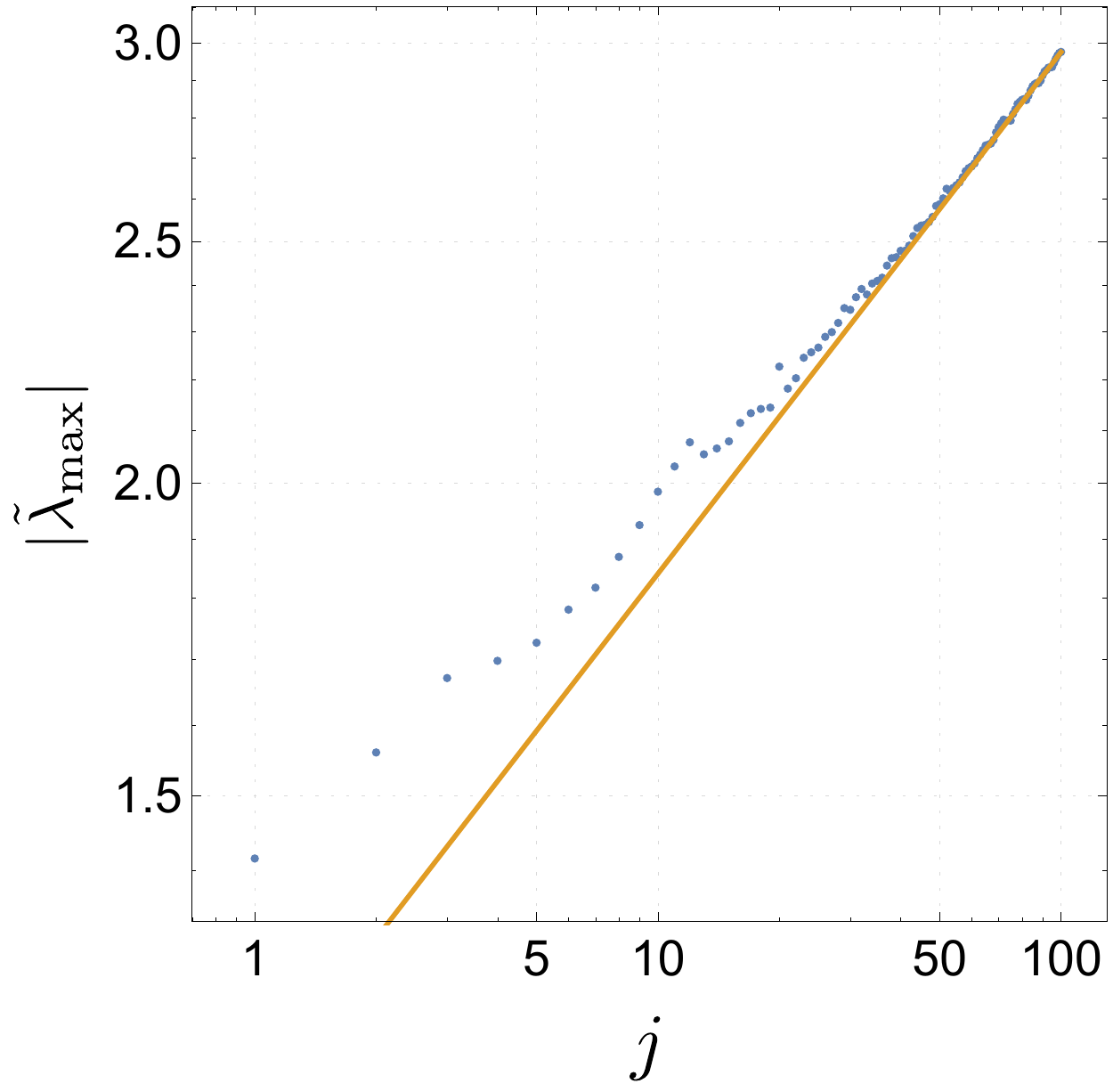}
    \\
    \includegraphics[height=0.25\textwidth]{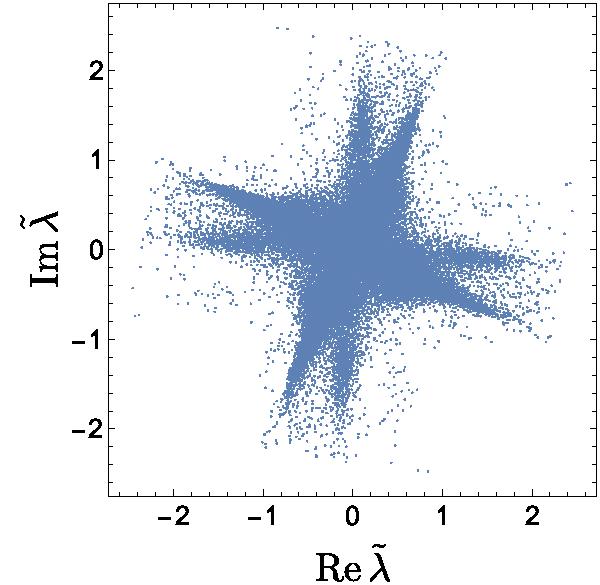}
    \hfill
    \raisebox{0.15\height}{\includegraphics[height=0.23\textwidth]{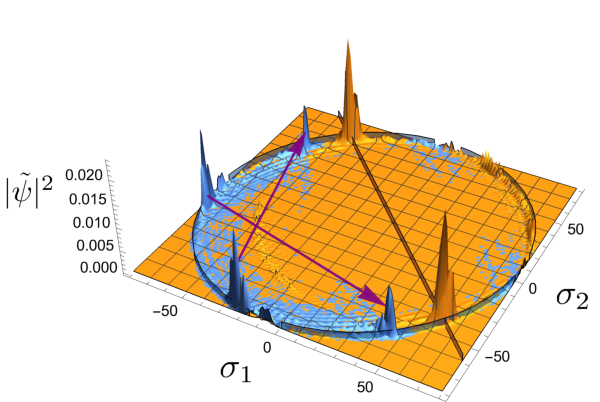}}
    \hfill
    \includegraphics[ height=0.25\textwidth]{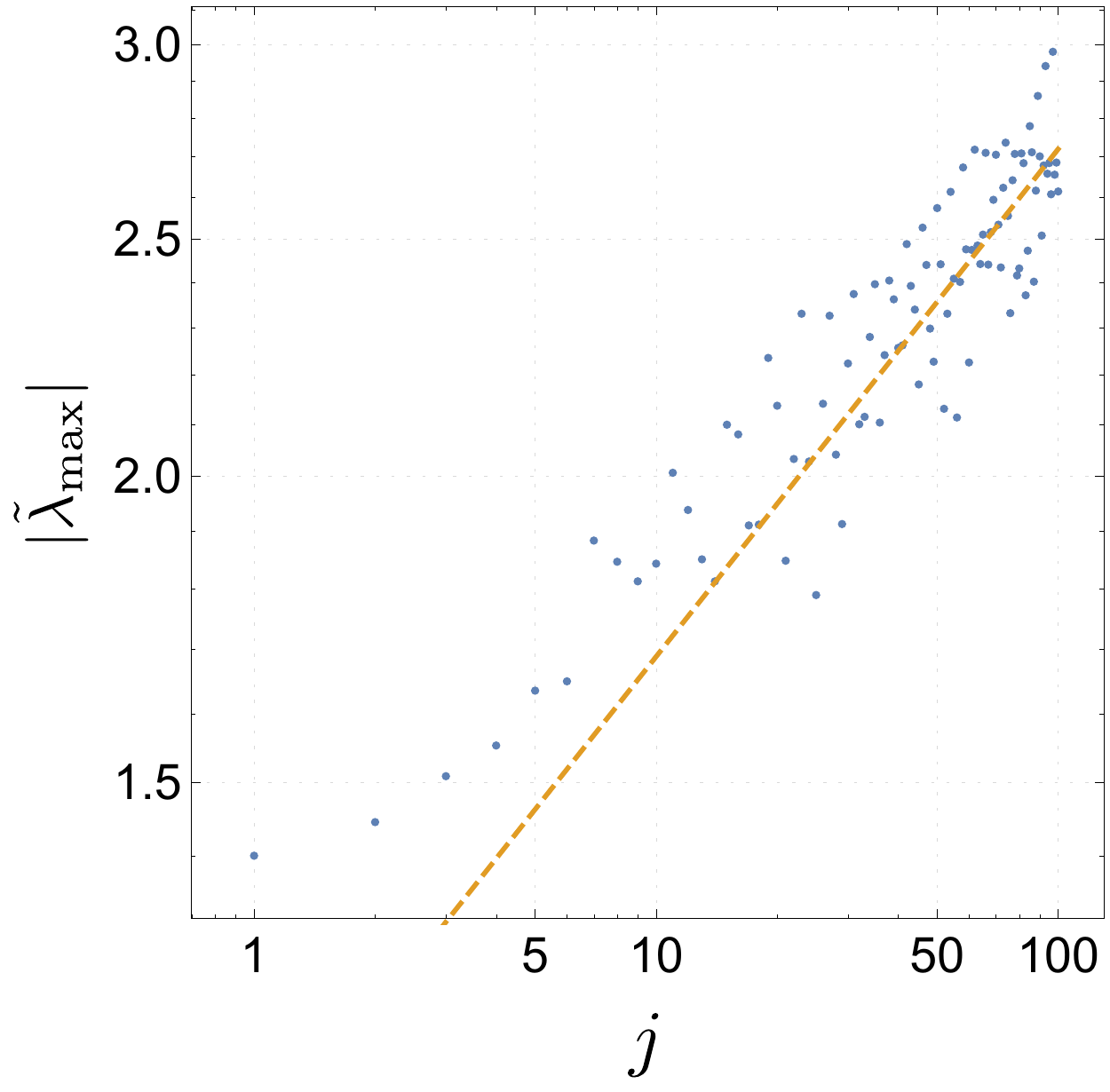}
    \caption[Spectrum of $\Ut$ for $T\teq 2$ in the cases of none, one and several manifolds. Additionally, the scalings of $\ldmax$ and the corresponding largest eigenvectors are shown.]{\conline To the left the spectrum of the dual operator for \(T\teq2\) and \(j\teq 100\) is presented in the complex plane. The right column shows the scaling of its largest eigenvalue in dependence of \(j\) with a numerical fit of \(\alpha_0\) where applicable. 
    The middle column depicts the eigenvector corresponding to the largest eigenvalue (for \(j\teq 80\)). Therein, the blue rim corresponds to the boundary of the classically allowed region as specified in eq.~\eqref{eq:spec:transitReg}.
    The parameters are chosen as \(J\teq 0.2\) and \(b^x\teq b^z\teq  0.3\) (first row) where no manifold is present. In the second row \(J\teq 0.6\) with \(b^x\teq b^z \teq 0.9\) represents the single manifold regime leading to \(\alpha_0\approx 0.21\). The purple bar shown with the eigenvector is the solution to \(\sigma_1+\sigma_2\teq g\) where \(g\) is given by \eqref{eq:spec:g} for \(p\teq 0\). In the last row \(J\teq 0.8\), \(b^x\teq b^z \teq 0.9\),  where  several manifolds exist. The scaling lies in between \(0.17\leq\alpha_0\leq0.23\), to guide the eye the shown dashed line corresponds to \(\alpha_0\approx0.21\).
    The middle figure in this case shows two different eigenvectors, colored   blue and orange, that correspond to the largest eigenvectors of  two different crosses shown on the left  figure. 
    The endpoints of the  purple arrows represent the semiclassical predictions for the  localization   centers of the first (shown in blue) eigenvector. The two arrows  correspond to the parameters $p_1=0, p_2=-1$ and $p_1=-1$, $p_2=0$, respectively, see  \ref{sec:apx:dualEvec} for details.  The intersection points of the purple bar with the  ellipse boundary  indicate localization centers of the second eigenvector (shown in orange). It corresponds to   $p_1\teq p_2 \teq 0$, as in the single manifold case.
    %vectors \(\vec{R}(0,-1)\) and \(\vec{R}(-1,0)\) of the more general solution presented in \ref{sec:apx:dualEvec}.
    }
    \label{fig:spec:smallJ}
    \label{fig:spec:T2diagSpec}
\end{figure}
For  two time steps, $T=2$,  the dual spectrum \(\{\tilde{\lambda}_i|i=1,\dots, (2j+1)^2\}\)  has a similar rotationally invariant distribution in the regime where no PO manifolds exist, see fig.~\ref{fig:spec:smallJ}. In sharp contrast,   a pronounced structure  emerges whenever   PO manifolds are present. To illustrate this, figure \ref{fig:spec:T2diagSpec} shows the dual spectrum  in the regime where either only one  or  several PO manifolds exist.
% \begin{figure}[tbh]
%     \includegraphics[width=0.4\textwidth]{KLM_dualSpec_j0d6bxbz0d9s100}
%     \hfill
%     \includegraphics[width=0.4\textwidth]{KLM_dualSpec_j0d8bxbz0d9s100}
%     \\
%     \includegraphics[width=0.4\textwidth]{KLM_dualEvec_j0d6bxbz0d9s80}
%     \hfill
%     \includegraphics[width=0.4\textwidth]{KLM_dualEvec_j0d8bxbz0d9s80}
%     \caption{Eigenvalue spectrum of \(\Ut\) for \(T\teq 2\) in the complex plane (upper row) for \(b^x\teq b^z \teq 0.9\) and \(J\teq 0.6\) (left) or \(J\teq 0.8\) (right) at \(j\teq 100\). The lower row depicts the absolute square of the eigenvector corresponding to the largest eigenvalue at \(j\teq 80\). The two dimensional representation takes into account the product structure of the Hilbert space for the two dual spins\Onote{tilde sigma?}. The blue rim corresponds to the boundary of the classically allowed region as specified in equation \eqref{eq:spec:transitReg}.}
%     \label{fig:spec:T2diagSpec}
% \end{figure}
The spectral distribution has a remarkable cross-like  shape(s) indicating an approximate  four-fold rotational symmetry, which becomes more and more  pronounced for the largest eigenvalues  as $j\to \infty$. This symmetry  singles out  sequences $N=4k$,  where, according to  \eqref{eq:spec.large:ev}, the sum   of the   largest eigenvalues adds up coherently. On the contrary, for $N\neq 4k$ the sum of the largest eigenvalues vanishes to the leading order in $j$, thus significantly reducing the magnitude of the spectral fluctuations.   To make a quantitative  prediction it is, therefore, natural to look at the largest \(\tilde{\lambda}_i\) as functions of $j$. Focusing on the regime where only one PO manifold exists, we find that  the phases of the four largest  dual eigenvalues are given by
\begin{equation}
    \arg{\tilde{\lambda}_{\text{max},l}}=(j+1/2) \Sman +\frac{\pi l}{2}+O(1/j)
    \qquad l\in \{1,2,3,4\}
    \,.\label{eq:fourLargestEig}
\end{equation}
As predicted, in the cases $N\teq 4k$ the  $\pi l/2$ parts in the phase cancel under summation of the eigenvalues.
Remarkably,  to the leading order in $j$, the phases are   determined  by the prime  action $\Sman$ of the PO manifold, \eqref{eq:po:smanBase:single}.
Such a connection is reminiscent of the Bohr-Sommerfeld quantisation rule  for the spectrum  of integrable Hamiltonian systems. Furthermore, the absolute values of the  largest eigenvalues scale algebraically with \(j\),
\begin{equation}
    |\ldmax|\propto j^{\alpha_0}(1+O(1/j))\,.
    \label{eq:spec:ldmaxAbsScale}
\end{equation}
This    explains the linear  dependence of \(\alpha(N)\) on  \( N\), \ie  \(\alpha(N)\sim \alpha_0 N\) in \eqref{eq:action:rhoSmanNScaling}. The same  scaling carries over  to the traces \(\Tr \Ut^N\)  even for  \(N\neq 4k\) where  a similar linear growth of \(\alpha\) with $N$ is observed,  but with a constant negative offset,  see figure \ref{fig:action:NdepScaling}.

In the regime of a single PO manifold, the contribution of the four largest  eigenvalues is sufficient  to  get  the total phase of the trace even for large powers in \(N\), improving with increased \(j\), see figure \ref{fig:spec:manPhaseOverSmallJ}. Therefore,  for $T=2$ the whole essential information about the spectral fluctuations in the system  is stored in two parameters: \(\Sman\) and \(\alpha_0\). Additional    PO manifolds contribute other quadruples of  eigenvalues $\tilde{\lambda}^{(\ell)}_{\text{max},l}$ with a similar  scaling of the absolute value $|\tilde{\lambda}^{(\ell)}_{\text{max},l}|=C_\ell j^{\alpha_0}$, but  (possibly)  different phases \(  (j+1/2)N\Sman^{(\ell)}+l\pi/2, \, l=1,\dots,4\), where $\Sman^{(\ell)}$ is the action of the respective PO manifold. As a result, the total 
contribution in the traces  of the evolution operator for $N=4k$  is given by:
\begin{equation}
\Tr \Uh^2= 4j^{N\alpha_0} \sum_\ell C_\ell e^{i(j+1/2)N\Sman^{(\ell)}}\left(1+O(1/j)\right),
\end{equation}
where the sum is over the distinct PO manifolds. 

% \begin{figure}[tbh]
%     \includegraphics[width=0.4\textwidth]{WLP_phaseOfTrace_N56T2_jd7bxbzd9}
%     \caption{Phase of the trace of the dual operator for differing spin quantum numbers \(j\) and \(J\teq 0.7\), \(b^x\teq b^z \teq 0.9\) where we consider \(T\teq 2 \) time-steps for \(N\teq 56\) particles. The (rescaled) contribution of the manifolds action (gray line) is clearly visible and works more accurately for larger \(j\).}
%     \label{fig:spec:manPhaseOverSmallJ}  
% \end{figure}

A straightforward inspection of the eigenvectors corresponding to the maximal eigenvalues of \(\Ut \) reveals their   remarkable localization properties, see fig. \ref{fig:spec:T2diagSpec}. 
These  eigenvectors comprise  two parts,
\begin{equation}
\tilde{\psi} =  \tilde{\psi}_q   +   \tilde{\psi}_p \,,
    \label{eq:spec:evecStruct}
\end{equation}
of which  \(\tilde{\psi}_q\)   is sharply localized in the \(|\sigma_1\rangle\otimes|\sigma_2\rangle\) basis while \(\tilde{\psi}_p\) is localized in the  momentum basis \(|\bar\sigma_1\rangle\otimes|\bar\sigma_2\rangle\), where
\begin{equation}
    |\bar\sigma\rangle=\frac{1}{\sqrt{2j+1}}\sum_{\sigma=1}^{2j+1}e^{i2\pi\sigma\bar\sigma/(2j+1)}|\sigma\rangle\,.
\end{equation}

%%%%%%%%%%%%%%%%%%%%%%%%%%%%%%%%%%%%%%%%%%

\subsection{Semiclassical Theory for $T=2$}
\label{sec:spec:evec}

\begin{figure}
 \centering
    \includegraphics[width=0.45\textwidth]{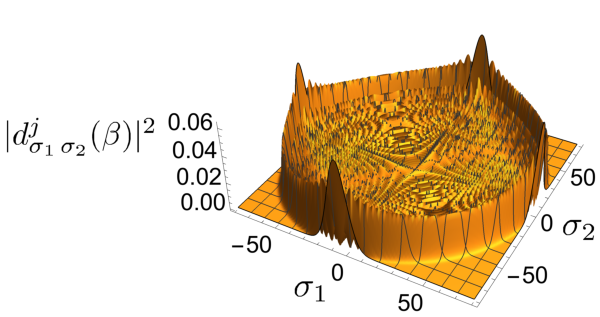}
    \caption[Absolute square of the Wigner small $d$-matrix \(d_{\sigma_1\,\sigma_2}^j(\beta)\).]{\conline Absolute square of the Wigner small $d$-matrix \(d_{\sigma_1\,\sigma_2}^j(\beta)\) for an angle \(\beta\approx 1.48\) (\(b^x = b^z = 0.9\)), and \(j\teq 80\).}
    \label{fig:spec:wignerSmallD}
\end{figure}

To understand the  form  of \(\tilde{\lambda}_{\text{max},l}\) and the localization properties of the correponding eigenvectors, let us first recall   the   product structure   of the dual operator \(\Uti\Utk\). 
 For \(\K \teq 0\) the form  \eqref{eq:dual:utk}  of the kick part \(\Utk\) is reminiscent of the kernel of the  Fourier transformation, such that the correspondence becomes exact if \(J\teq \pi/4\). The action of this part  on a  coherent state,  localized in  both momenta and coordinates, can be interpreted as an exchange of position and momenta values.
The interaction part \(\Uti\), eq.~\eqref{eq:dual:uti}, is instead given by a product of transition elements of a unitary rotation induced by  a constant magnetic field
\begin{equation}
    \fl
    \langle \sigma_1 \sigma_2 | \Uti | \sigma_1 \sigma_2\rangle
    = \langle \sigma_1| \eu^{-2\iu \, \vec{b}\cdot \hat{\vec{S}}}\, | \sigma_2 \rangle
    \langle \sigma_2| \eu^{-2\iu \, \vec{b}\cdot \hat{\vec{S}}}\, | \sigma_1 \rangle
    =\eu^{-\iu (\sigma_1 + \sigma_2)(\alpha+\gamma-\pi)}(d_{\sigma_1 \sigma_2}^j (\beta))^2
    \,.
    \label{eq:spec:t2Wigner}
    %pi in exponent stems from exchange in d_{12}
\end{equation}
Here \(\alpha,\beta,\gamma\) (with \(\alpha\teq\gamma\)) are the Euler angles given in \eqref{eq:spec:eulerAnglesDef}
 and 
\begin{equation}
   d_{\sigma\,\sigma'}^j(\beta)=\langle \sigma| \eu^{-\iu \beta \hat{S}^y} | \sigma' \rangle
   \label{eq:spec:wignerSmallD}
\end{equation}
is Wigner's  small $d$-matrix.
Conveniently, the uniform semiclassical limit of \(d_{\sigma\,\sigma'}^j\) is well known  \cite{braunRot}. When $j\to\infty$, the function \(d_{\sigma\,\sigma'}^j\) is supported within  the  elliptic region,
\begin{equation}
     \sigma'^2 + \sigma^2 - 2 \sigma' \sigma \cos{\beta} \leq (j+1/2)^2 \sin^2{\beta}\,,
    \label{eq:spec:transitReg}
\end{equation}
 where it    scales as  \(d_{\sigma\,\sigma'}^j\sim j^{-1/2}\) at a finite distance from the boundary and exponentially decays   outside of  the region  \eqref{eq:spec:transitReg}, see fig.~\ref{fig:spec:wignerSmallD}.  In the  semiclassical limit any  eigenvector of the dual matrix \(\Ut\) must, therefore, reside in the classically allowed  region given by \eqref{eq:spec:transitReg}.
 The largest values of  \(d_{\sigma\,\sigma'}^j\) are attained along the ellipse boundary. Here   one generically finds \(d_{\sigma\,\sigma'}^j\sim j^{-1/3}\) while in the vicinity of the four tangent points of the boundary, where in addition either \(\sigma\) or \(\sigma'\) take on values of \(\pm j\), the scaling is \(d_{\sigma\,\sigma'}^j\sim j^{-1/4}\). As we show below, such  enhanced  scaling at the boundary of  \eqref{eq:spec:transitReg} is responsible for large  spectral fluctuations in the model at $T=2$.

To analyze the spectrum of   \(\Ut\) it turns out to be rather instructive to treat its second power which  can be  represented as the product
\begin{equation}
    \Ut^2=\Uti\, \Uto
\end{equation}
of the diagonal matrix \(\Uti\) and an ``almost permutation''    $\Uto=\Utk\Uti\Utk$.  More specifically,  we show in  \ref{sec:apx:dualEvec} that, for  the regime where only  a single PO manifold exists,  the second factor  can be  split into the product  of two matrices $\Uto=(2j+1)\Utp\Utg$, where $\Utp$ is  the following truncated permutation,
 \begin{equation}\langle \sigma_1\sigma_2|\Utp|\sigma'_1\sigma'_2 \rangle = \delta_{\sigma_1+\sigma'_1, g_1 } \, \delta_{\sigma_2+\sigma'_2, g_2},
 \end{equation}
   and    $\Utg$ is a band diagonal matrix   whose elements are of order $1$ near  the diagonal and scale as $j^{-1}$ away of it.  Here, the  constants     \(  g_1, g_2\) are  given by
   \begin{equation}
    g_i= \left\lfloor{(2j+1)(\pi(1+2p_i) -2\gamma)}/{8J}\right\rfloor\,,   \qquad i \in \{1,2\},
    \label{eq:spec:g}
   \end{equation}
 with  $p_i$'s  being   integers  such that \(-2j\leq g_i\leq 2j\) holds. In the regime of only a single PO manifold the last condition determines $p_i$ (resp. $g_i$) uniquely \ie $p_1=p_2=p$ (resp. $g_1=g_2=g$). For the sake of  simplicity of exposition we  focus below  on this particular case and later briefly comment on the extension   of the results to the regimes where multiple PO manifolds exist.   
   
To simplify the problem  further  we  substitute  $\Utg$ with the unity matrix   and consider the spectrum of $(2j+1)\Uti \Utp$ instead. 
Recalling the diagonal structure of $\Uti$ it is  straightforward to see that   the eigenvectors of $(2j+1)\Uti\Utp$ take a simple form:
  
 \begin{equation}
  \psi_{(\sigma_1,\sigma_2)}= C_1|\sigma_1,\sigma_2\rangle \pm C_2|g-\sigma_1, g-\sigma_2\rangle,\label{eq:eigenV}
 \end{equation}
 with the corresponding eigenvalues $\Lambda_{(\sigma_1,\sigma_2)}$ given by 
 \begin{equation}
  \frac{\Lambda_{(\sigma_1, \sigma_2)}}{2j+1}= \pm\left(\langle \sigma_1,\sigma_2|\Uti |\sigma_1,\sigma_2\rangle \langle g-\sigma_1,g-\sigma_2 |\Uti |g-\sigma_1, g-\sigma_2\rangle\right)^{\frac{1}{2}}\! .
 \end{equation}
 This in turn can be written down in terms of Wigner's $d$-functions as 
 \begin{equation}
  \Lambda_{(\sigma_1,\sigma_2)}= \pm e^{-i(2\gamma-\pi) g}(2j+1) d^j_{\sigma_2,\sigma_1}(\beta) \, d^j_{g-\sigma_2, g-\sigma_1} (\beta), \label{eq:eigenVal}
 \end{equation}
 where we have explicitly  separated the complex phase from the amplitude. Having at hand the approximate  spectrum (\ref{eq:eigenV}, \ref{eq:eigenVal}) of the operator $\Ut^2$ we can straightforwardly write down  the corresponding eigenvalues and eigenvectors  for  $\Ut$,
\begin{equation}
    \psi \approx \psi_{(\sigma_1,\sigma_2)}\pm \Lambda^{-1/2}_{(\sigma_1,\sigma_2)}\Ut \psi_{(\sigma_1,\sigma_2)}, \qquad \tilde{\lambda} \approx \pm \Lambda^{1/2}_{(\sigma_1,\sigma_2)}\,.
    \label{eq:SpecFirstPower}
\end{equation}
The first term    $\psi_{(\sigma_1,\sigma_2)}$ is sharply localized in the \( (\sigma_1,\sigma_2)\) space. In contrast, the second term  \( \Ut \psi_{(\sigma_1,\sigma_2)}\)  is localized in momentum space \( (\bar\sigma_1,\bar\sigma_2)\) due to the presence of the $\Utk$ factor  in $\Ut$. This is in agreement with the previous numerical observation \eqref{eq:spec:evecStruct}. 
 (It is important to emphasize  that the actual eigenstates of  $\Ut^2$ have a finite support, while the eigenstates of the approximation  $\Uti\Utp$ are point-like localized.)

\begin{figure}
    \centering
    \includegraphics[height=0.25\textwidth]{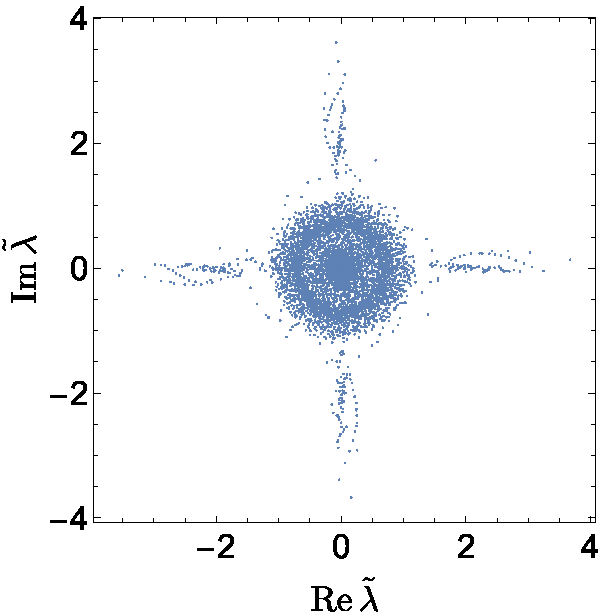}
    \hfill
    \raisebox{0.15\height}{\includegraphics[height=0.23\textwidth]{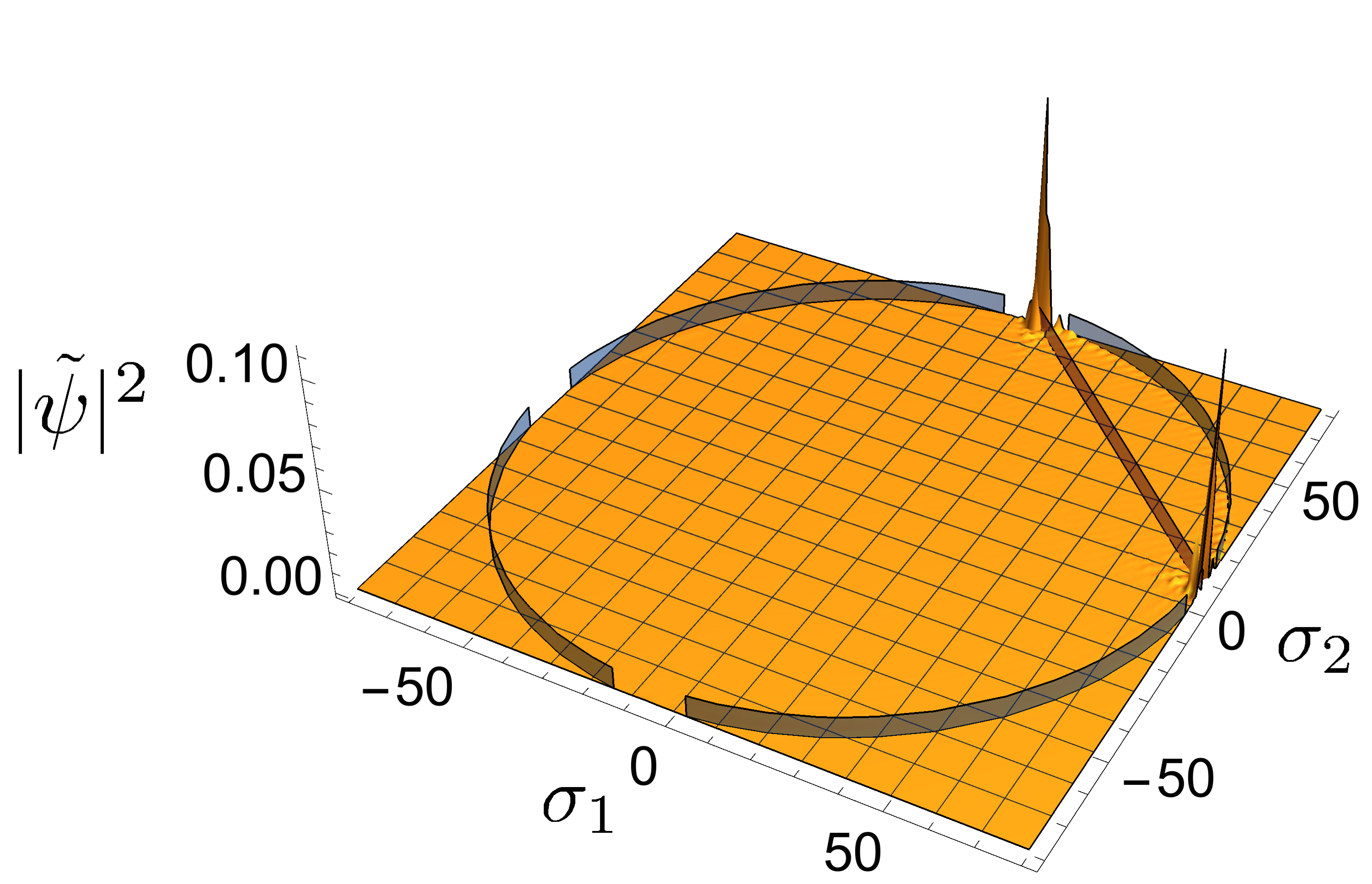}}
    \hfill
    \includegraphics[height=0.25\textwidth]{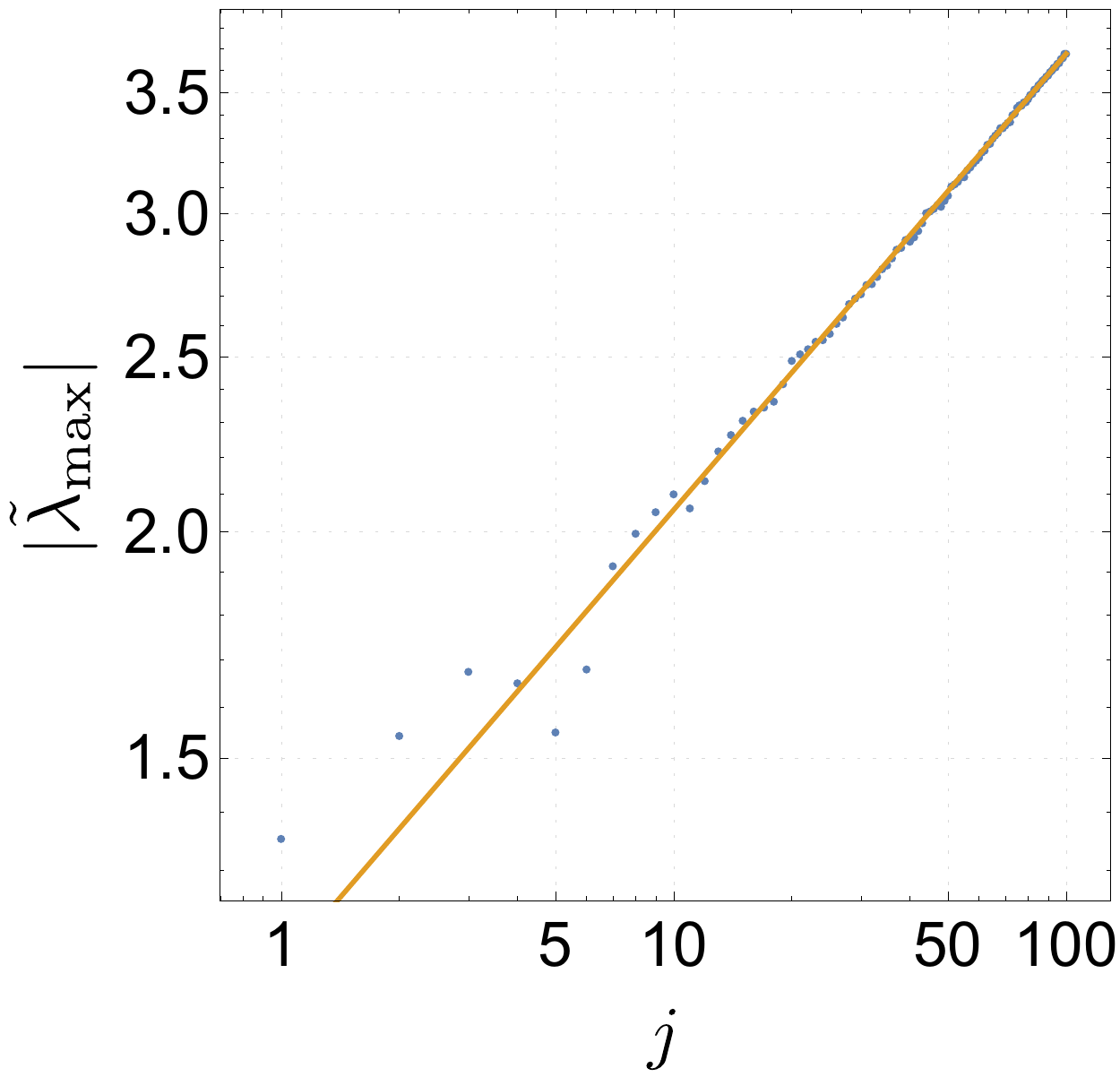}
    \caption[Spectrum, eigenvector and $\ldmax$ scaling for $\Ut$ ($T\teq 2$) for parameters where the eigenvector localizes at the tangent points.]{\conline From left to right: spectrum of the dual operator (\(j\teq 100\)), corresponding largest eigenvector (\(j\teq 80\)) and scaling of the largest eigenvalue in dependence of \(j\) (numerical slope \(\alpha_0\teq 0.252\pm0.004\)). Compare also figure \ref{fig:spec:T2diagSpec} for further information. Parameters are chosen as \(T\teq2\), \(J\teq 0.45\), \(b^x\teq 0.789802\) and \(b^z\teq 0.483691\) such that the eigenvector localizes at the tangent points, see text.}
    \label{fig:spec:onTangentPoint}
\end{figure}

In order to find the largest eigenvalues of $\Uti\Utp$  (resp. $\Ut$) we need to look for  $(\sigma_1,\sigma_2)$ such that  $|\Lambda_{(\sigma_1,\sigma_2)}|$  reaches its maximum value. By eq.~\eqref{eq:eigenVal} this happens  whenever  both  $(\sigma_1,\sigma_2)$ and $(g-\sigma_1,g-\sigma_2)$  belong to the boundary of the ellipse \eqref{eq:spec:transitReg}. In other words, the localization points of the corresponding eigenvectors are located at the intersection points between the line  $\sigma_1+\sigma_2=g$ and the ellipse boundary. Figure \ref{fig:spec:T2diagSpec} shows such an eigenvector  $\tilde\psi$ of $\Ut$ corresponding to its  largest eigenvalue as well as the respective line together with the ellipse boundary. As can be observed, the  localization points of $\tilde\psi$ are, indeed,  in a good agreement with the above prediction.
By  eqs.~(\ref{eq:eigenVal}, \ref{eq:SpecFirstPower}) the phases of the four  largest eigenvalues  of $\Ut$  are given by 
\begin{equation}
    (2\gamma-\pi) g/2 +\frac{l\pi}{2}= \left(2J\chi^2 (j+1/2)+\frac{l\pi}{2}\right) \mod \, 2\pi, \quad l=1,2,3,4
    \,,
\end{equation}
where on the left hand side we used the identity $2J\chi=(\pi/2-\gamma) \mod\, \pi$, see \eqref{eq:spec:eulerAnglesDef}. After taking into account the expression \eqref{eq:po:smanBase:single} for  the actions of the PO manifolds this  immediately yields the previously, empirically found eq.~\eqref{eq:fourLargestEig}.  The same approach  can be used to evaluate  the absolute value  of $\tilde\lambda$. By eqs.~(\ref{eq:eigenVal}, \ref{eq:SpecFirstPower})  we have   
$ |\Lambda_{(\sigma_1,\sigma_2)}|^{1/2}\sim j^{\alpha_0}$,
 with  $\alpha_0 =1/4$ if ${\sigma_1,\sigma_2}$ belongs to the tangent points of the ellipse and   $\alpha_0 =1/6$, otherwise. In fig.~\ref{fig:spec:onTangentPoint} we check this prediction for specially tuned parameters such that the localization points are at the tangent points of the ellipse boundary.
In this case the maximum possible scaling $\alpha_0 =1/4$ is clearly observed, see fig.~\ref{fig:spec:onTangentPoint} (right). On the other hand, for generic parameters   the scaling exponent is typically above the naive prediction $1/6$. This is   probably a  consequence of the fact that some (small)  portion  of the eigenstate $\tilde \psi$ is always localized at the  tangent points of the ellipse. A detailed investigation of this question would require taking into account the precise structure of $\Utg$ which is beyond the scope of the present paper.

 So far, we considered the case of  the single manifold regime.  In the parameter regime for multiple PO manifolds  several combinations of    different integers $(p_1, p_2)$ exist, see  \ref{sec:apx:dualEvec},  such that   the corresponding $g_1, g_2$ satisfy the conditions $-2j \leq g_i\leq 2j$, $i=1,2$.  As a result,  the matrix $P$ is provided by a sum of permutations  -- each one  corresponds  to some  particular solution $(p_1,p_2)$.   To find the spectrum of  eigenvalues   of $\Ut$ one follows the same  procedure as in the single manifold case. Accordingly,  for the largest eigenvalues  of $\Ut$ both points $(\sigma_1, \sigma_2)$ and  $ (g_1-\sigma_1, g_2- \sigma_2)$   should belong to the boundary of  the ellipse \eqref{eq:spec:transitReg}. This condition defines a  pair  of  points on the ellipse boundary for each solution $(p_1, p_2)$.    All these  points serve as centers  of localization for the corresponding eigenvectors $\tilde \psi$, see fig.~\ref{fig:spec:T2diagSpec}.

%%%%%%%%%%%%%%%%%%%%%%%%%%%%%%%%%%%%%%%%%%%%%%%%%%%%%%%%%%
% NL-KIC
%%%%%%%%%%%%%%%%%%%%%%%%%%%%%%%%%%%%%%%%%%%%%%%%%%%%%%%%%%
\section{ Kicked Spin Chain Model for $\K \neq 0$}
\label{sec:nlkic}
\begin{figure}[tbp]
\includegraphics[width=0.24\textwidth]{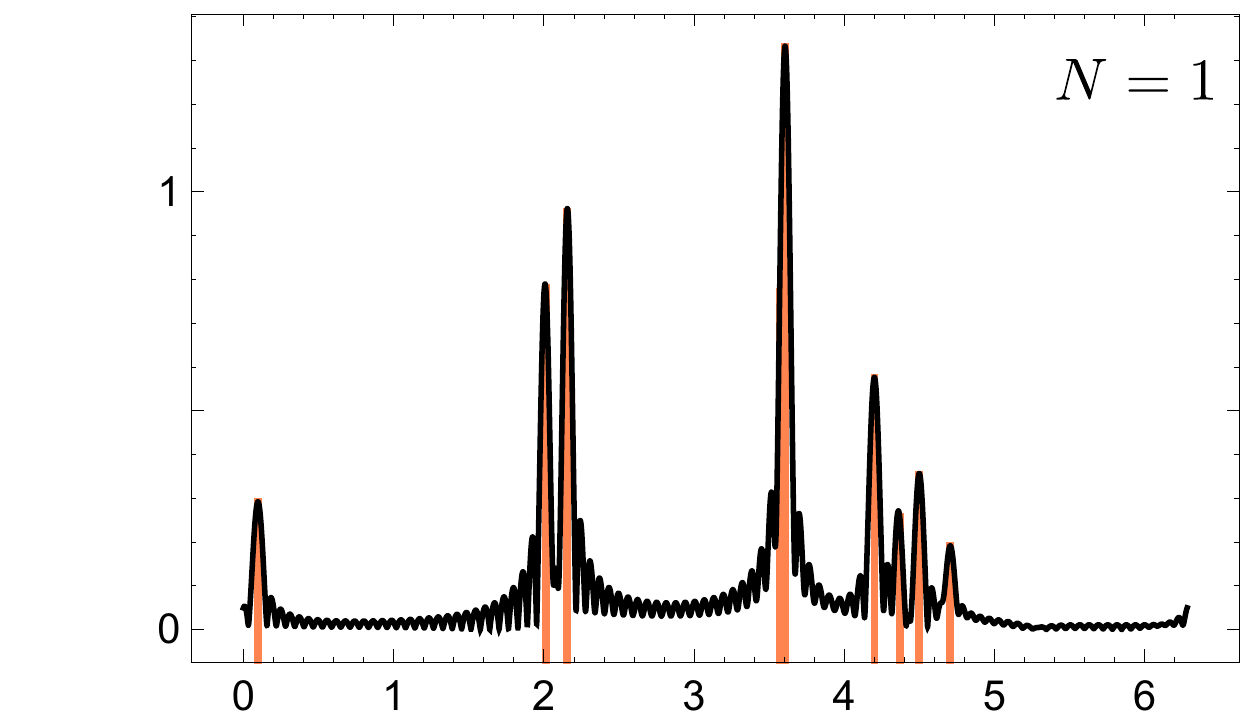}
\hfill
\includegraphics[width=0.24\textwidth]{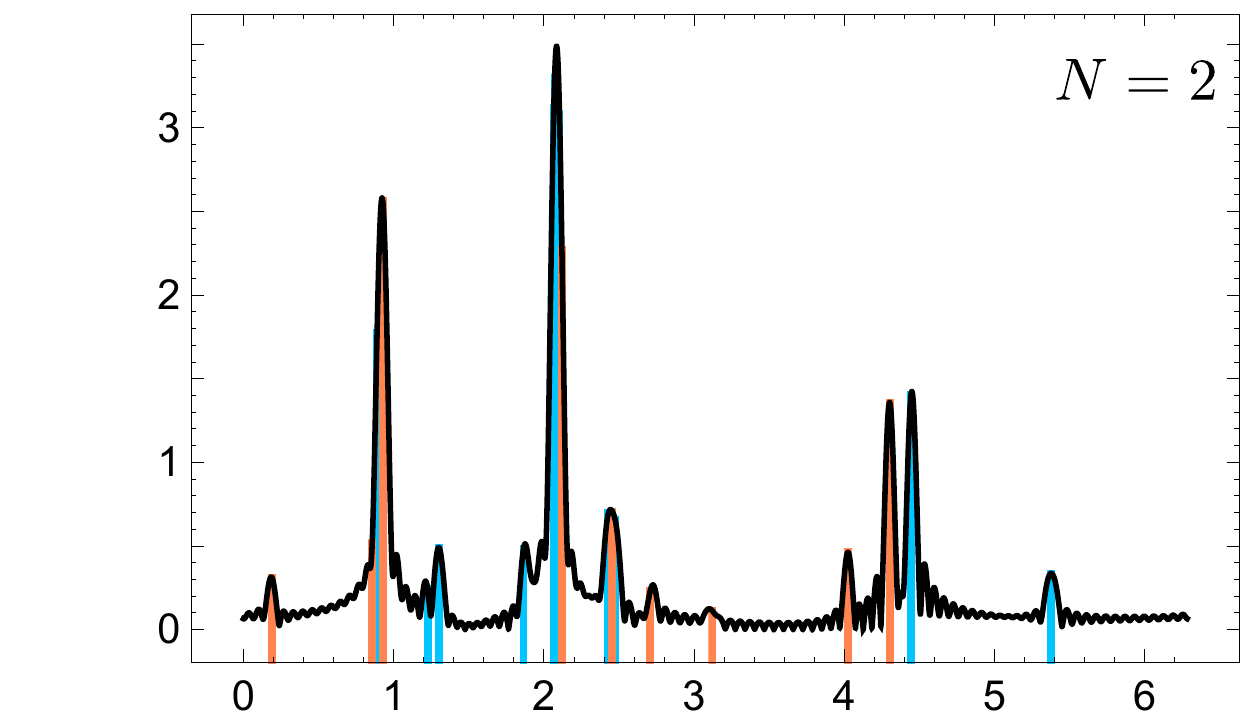}
\hfill
\includegraphics[width=0.24\textwidth]{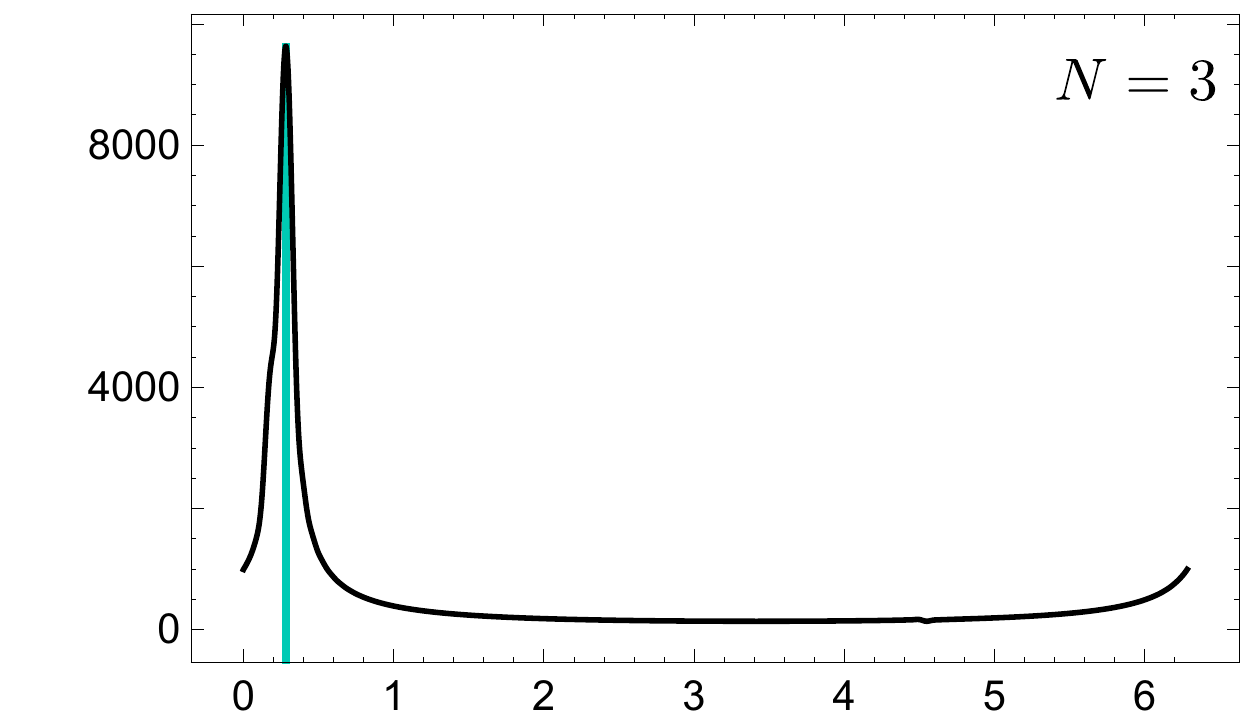}
\hfill
\includegraphics[width=0.24\textwidth]{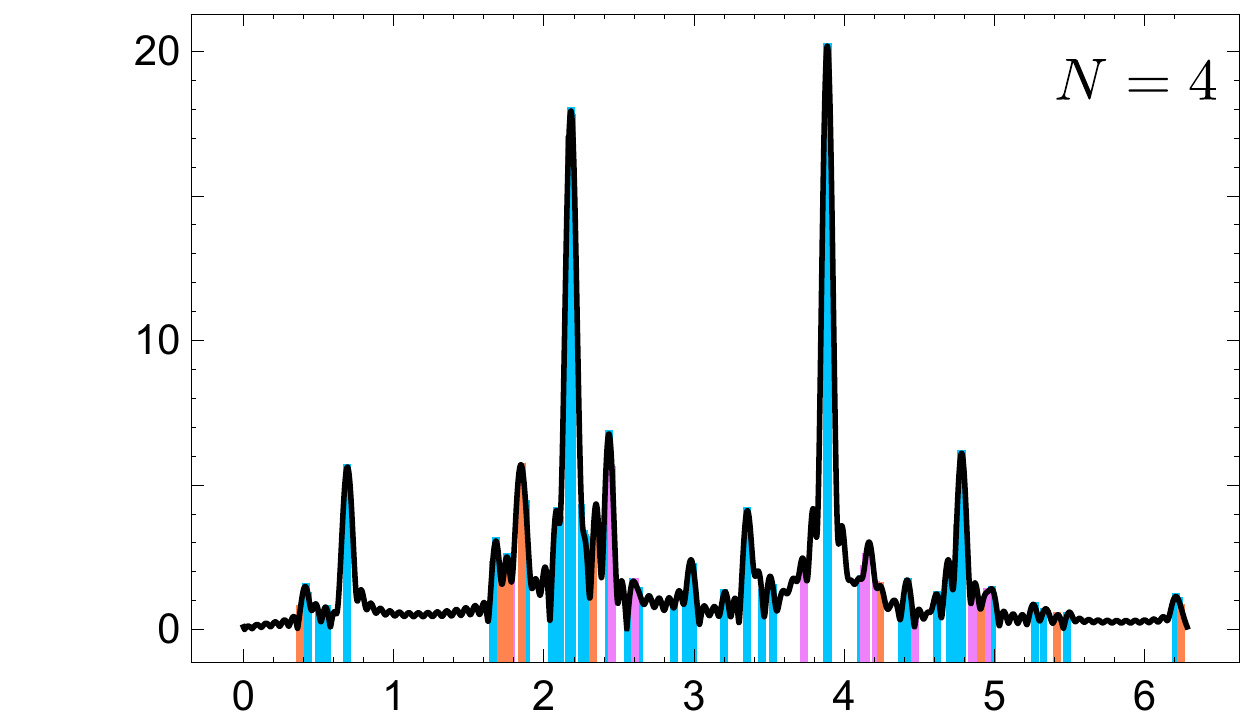}
\\
\includegraphics[width=0.24\textwidth]{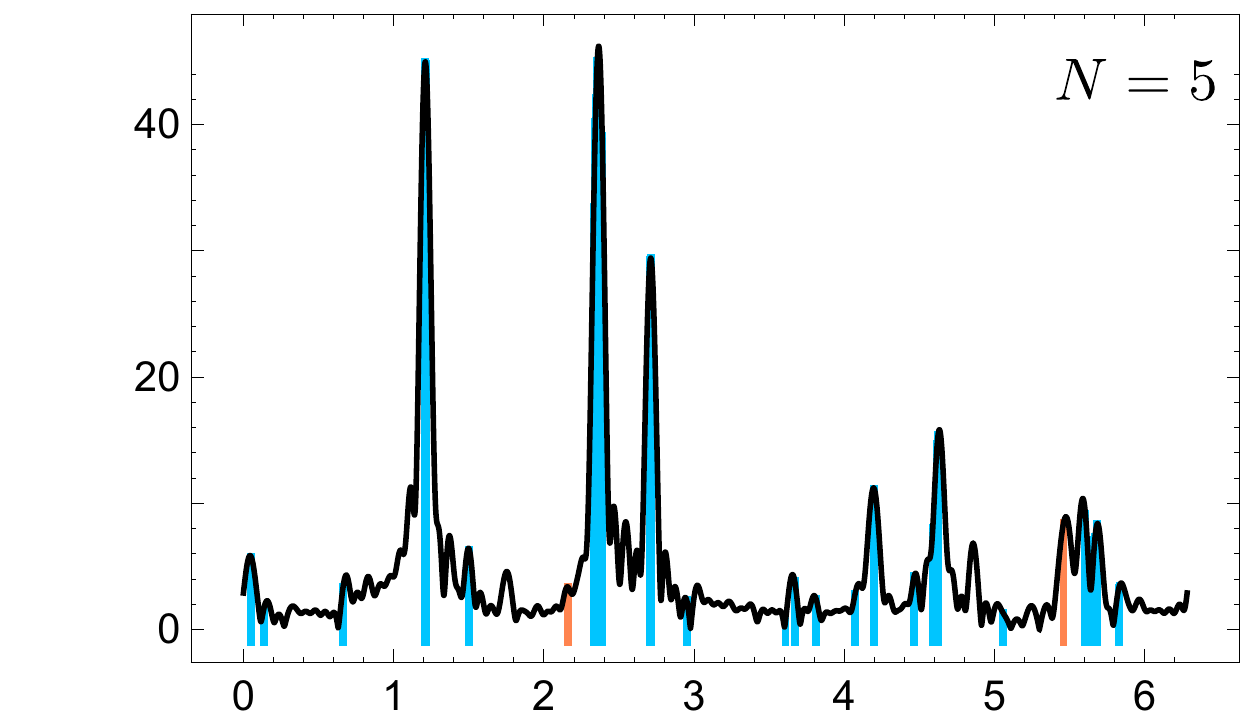}
\hfill
\includegraphics[width=0.24\textwidth]{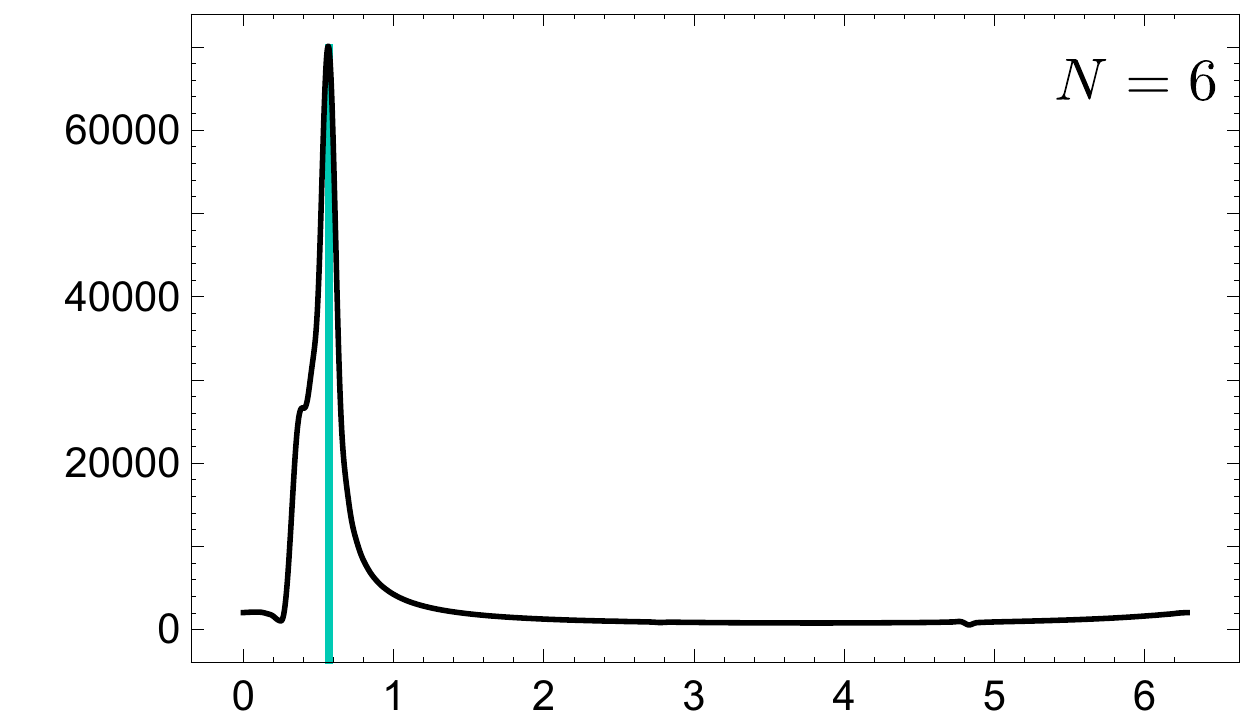}
\hfill
\includegraphics[width=0.24\textwidth]{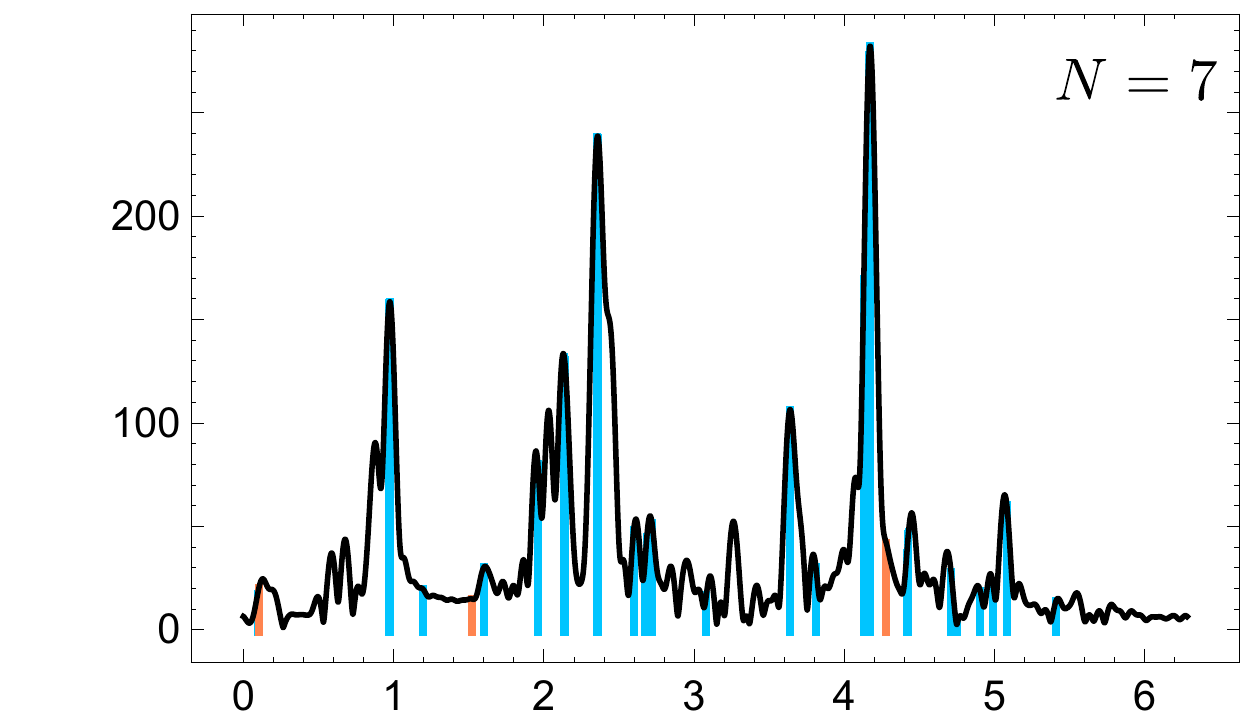}
\hfill
\includegraphics[width=0.24\textwidth]{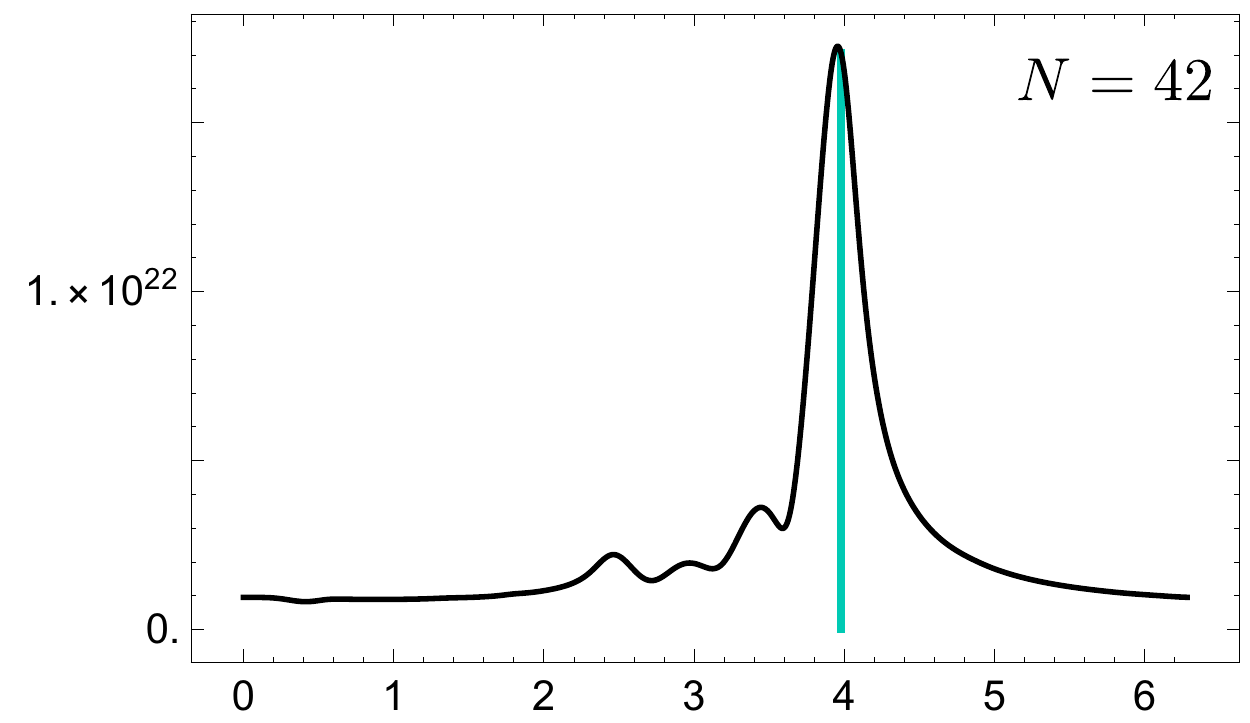}
\caption[Action spectrum for various $N$ and $T\teq 2$ with non-zero $\K $ such that manifolds occur for $N\teq 3k$.]{
    \conline Absolute value \(|\rho(\Sact)|\) of the action spectrum for non-zero $\K $ over \(\Sact\) for \(T\teq 2\) using \(\jcut\teq 100\). The particle number is indicated in the upper right corners and for \(N\teq 5\), \(N\teq 7\) respectively, only orbits with \(\Aga>0.25\), \(\Aga>0.5\) respectively, are shown. The system parameters are \(J\teq\sqrt{\pi}/3\), \(\K \teq J/2\) and \(b^x\teq b^z\teq 0.9\). For the color coding see figure \ref{fig:action:sftT2}.
    }
\label{fig:nlkic:action3fold}
\end{figure}

\begin{figure}
    \centering
    \includegraphics[height=0.25\textwidth]{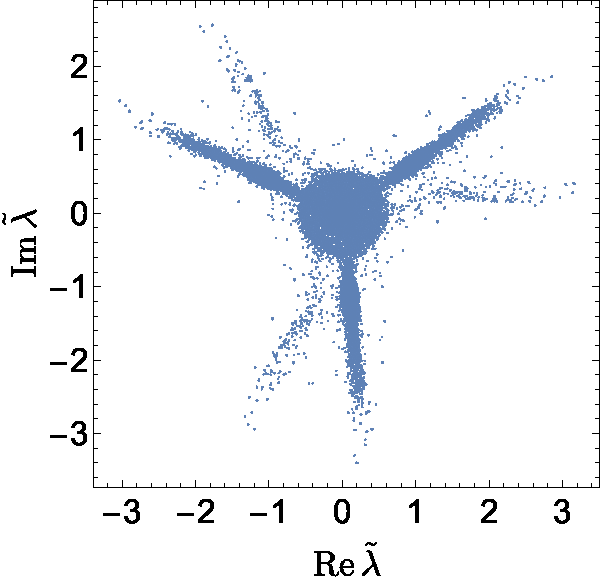}
    \hfill
    \raisebox{0.15\height}{\includegraphics[height=0.23\textwidth]{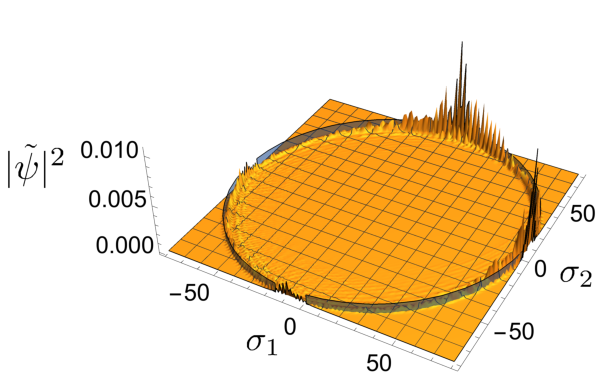}}
    \hfill
    \includegraphics[height=0.25\textwidth]{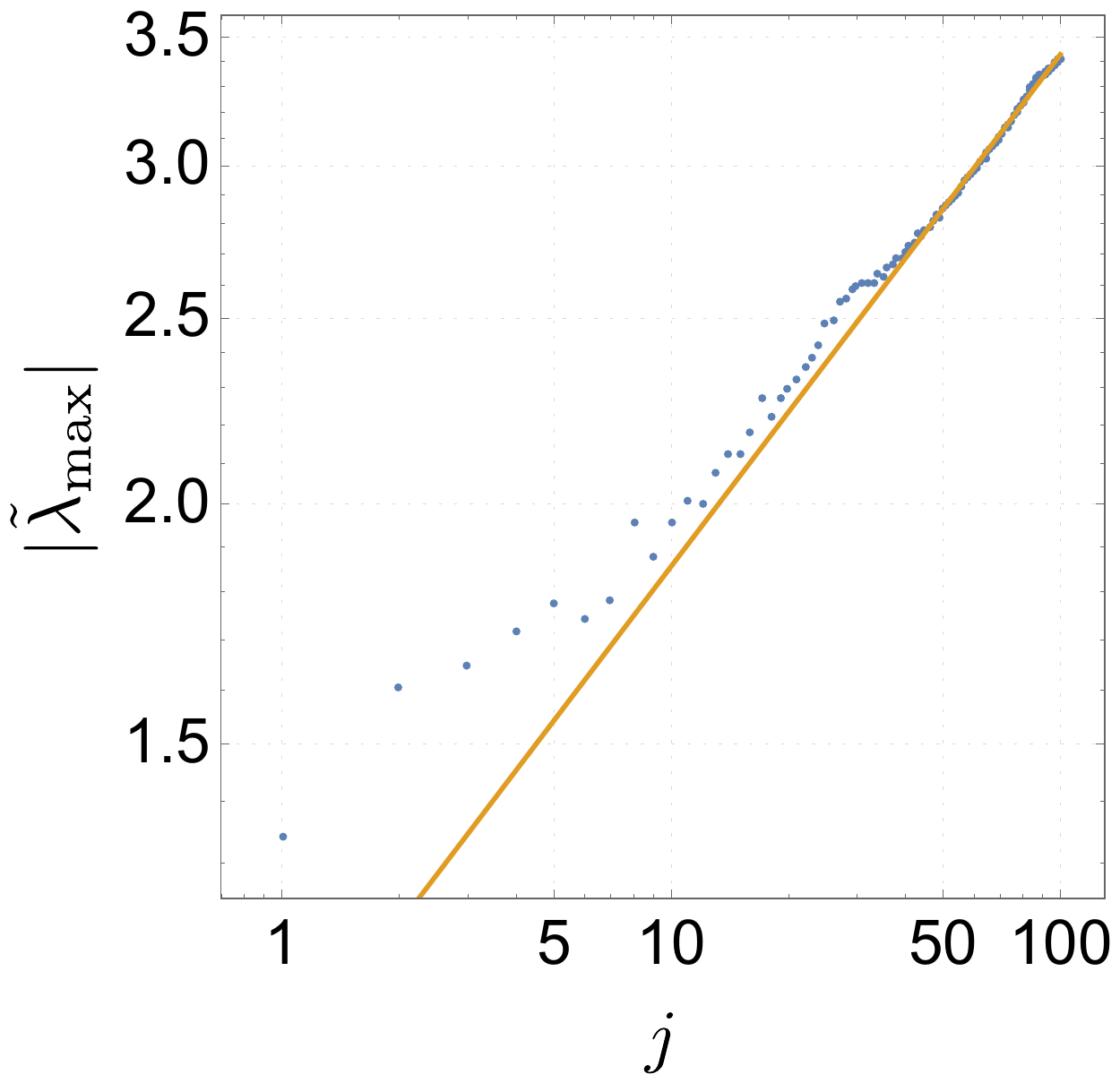}
    \caption[Spectrum, eigenvector and $\ldmax$ scaling of $\Ut$ for $T\teq 2$ and $K\neq 0$ such that the spectrum possesses a 3-fold symmetry.]{\conline Spectrum of the dual operator containing a non-linear part for \(j\teq 100\) (left panel), corresponding largest eigenvector (\(j\teq 80\)) (middle) and scaling of the largest eigenvalue in dependence of \(j\) (numerical slope \(\alpha_0\approx 0.167\)). Parameters are chosen as \(T\teq2\), \(J\teq \sqrt{\pi}/3\), \(\K \teq \sqrt{\pi}/6\) and \(b^x=b^z=0.9\).}
    \label{fig:nlkic:dualScalings:3fold}
\end{figure}

\begin{figure}[tbp]
\includegraphics[width=0.24\textwidth]{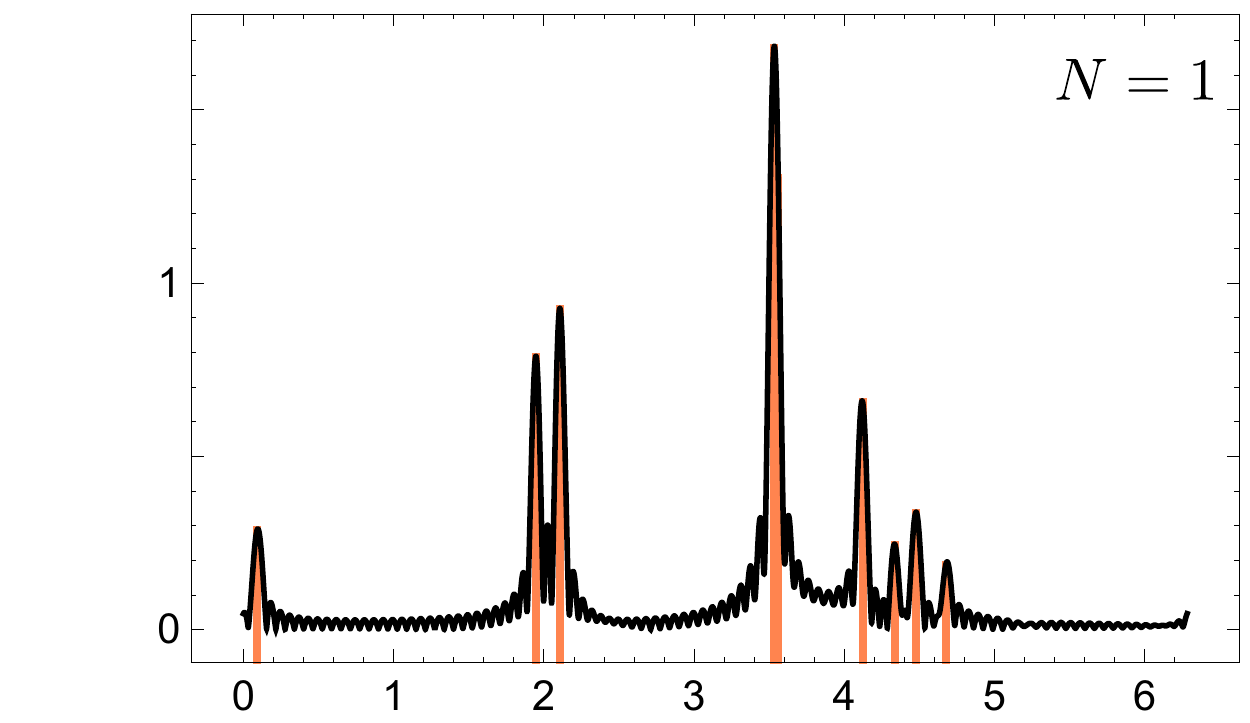}
\hfill
\includegraphics[width=0.24\textwidth]{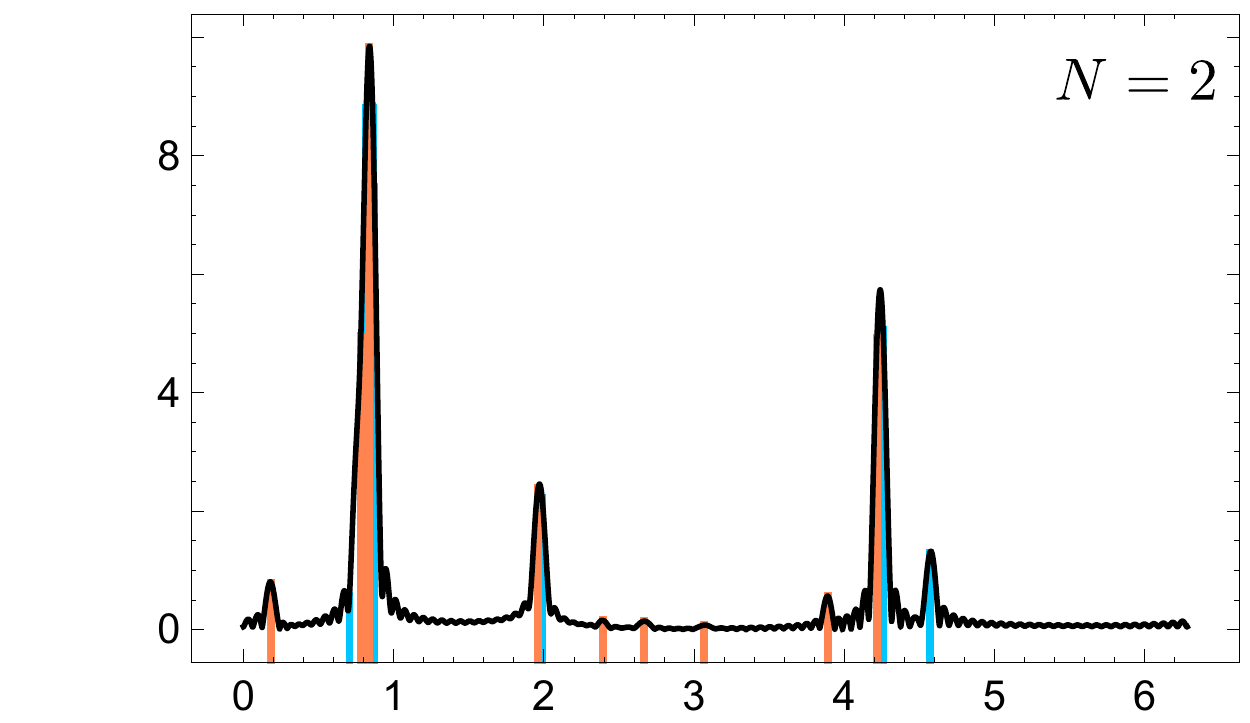}
\hfill
\includegraphics[width=0.24\textwidth]{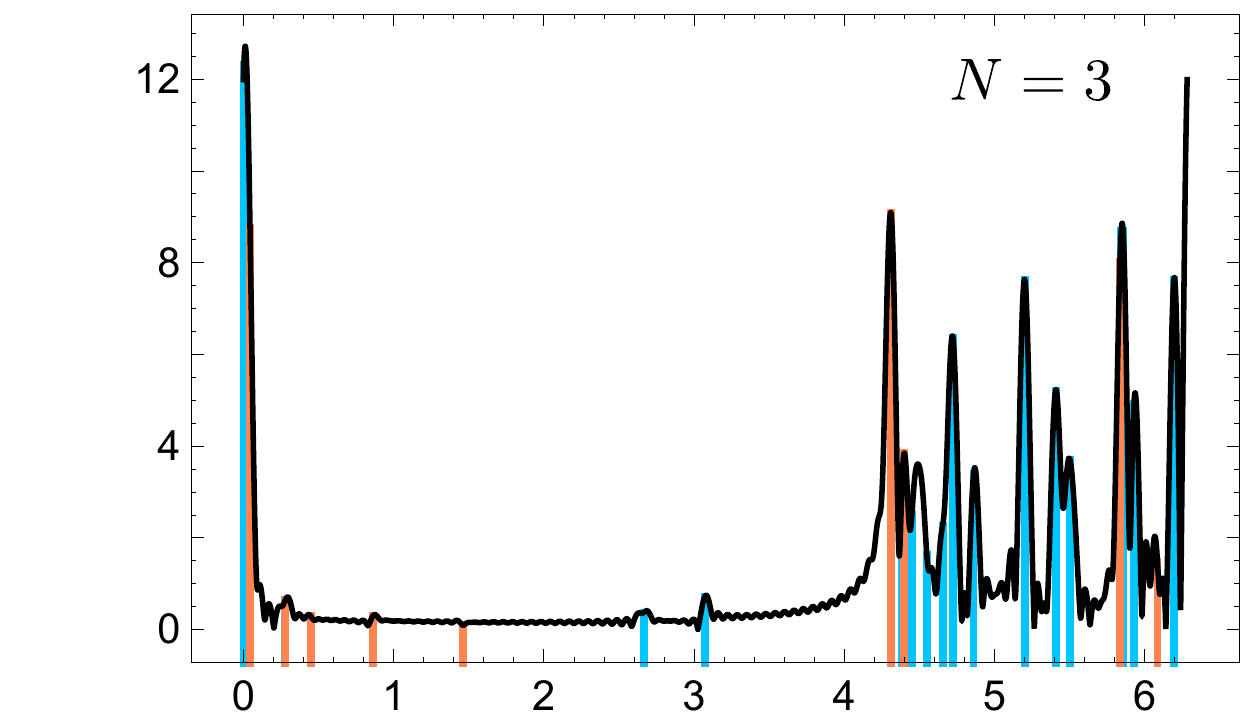}
\hfill
\includegraphics[width=0.24\textwidth]{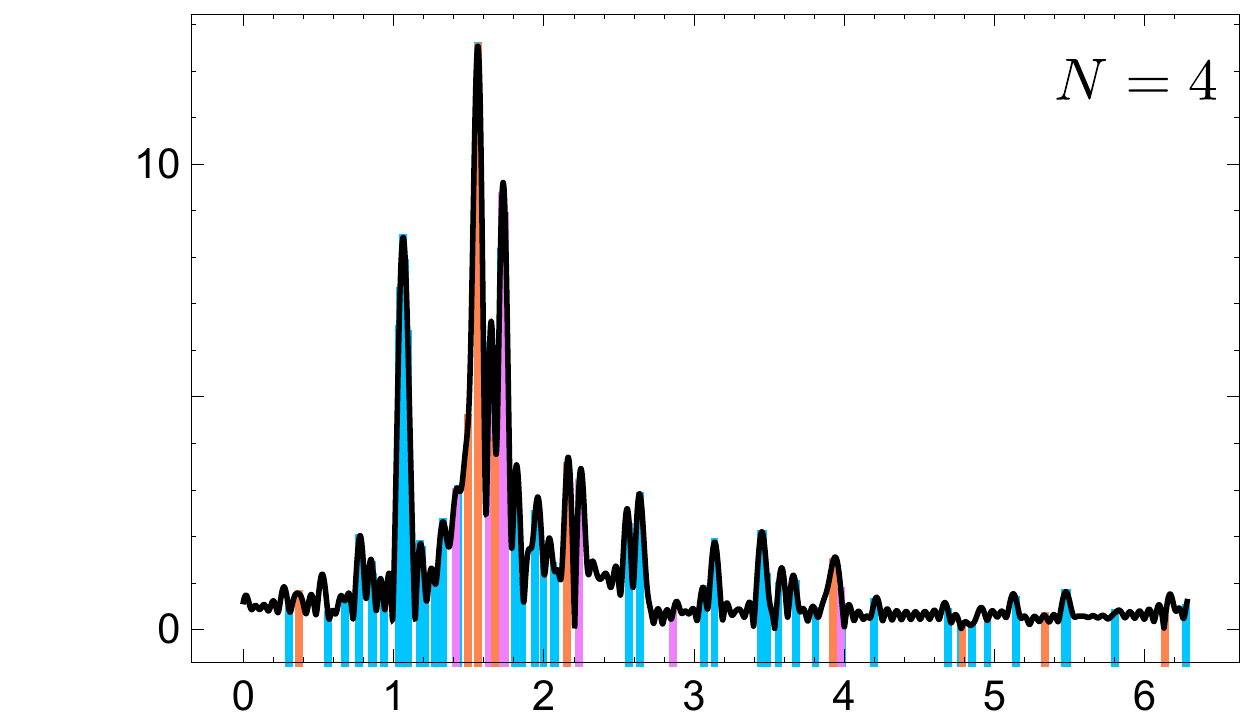}
\\
\includegraphics[width=0.24\textwidth]{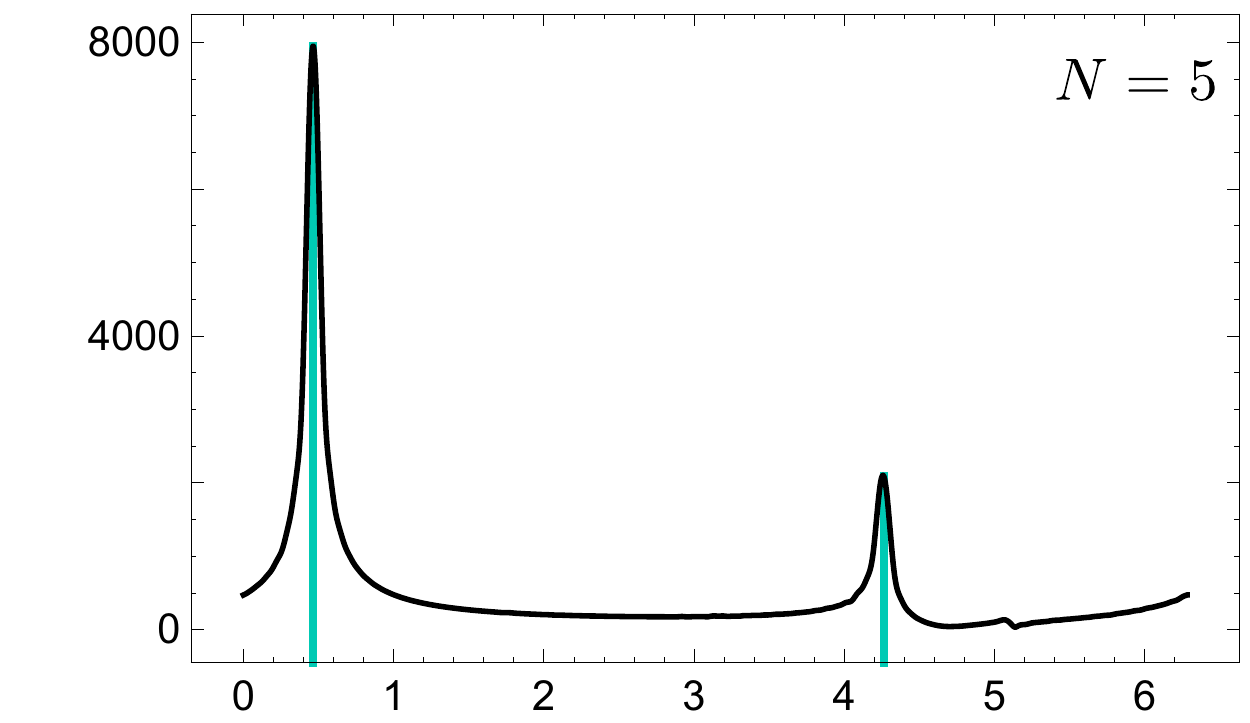}
\hfill
\includegraphics[width=0.24\textwidth]{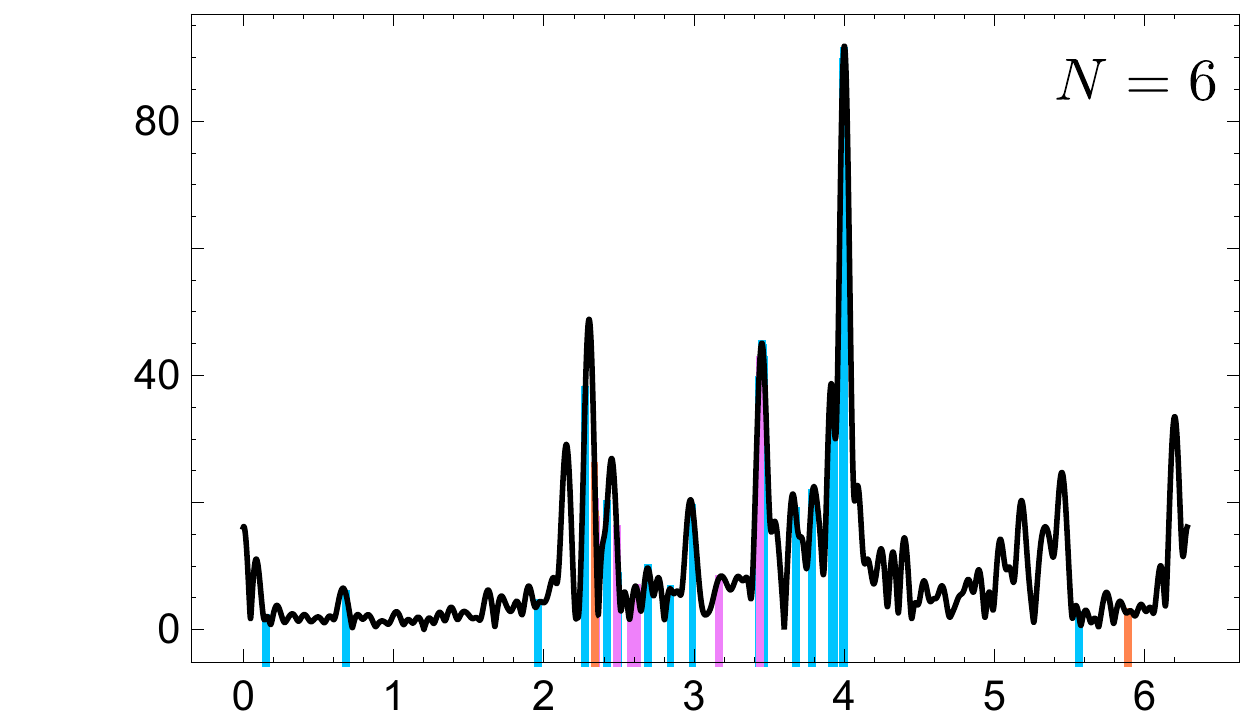}
\hfill
\includegraphics[width=0.24\textwidth]{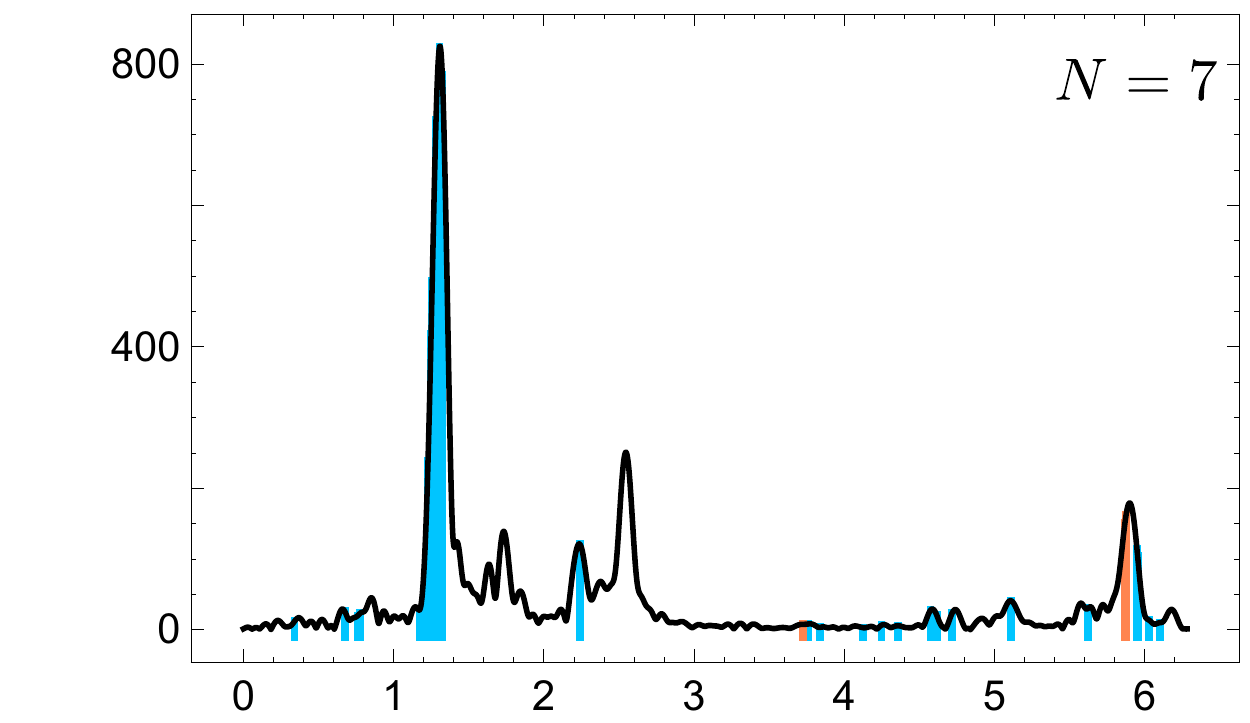}
\hfill
\includegraphics[width=0.24\textwidth]{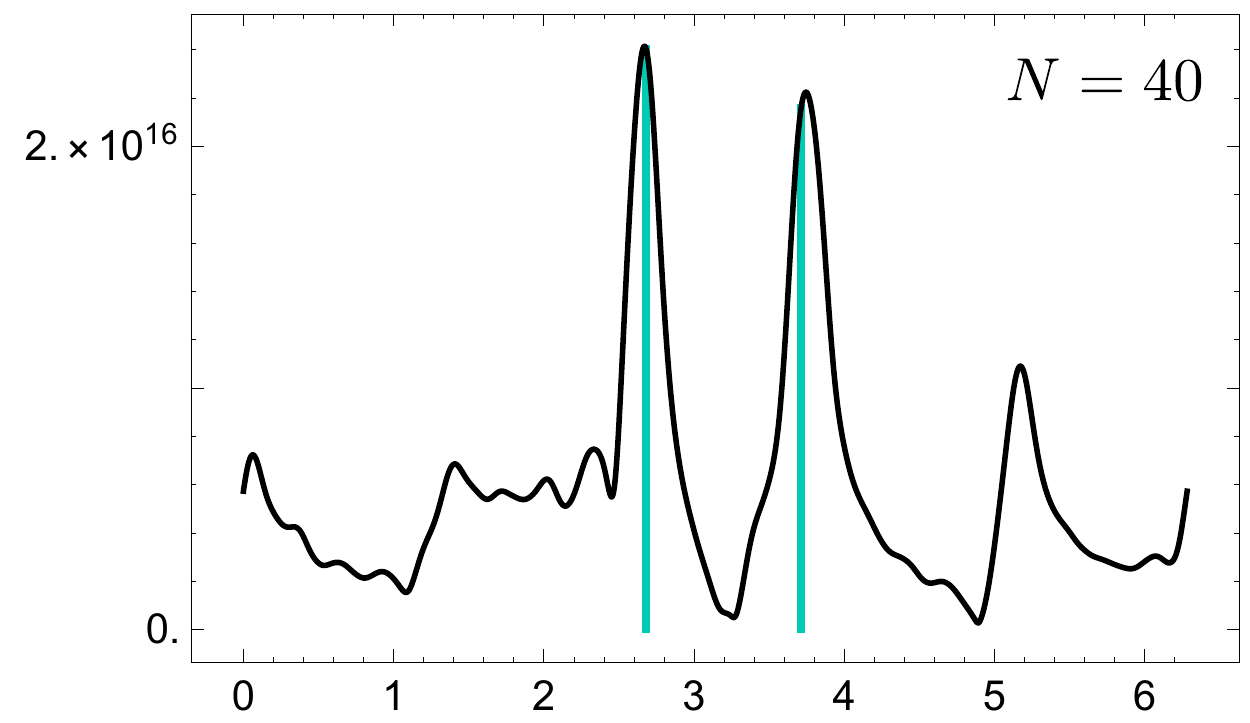}
\caption[Action spectrum for various $N$ and $T\teq 2$ with non-zero $\K $ such that manifolds occur for $N\teq 5k$.]{
    \conline Absolute value \(|\rho(\Sact)|\) of the action spectrum over \(\Sact\) for \(T\teq 2\) using \(\jcut\teq 100\). The particle number is indicated in the upper right corners and for \(N\teq 6,7\) only orbits with \(\Aga>0.5\) are shown. The other system parameters are \(J\teq0.5\), \(\K \teq (\sqrt{5}+1)/8\) and \(b^x\teq b^z\teq 0.9\). For the color coding see figure \ref{fig:action:sftT2}.
    }
\label{fig:nlkic:action5fold}
\end{figure}

\begin{figure}
    \centering
    \includegraphics[height=0.25\textwidth]{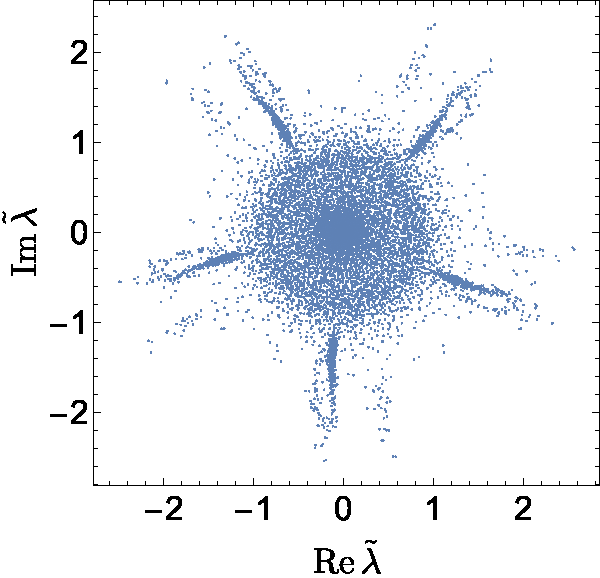}
    \hfill
    \raisebox{0.15\height}{\includegraphics[height=0.23\textwidth]{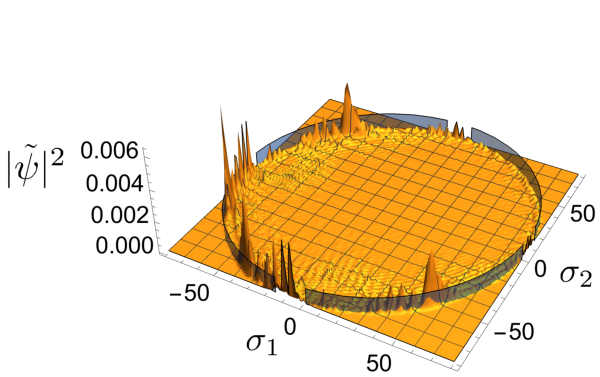}}
    \hfill
    \includegraphics[height=0.25\textwidth]{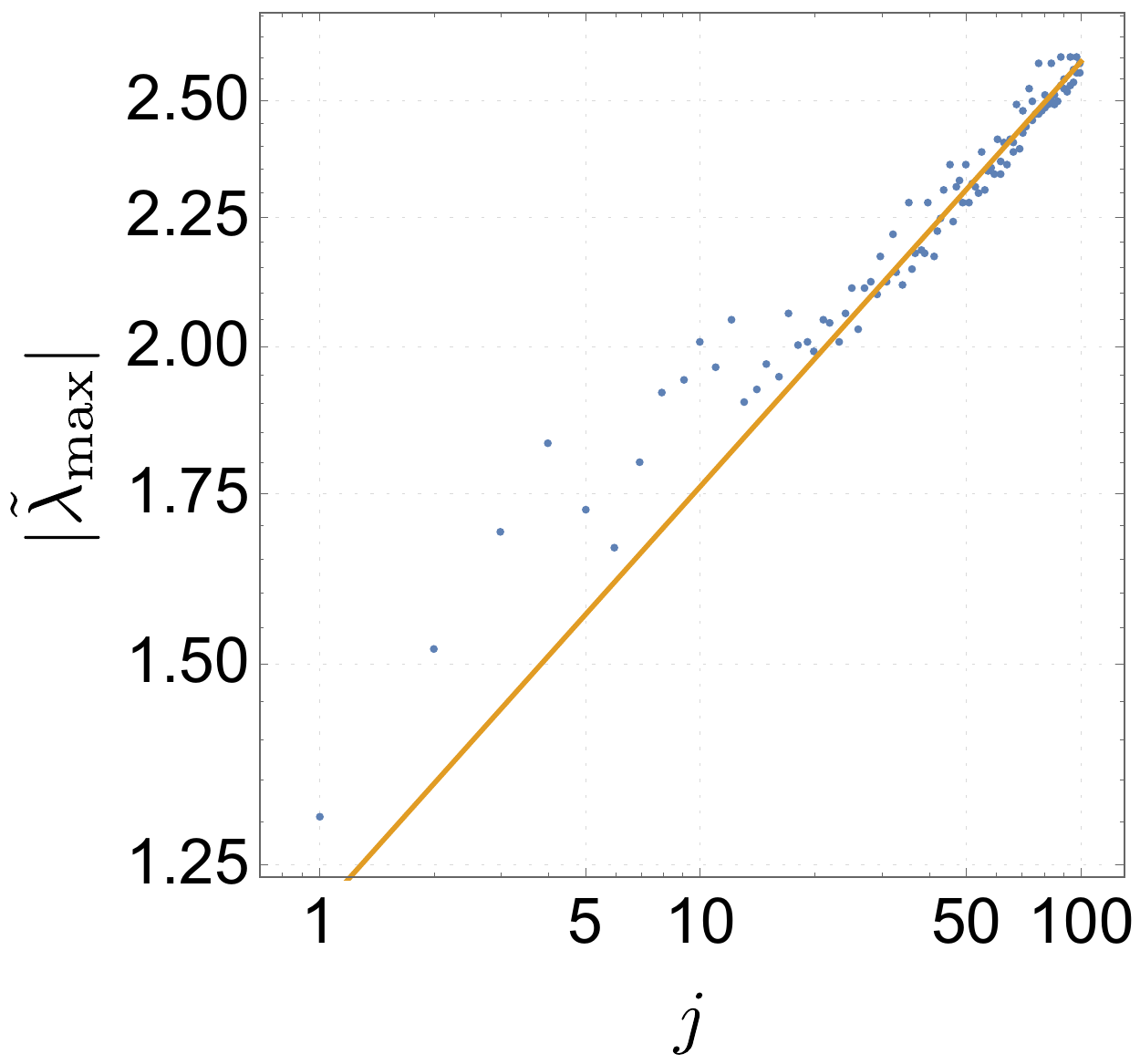}
    \caption[Spectrum, eigenvector and $\ldmax$ scaling of $\Ut$ for $T\teq 2$ and $\K \neq 0$ such that the spectrum possesses a 5-fold symmetry.]{\conline Spectrum of the dual operator containing a non-linear part for \(j\teq 100\) (left panel), corresponding largest eigenvector (\(j\teq 80\)) (middle) and scaling of the largest eigenvalue in dependence of \(j\) (numerical slope \(\alpha_0\approx 0.167\)). Parameters are chosen as \(T\teq2\), \(J\teq 1/2\), \(\K \teq (\sqrt{5}+1)/8\) and \(b^x=b^z=0.9\).}
    \label{fig:nlkic:dualScalings:5fold}
\end{figure}

%Figures stored in NonLinearKIC(att2_exp1)
So far, we considered  a particular case of the Kicked Spin Chain model  as we set $\K =0$  in  the interaction part of the   Hamiltonian \eqref{eq:kic:hi}. In this section  we allow  an arbitrary strength  \(\K \)  of the  quadratic term. The kick part of the dual operator is now reminiscent of the kernel of the so called fractional Fourier transformation \cite{namias,candan}. As we show below,  the core result of the previous sections -- the emergence  of the anomalously large spectral fluctuations for  chain lengths $4 k$, $k\in \mathbb{N}$  -- reappears here again with a peculiar twist.  For spin chains of the length $N=N_0 k$,  $k\in \mathbb{N}$ the anomalously  large fluctuations, dominated by PO manifolds, emerge  
when the ratio between interaction and torsion strength $\mu=\K /J$ attains    the following set of values:
\begin{equation}
   \mu=-\cos{\frac{2\pi p}{N_0}}
    \qquad
    p\in\{ 1,2,\ldots,N_0-1\}\,.
    \label{eq:nlkic:mu}
\end{equation} In other words, the model possesses large spectral fluctuations for spin chains of lengths $N=N_0 k$  with an arbitrary $N_0$ when  the parameter $\mu$ is tuned according to  \eqref{eq:nlkic:mu}. From this perspective the previous $\K =0$ case is merely  a special one corresponding to $N_0=4$.   
To illustrate this we show in  figure \ref{fig:nlkic:action3fold}  the action spectrum for \(T=2\) at a ratio of \(\mu \teq 1/2\) where we find strong peaks for every \(N\teq 3k\).
The spectrum of the  corresponding dual operator \(\Ut\) shows  a three-fold symmetry, see figure \ref{fig:nlkic:dualScalings:3fold}. As one can observe, the absolute value of the largest eigenvalue scales algebraically with $j$, which explains the large spectral fluctuations in that case. Furthermore, the corresponding  eigenvector looks  structurally  similar to what occurred in the \(\K \teq 0\)  case. As a further example, for \(N_0\teq 5\) and \(\mu\teq (\sqrt{5}+1)/4\)  one has large spectral fluctuations  for all chains with \(N\teq 5k\), see fig. \ref{fig:nlkic:action5fold}. The dual operator in this case, see figure \ref{fig:nlkic:dualScalings:5fold}, has a 5-fold symmetry.

To explain condition \eqref{eq:nlkic:mu}  we turn back to the classical dynamics of the model. Looking at  periodic orbits of the system  for $\K \neq 0$ we again find 
\begin{equation}
( \underline{R}_I(4J \chi_n) \underline{R}_{\vec{b}} )^2\teq \mathds{1}, \qquad n=1,\dots N_0
\label{eq:nlkic:POManif}
\end{equation}  as conditions for the existence of PO manifolds,  where we introduced
$\chi_n = p_{n-1}+2 \mu p_n +p_{n+1}$, 
compare also with \eqref{eq:kic:classRot}. The above conditions fix uniquely (up to an addition of factors $(4J)^{-1} 2\pi k_n $,   $ k_n\in \mathbb{Z}$) the variables $
    \chi_n$, but not necessarily the $p_n$.  The PO manifolds  emerge whenever the
$N_0$ conditions  \eqref{eq:nlkic:POManif} do not resolve the set $p_1, \dots p_{N_0}$ uniquely.
For instance, in the case of  $3$ spins and  \(\mu = 1/2\) one finds that all \(\chi_n\) are identical to \(p_1+p_2+p_3\). This linear dependence explains the emergence of classical PO manifolds. We can  extend this line of reasoning to   arbitrary $N$ and $\mu$ based on  the cyclic \(N_0\times N_0\) dimensional band matrix
\begin{equation}
    \Zmat=
    \left(\begin{array}{ccccc}
        2 \mu & 1 & 0 \cdots & 0 & 1 \\
        1 & 2 \mu & 1 & 0 & \cdots \\
        0 & 1 & 2 \mu  & \cdots & 0 \\
        \vdots & \ddots & \ddots & & \vdots \\
        1 & 0 &\cdots & 1 & 2 \mu
    \end{array}\right)\,
    \label{eq:nlkic:Zmat}
\end{equation}
connecting the $\chi_n$ and $p_n$ variables via $\vec{\chi}\teq \Zmat \vec{p} $.
PO manifolds appear whenever \(\Zmat\) is not of full rank. This happens if one of the   eigenvalues of \(\Zmat\),  
\begin{equation}
    z_n = 
    2\mu+2\cos{\frac{2\pi n}{N_0}}
    \qquad
    n\in\{1,2,\ldots,N_0\}\,.
    \label{eq:nlkic:zev}
\end{equation}
satisfies the condition  $z_n=0$ for some $n$. As one can easily see, this  immediately implies   \eqref{eq:nlkic:mu}. All eigenvalues in   \eqref{eq:nlkic:zev} are  doubly degenerate, except $z_{N_0}$ and $z_{N_0/2}$ (for  even $N_0$). Accordingly,   for all $\mu\neq\pm 1$ from the set  \eqref{eq:nlkic:mu}  we have a freedom to choose two (continuous)  parameters $\eta^{(i)}_1,\eta^{(i)}_2$ at each time step $i=1,2$ such that $p^{(i)}_n(\eta^{(i)}_1,\eta^{(i)}_2)$,  while the  $\chi^{(i)}_n$  are  independent of $\eta^{(i)}_1,\eta^{(i)}_2$. This yields 4-dimensional PO manifolds parametrized by $\eta^{(1)}_1,\eta^{(1)}_2, \eta^{(2)}_1,\eta^{(2)}_2$.    
In the case  $\mu= 1$  the corresponding eigenvalue $z_{N_0/2}$ is non-degenerate and the dimension of the PO manifold is $2$ rather  than $4$  while     the spectrum   of the  dual operator    is distributed  isotropically. Finally, for $\mu=-1$ the parameters $\chi_n$  satisfy $\sum\chi_n=0$ (by definition) which immediately implies  $4J\chi_n=0 \,\mod\, 2\pi$ for each $n$. A simple substitution of this value back  into \eqref{eq:nlkic:POManif} shows that for $\mu=-1$ this equation  might hold only if  $\underline{R}_{\vec{b}}$ is a rotation by $\pi$ itself. Therefore, for a generic value of the magnetic field
 and $\mu=-1$ PO manifolds do not exist.

\begin{figure}
    \centering
    \includegraphics[height=0.25\textwidth]{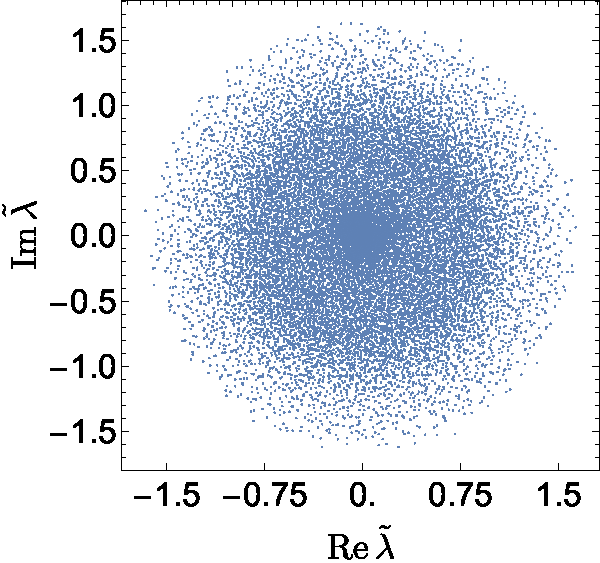}
    \hfill
    \raisebox{0.15\height}{\includegraphics[height=0.23\textwidth]{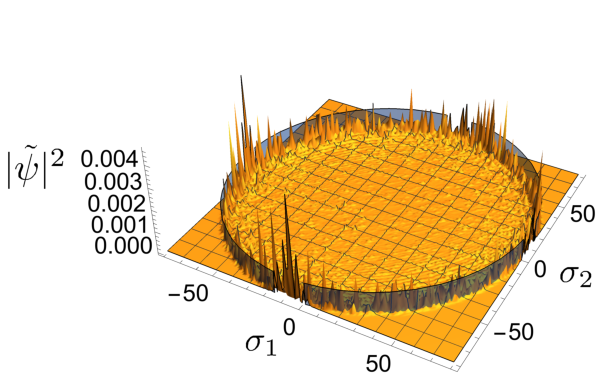}}
    \hfill
    \includegraphics[height=0.25\textwidth]{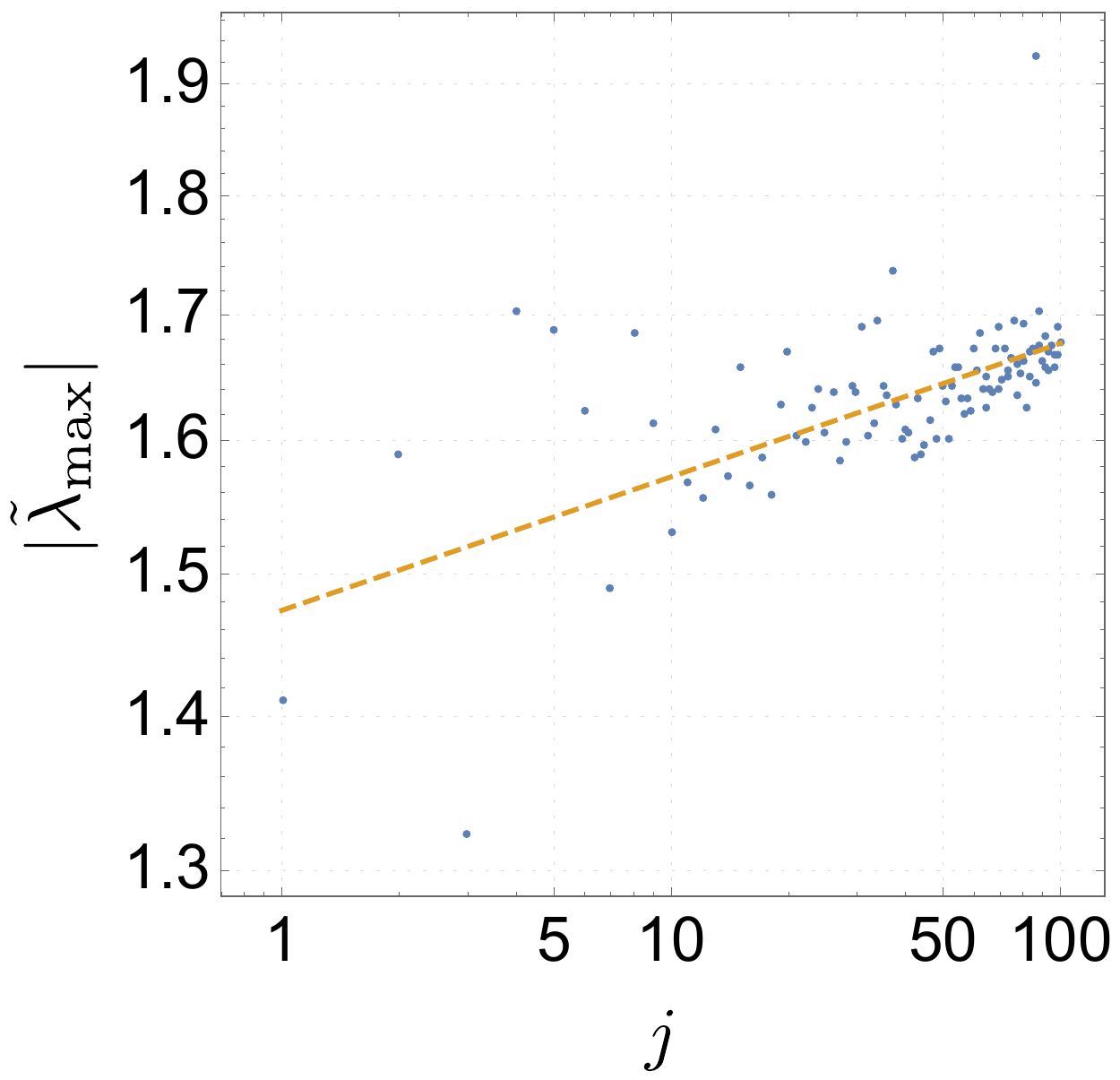}
    \caption[Spectrum, eigenvector, and scaling for $\Ut$ at $T\teq 2$ with $\mu>1$ ($\K \neq 0$) such that no manifolds are expected.]{\conline Spectrum of the dual operator for large \(\mu\teq 12\) (left panel, $j\teq 100$). Besides \(\K \teq 12 J\) all other parameters are identical to figure \ref{fig:nlkic:dualScalings:3fold}. The slope indicated in the rightmost panel is \(\alpha_0\approx 0.03\). The middle panel shows the eigenvector corresponding to the largest eigenvalue at $j\teq 80$.}
    \label{fig:nlkic:dualScalings:largeMu}
\end{figure}

 Remarkably, as  $N_0$ runs through all integer numbers,  the set of $\mu$ values defined by \eqref{eq:nlkic:mu} becomes dense in the interval $[-1,1]$. Informally speaking  this implies that in the parameter space we are always ``arbitrary close" to PO manifolds for  $\mu \in [-1,1]$. This in turn  suggests that relatively  large spectral fluctuations should be observed for  any set of parameters with $|\mu| < 1$.  Indeed, for such  parameters we observe a non-trivial scaling  $j^\alpha$ of the largest dual eigenvalues, with the values of $\alpha$ similar to  the $\K =0$ case. On the other hand, for $|\mu|>1$ this scaling turns out to be  close to zero, see fig. \ref{fig:nlkic:dualScalings:largeMu}.

%%%%%%%%%%%%%%%%%%%%%%%%%%%%%%%%%%%%%%%%%%%%%%%%%%%%%%%%%
% Conclusion
%%%%%%%%%%%%%%%%%%%%%%%%%%%%%%%%%%%%%%%%%%%%%%%%%%%%%%%%%
\section{Conclusion}
\label{sec:conclusion}

\begin{figure}
 \centering
\includegraphics[width=0.52\textwidth]{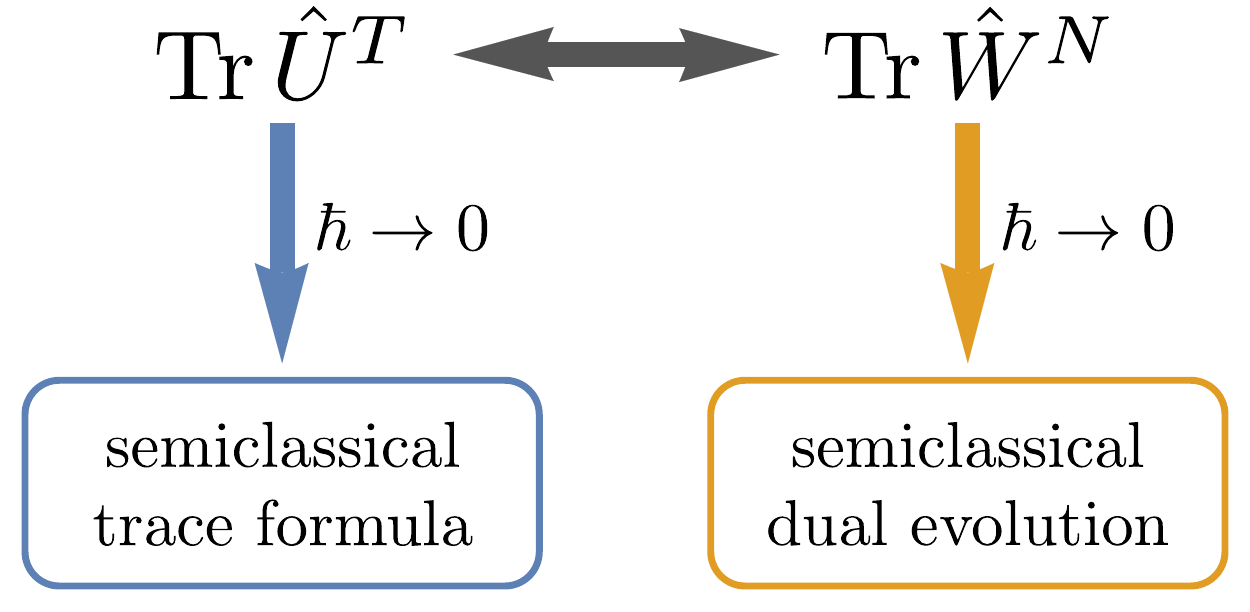}
\caption[Sketch of dual semiclassics for chain-like kicked systems.]{\conline Visualization of the two possible ways of semiclassical analysis in chain-like kicked systems. On the left hand side the standard approach of quantum chaos is shown, here the semiclassical trace formula is directly applied to the unitary time evolution $\hat{U}^T$ of the $N$-spin chain. The right hand side illustrates the dual approach. In this case the traces of the time evolution operator are first rewritten through traces of spatial (\ie along the chain) evolution $\Ut^N$. For the second step the semiclassical analysis is applied to the spectrum of the non-unitary operator  $\Ut$.}
\label{fig:conclusion}
\end{figure}

Although many-body systems played a pronounced role in the foundation of quantum chaos its  later on development has been  mainly restricted  to few particle systems.
This limitation is seemingly related to one of the key semiclassical tools, the trace formula, which connects  traces of  quantum evolution operators with periodic orbits (POs) of the underlying classical system.  For systems with   few degrees  of freedom this approach is applicable to  a very wide range of  time scales, including  the Heisenberg times,    where the phenomenon of spectral  universality holds. On the other hand, an increase in  the number of particles $N$ leads to an  exponential proliferation of POs on the classical side of the problem  and, simultaneously,  to an exponential growth of the effective  Hilbert space dimension (resp.  density of states) on the quantum side. Thus, it becomes apparent  that the conventional quantum chaos  path, illustrated  on the left hand side of figure \ref{fig:conclusion}, should fail, in general,   to reproduce correctly the classical-quantum correspondence in a  limit where    both $N$ and $\hbar^{-1}$  grow simultaneously. 

Still, as we show in the present paper, the situation is not entirely hopeless, as long  as  one is interested in the short time scales of many-body evolution.  The key ingredient of our approach   is the  duality relation which connects traces  of the unitary evolution $\Uh^T$  to those of the non-unitary operator $\Ut$. Crucially,    the dimension of $\Ut$ is independent of $N$ and remains small  for short evolution times. This drastically reduces the complexity of the problem from the numerical point of view. What is even  more important,  the duality relation opens up a second path, illustrated on the right hand side of figure \ref{fig:conclusion}, suitable to address large $N$ systems. Instead of treating the spectrum of the original unitary evolution $\hat{U}$ we can apply semiclassical techniques to the dual operator $\Ut$.

We   focus on the trace formula for a model of a long, interacting, kicked spin chain  in the regime of very short times  $T=1$ and $T=2$ while the spin quantum number $j$ plays the role of the inverse Planck's constant. The most significant result of our study  is the observation  of extremely large   spectral fluctuations for model  parameters  where non-isolated 4-dimensional   manifolds of POs appear.
This POs feature a short spacial period and can be interpreted as signatures of collective dynamics.
In  particular, this happens   if  $T=2$ and  the ratio $\mu$ between the inter-spin coupling strength and the on-side torsion is tuned to  satisfy the relation \eqref{eq:nlkic:mu}. Furthermore,  provided that the spin chain length $N$ is large and divisible by the  spatial  prime period of the PO manifolds,  the trace formula is completely  dominated by them   while all isolated POs are suppressed. 
As our analysis shows,  the contributions  of  PO manifolds   to the trace formula comes with  large  prefactors   $|\Aga|\sim j^{\alpha(N)}$ which exponentially  grow    with $N$:  $\alpha(N)=\alpha_0 N$. This explains the dominance of such structures over isolated POs, where $\alpha_0=0$.
The exponent $\alpha_0$  weakly depends on the system parameters and ranges between $0$ and $1/4$.  A similar growth is observed in fully integrable system with $\alpha_0=1/2$.  Informally speaking, this result puts   our model  somewhere in between fully integrable   and fully chaotic systems (where $\alpha_0=0$), as far as long  range spectral fluctuations are of concern. 

The above statements nicely  illustrate a pronounced difference between the pure semiclassical limit $j\to\infty$ with fixed $N$ and one where $N$ tends to infinity together with $j$. In the first case $\alpha(N)$ would be bounded by half of the marginal directions of the PO manifold's stability matrices, \ie it would not grow with $N$. 
The linear growth of $\alpha(N)$  in the double limit case  can be traced down to a growing number of  quasi-marginal directions, whose contribution to the trace formula is hard to evaluate for finite $j$. The duality approach accounts them in a systematic and quantitative way based on features of the dual operator's spectrum.
%This approach is the more important the closer the limit tips toward the second extreme of fixed $j$ but $N\to\infty$. In this case, only briefly touched in the paper, the traces are determined solely by the largest eigenvalues of $\Ut$. One can not expect that the semiclassical trace formula still holds in this case.

What remains an open question is the dependence of the spectral  fluctuations   on  $T$. So far, the analysis has been  limited to the two  shortest  times $T=1,2$ and it would be of interest  to assert  how (or whether)   large spectral oscillations   exist for larger  $T$. Another type of open questions left beyond the scope of the present paper concerns the generality of the phenomenon. For instance, whether it appears in other kicked models, or  in Hamiltonian systems with continuous time evolution. More specifically,  one would like to understand whether the existence of PO manifolds is a necessary/sufficient condition for large spectral oscillations in general. 

%%%%%%%%%%%%%%%%%%%%%%%%%%%%%%%%%%%%%%%%%%%%%%%%
% Acknowledgements
%%%%%%%%%%%%%%%%%%%%%%%%%%%%%%%%%%%%%%%%%%%%%%%%
\subsection*{Acknowledgements}
We are grateful  to U.~Smilansky for instructive and stimulating discussions.

%%%%%%%%%%%%%%%%%%%%%%%%%%%%%%%%%%%%%%%%%%%%%%%%
% Appendix
%%%%%%%%%%%%%%%%%%%%%%%%%%%%%%%%%%%%%%%%%%%%%%%%
%%%%%%%%%%%%%%%%%%%%%%%%%%%%%%%%%%%%%%%%%%%%%%%%%%%%%%%%%%%%%
% Appendix
%%%%%%%%%%%%%%%%%%%%%%%%%%%%%%%%%%%%%%%%%%%%%%%%%%%%%%%%%%%%%
\appendix

%%%%%%%%%%%%%%%%%%%%%%%%%%%%%%%%%%%%%%%%%%%%%%%%%%%%%%%%%%%%%%
% Classical Action
%%%%%%%%%%%%%%%%%%%%%%%%%%%%%%%%%%%%%%%%%%%%%%%%%%%%%%%%%%%%%%
\section{Classical Action}
\label{sec:apx:classAct}

% e use
% \begin{equation}
% \Sga=\oint_\gamma \vec{p}\cdot \mathrm{d}\vec{q}-\int\! H(\vec{q},\vec{p})\, \mathrm{d}t
% \label{eq:po:abstractS}
% \end{equation}
% with \(\vec{p}\) and \(\vec{q}\) containing the classical coordinates and momenta of all \(N\) spins, as given by \eqref{eq:nInQP}, and the classical Hamiltonian \eqref{eq:kic:hamiltonian}.
% Except for the action of \(\underline{R}_{\vec{b}}\) the calculations are elementary and relegated to \ref{sec:apx:classAct}.
The classical action, as used in \eqref{eq:po:abstractSScale}, of the system contains two contributions, a part stemming from the interaction (\(\Sact_I\)) and one from the local kicking (\(\Sact_K\)). Along the orbit, or similarly on any other trajectory, it may be split according to
\begin{equation}
\Sga=\sum_{t=0}^{T_\gamma-1}
\Sact_{K} \Big(\vec{q}(t),\vec{p}(t);\vec{q}(t+\epsilon),\vec{p}(t+\epsilon)\Big)
+\Sact_{I} \Big(\vec{p}(t+\epsilon)\Big)
\label{eq:apx:actSplitUp}
\end{equation}
where the kick is restricted to times \(t\) to \(t+\epsilon\) with \(\epsilon\to 0\) and \(\vec{q},\vec{p}(t+\epsilon)\) are the positions of the spins directly after its application.

As long as the rotation of the spins is around the \(z\)-axis as in the interaction part, only the \(q\) component changes while \(p\) remains constant. This makes the evaluation of \(\Sact_I\) straightforward and for the contribution to a single time step we find
\begin{eqnarray}
\Sact_I\Big(\vec{p}(t+\epsilon)\Big) &=& \int_{\vec{q}(t+\epsilon)}^{\vec{q}(t+1)}\vec{p}\cdot\dd\vec{q}-\int_{t+\epsilon}^{t+1} H(\vec{q},\vec{p}) \,\dd \tau
\nonumber
\\
&=&\sum_{n=1}^{N_\gamma} \Delta q_n(t+\epsilon) p_n(t+\epsilon) + \Delta t H_I(\vec{p}(t+\epsilon))
\label{eq:apx:si}
\\
\nonumber
&=& 4 \sum_{n=1}^{N_\gamma} \left(
J p_{n+1}(t+\epsilon) p_{n}(t+\epsilon)
+\K (p_n(t+\epsilon))^2 \right)\,,
\end{eqnarray}
wherein \(\Delta t\to 1\) and \(\Delta q_n(t+\epsilon)\teq 4 J (p_{n-1}(t+\epsilon) +p_{n+1}(t+\epsilon)) + 8 \K p_{n}(t+\epsilon)\). Throughout one type of dynamics ``energy'' is conserved and thus \(H_I\) is constant along the trajectory segment. 
From a conceptual point of view $\dd p_i\dd q_i$ are the area elements on the Bloch sphere and the integrals thus measure the area swept by the spin vectors $\vec{n}_i$.

The kicking part is given by the Larmor rotation of all spins about \(\vec{b}\) around the same angle. Is is local for every spin and its action is thus a sum of single spin actions. But, besides the integrable case (\(b^x\teq 0\)) both \(p\) and \(q\) change. However, we may change our coordinate system \((q,p)\to(Q,P)\) into a basis where the rotation is around the \(Z\)-axis instead of \(\vec{b}\).
The transformations are given by
\begin{eqnarray}
q(Q,P)=\arctan\frac{\sqrt{1-P^2} \sin{Q}}{\sqrt{1-P^2} \cos{\varphi} \cos{Q}-P \sin{\varphi}}\,,
\label{eq:apx:qpQPtrafoq}
\\
p(Q,P)=P \cos{\varphi}+\sqrt{1-P^2} \sin{\varphi} \cos{Q}
\label{eq:apx:qpQPtrafop}
\end{eqnarray}
with respect to the angle \(\varphi\) between the magnetic field and the \(z\)-direction. The inverse transformation is given by \(\varphi\to -\varphi\).
Neglecting particle indices we may cast the integral part of \(\Sact_K\) into
\begin{eqnarray}
\nonumber
\int_{q(t)}^{q(t+\epsilon)}\! p\,\dd q
&=&\int_{t}^{t+\epsilon} p(\tau) \dot{q}(\tau)\,\dd\tau
\\
&=&\int_{t}^{t+\epsilon} p(Q(\tau),P) \left(\partial_{\tau} q(Q(\tau),P) \right)\,\dd\tau
\label{eq:apx:skInt1}
\\
\nonumber
&=&\int_{Q(t)}^{Q(t+\epsilon)} p(Q,P) \left(\partial_{Q} q(Q,P) \right) \,\dd Q
\\
\nonumber
&=& \Phi(Q(t+\epsilon ),P)-\Phi(Q(t),P)
\end{eqnarray}
for which we use that \(P\) is constant under rotation. The change in angle, \(Q(t+\epsilon)=Q(t)+\Delta Q\), is given by the rotation matrix \(\underline{R}_{\vec{b}}\) and independent of \(\epsilon\)..
After some calculation the antiderivative \(\Phi\) may be found as
\begin{eqnarray}
    \Phi(Q,P)& &= QP + \arctan{w_-(Q,P)}-\arctan{w_+(Q,P)}
    \label{eq:apx:antiDeriv}
    \\
    \nonumber
    & &
    \text{with}\quad
    w_{\pm}=\frac{P\cos{\varphi}+\sqrt{1-P^2}\sin{\varphi}\pm 1}{P\pm \cos{\varphi}}\tan{\frac{Q}{2}}\,.
\end{eqnarray}
While using this equation one has to keep track of increased winding numbers when \(Q\) passes from \(+\pi\) to \(-\pi\).
The remaining part of \(\Sact_K\) is the (time) integral over \(H_K\). Again, along the segment \(H_K\) is constant and may be evaluated at an arbitrary point. Although the time interval of the kick tends to zero the delta distribution is adjusted such that the integral remains of unit measure. This part will compensate the \(P\,\Delta Q\) contribution from the previous integral.
As a side remark, for \(\varphi\teq 0\) we find \(\Sact_K\teq 0\).

\section{Kick Action for 2 Time Steps}
\label{sec:apx:t2action}

For periodic orbits with \(\Tprim\teq 2\) the kick action \(\Sact_K^{(n,1)}\) for the \(n\)-th spin at the first time step is identical to \(\pm\Sact_K^{(n,2)}\) at the other time step. It can therefore either add up to double its value or cancel all together. In the case of the 4D manifolds this cancellation, occurring for all of the spins, leads to their simple action formula \eqref{eq:po:smanBase}.
To understand this property we have to look at trajectories (not necessarily parts of periodic orbits) connecting two different values of \(p\), \(p^i\to p^f\), under the action of \(\underline{R}_{\vec{b}}\). For simplicity, we restrict our argument to a single spin. Generically, there are either none or two, and only two, trajectories \(z_{1,2}\),
\begin{equation}
    (q^i_{1,2}, p^i) \to (q^f_{1,2}, p^f)\,,
\end{equation}
connecting the initial and final momenta (compare with the spin rotation about the $y$-axis relevant for the evaluation of the Wigner $d$-function in \cite{braun}). As we show subsequently the action along the two trajectories fulfills
\begin{equation}
    \Sact_K(z_1) = -\Sact_K(z_2)\,.
    \label{eq:apx:actionSymT2}
\end{equation}
This is important as for any \(\Tprim\teq 2\) periodic orbit the spin, in the second time step has to return via \(p^f\to p^i\) along any one of the two possible trajectories \(z'_{1,2}\):
\begin{equation}
    (-q^f_{1,2},p^f)\to (-q^i_{1,2},p^i)
\end{equation}
which are time reversed reflections of \(z_{1,2}\) flipped perpendicular to the \(xz\)-plane. Due to symmetry we find the associated actions \(\Sact_K(z_{1,2})\teq \Sact_K(z'_{1,2})\), compare equations \eqref{eq:apx:qpQPtrafoq} and \eqref{eq:apx:antiDeriv}. 
A periodic orbit with the first kick segment given by, for example, $z_1$ may close either via $z_1'$ or $z_2'$ as its second segment. In the first case the actions of the kicks will add up, it is further easy to show that the orbit wil then be highly symmetric with all its four points in the same plane orthogonal to the field. On the contrary, if the orbit is composed of $z_1$ and $z_2'$ the overall kick action is zero.

To prove \eqref{eq:apx:actionSymT2} we point out that the action is path independent and we may safely use the Euler decomposition, see \eqref{eq:spec:eulerDecomp}, of the rotation into \(z,x\) and \(z\)-rotations. The \(z\) contributions lead to vanishing actions and only the \(x\) part has to be dealt with. Denoting the corresponding segments of the two trajectories by either \((q^i_{1x},p^i)\to (q^f_{1x},p^f)\) or \((q^i_{2x},p^i)\to (q^f_{2x},p^f)\) one may from purely geometrical reasons conclude that \(q^i_{2x}\teq \pi - q^i_{1x}\). In other words, the second possible trajectory segment connecting two different \(p\) values under \(\underline{R}_x\) is obtained by reflection at the \(yz\)-plane. Using the rotated coordinate system \(Q,P\) aligned to the field, see \ref{sec:apx:classAct}, we find that \(Q^i_{1x}\teq Q^i_{2x}\), \(P^i_{2x}\teq -P^i_{1x}\) corresponds to this reflection. Looking at \eqref{eq:apx:antiDeriv} for \(\varphi\teq\pi/2\) it is straightforward to see that \(P\to-P\) leads to a sign change in \(\Phi\), which concludes the proof of \eqref{eq:apx:actionSymT2}.

A generic PO consists of both types of spins, those  for which the (local) kick action cancels as well as those where it adds up leading still to a non-trivial result for the overall $\Sact_K $.
What remains to be argued is that for the manifolds' orbits only the cancelling type occurs. To make this plausible, let us again look at a single spin \(\vec{n}_1\) of the manifold. It is mapped under time evolution onto 
\begin{equation}
   \vec{n}_2 = \underline{R}_I\underline{R}_{\vec{b}}\vec{n}_1\,, 
\end{equation}
where by construction of the manifold we may assume \(\underline{R}_I\) to be a fixed, given matrix independent of our concrete choice of \(\vec{n}_1\). proving our statement by contradiction, let us assume that the new vector belongs to those mirror reflected trajectories that have identical action. In this case it may also be obtained as 
\begin{equation}
   \vec{n}_2 = \underline{P}_y\underline{R}_{\vec{b}}\,\vec{n}_1\,,
\end{equation}
where \(\underline{P}_y\) denotes the reflection along the \(xz\)-plane. While these two equations can be satisfied for single vectors \(\vec{n}_1\) for the manifold it would have to be satisfied for the set of linear independent vectors residing on it. Thus, we would require that a rotation equates a reflection, \(\underline{R}_I\teq \underline{P}_y\), which can not be satisfied. Therefore, orbits on \(\Tprim\teq 2\) manifolds have to feature vanishing \(\Sact_K\) contributions wherever the dimension of the manifold (locally) does not collapse.
As a closing remark, while the construction of the point \(\vec{n}_2\) belonging to the cancelling trajectory is slightly more involved it necessarily involves a further reflection \(\underline{P}_x\) and two reflections can be expressed by a rotation.

%%%%%%%%%%%%%%%%%%%%%%%%%%%%%%%%%%%%%%%%%%%%%%%%%%%%%%

\begin{equation}
    \fl
    \langle \sigma_1 \sigma_2 | \Uti | \sigma_1 \sigma_2\rangle
    = \langle \sigma_1| \eu^{-2\iu \, \vec{b}\cdot \hat{\vec{S}}}\, | \sigma_2 \rangle
    \langle \sigma_2| \eu^{-2\iu \, \vec{b}\cdot \hat{\vec{S}}}\, | \sigma_1 \rangle
    =\eu^{-\iu (\sigma_1 + \sigma_2)(\alpha+\gamma-\pi)}(d_{\sigma_1 \sigma_2}^j (\beta))^2
    \,.
    %\label{eq:spec:t2Wigner} already defined in the body of the paper
    %pi in exponent stems from exchange in d_{12}
\end{equation}
Here \(\alpha,\beta,\gamma\) (with \(\alpha\teq\gamma\)) are the Euler angles given in \eqref{eq:spec:eulerAnglesDef}
 and 
\begin{equation}
   d_{\sigma\,\sigma'}^j(\beta)=\langle \sigma| \eu^{-\iu \beta \hat{S}^y} | \sigma' \rangle\,,
   %\label{eq:spec:wignerSmallD} already defined in the body of the paper
\end{equation}

% Boris explanations
%%%%%%%%%%%%%%%%%%%%%%%%%%%%%%%%%%%%%%%%%%%%%%%%%%%%%%
 \section{Dual matrix spectrum}
 \label{sec:apx:dualEvec}
 
 In this appendix we provide an approximation for the spectrum of the dual evolution $\Ut$.
 Rather than consider the dual operator itself it is instructive to analyze the spectrum of its square $ \Ut^2= \Uti \Uto$, with $\Uto=\Utk\Uti \Utk$. The idea is that the operator $\Uto$  can be thought of as an approximate permutation. To see this we notice that its matrix elements can be written down as 
 
 \begin{eqnarray}\langle \sigma_1 \sigma_2|\Uto |\sigma'_1 \sigma'_2\rangle &=&  \sum_{m_1=-j}^j\sum_{m_2=-j}^j \exp{\left[\frac{-\iu 4J(m_1\sigma_1+m_2\sigma_2) }{j+1/2} \right]}
 \langle m_1| \eu^{-2\iu \, \vec{b}\cdot \hat{\vec{S}}}\, | m_2 \rangle\cdot \nonumber\\
 &&\cdot  
 \exp{\left[\frac{-\iu 4J(m_1\sigma'_1+m_2\sigma'_2) }{j+1/2} \right]}
 \langle m_2| \eu^{-2\iu \, \vec{b}\cdot \hat{\vec{S}}}\, | m_1 \rangle
 .\end{eqnarray}
 By using $\hat{S}^z$ operators it can be rewritten as 
 
  \begin{eqnarray}&&\langle \sigma_1 \sigma_2|\Uto |\sigma'_1 \sigma'_2\rangle =  \sum_{m_1=-j}^j\sum_{m_2=-j}^j \langle m_1| \eu^{-\iu\kappa_1\hat{S}^z}|m_1\rangle 
 \langle m_1| \eu^{-2\iu \, \vec{b}\cdot \hat{\vec{S}}}\, | m_2 \rangle\cdot\nonumber \\ 
 &&\cdot\langle m_2| \eu^{-\iu\kappa_2\hat{S}^z}|m_2\rangle
 \langle m_2| \eu^{-2\iu \, \vec{b}\cdot \hat{\vec{S}}}\, | m_1 \rangle
 =\Tr\left(\eu^{-\iu \Phi \, (\vec{n}\cdot \hat{\vec{S}})}\right)\nonumber
 ,\label{matrellO}\end{eqnarray}
 where  \[  \qquad \kappa_1=\frac{2\pi a (\sigma_1+ \sigma'_1)}{2j+1}, \quad  \kappa_2=\frac{2\pi a(\sigma_2+  \sigma'_2)}{2j+1}, \qquad a=4J/\pi, \]
 and  the operator 
   \[\eu^{-\iu \Phi \, (\vec{n}\cdot \hat{\vec{S}})}:=\eu^{-\iu\kappa_1\hat{S}^z}  \eu^{-2\iu \, \vec{b}\cdot \hat{\vec{S}}}\eu^{-\iu\kappa_2\hat{S}^z}
 \eu^{-2\iu \, \vec{b}\cdot \hat{\vec{S}}},\]
 describes  rotation around some axis $\vec{n}$  by an angle $\Phi$. 
 From the last representation it follows
 \begin{equation}\langle \sigma_1 \sigma_2|\Uto |\sigma'_1 \sigma'_2\rangle = \sum_{m=-j}^j \eu^{-\iu m\Phi}= \frac{\sin(j+1/2)\Phi}{ \sin \Phi/2 }.\label{matrell}\end{equation}
  
 The rotation angle $\Phi:=\Phi (\kappa_1,\kappa_2)$ can be straightforwardly determined through the relationship: 
 \[\Tr\left( \mat{R}_z(\kappa_1) \mat{R}_{\vec{b}}(2\vec{b})\mat{R}_z(\kappa_2) \mat{R}_{\vec{b}}(2\vec{b})\right) = 1+2\cos \Phi   \]
 with $R_z(\kappa_1)$, $R(2 \vec{b})$ being rotations along $z$ and $\vec{b}$ directions, respectively. 
 At this point it is convenient to  use the Euler decomposition $R(2\vec{b})= R_z(\alpha) R_x(\beta)R_z(\alpha) $ leading   to   
  \begin{equation}
   \Tr\left( R_z(\kappa_1+\theta) R_x (\beta)   R_z(\kappa_2+\theta) R_x(\beta)\right) = 1+2\cos \Phi(\kappa_1+\theta,\kappa_2+\theta), 
  \end{equation}
  with $2\alpha=\theta$. This allows to evaluate  the function $\Phi(x,y)$ explicitly:
\begin{eqnarray*}
  2\cos \Phi(x,y)&=& \cos x \cos y (1+\cos^2 \beta)\\&-&2\sin x \sin y \cos \beta - (\cos x +\cos y+1) \sin^2 \beta .
 \end{eqnarray*}
 
 Note that the matrix   elements (\ref{matrell}) are of  the order $ 2j+1$ if $\Phi\approx 0$ and of the order $1$, otherwise. The solutions of the  equations $\Phi(\kappa_1+\theta,\kappa_2+\theta) = 0$ are provided by all $\kappa_1,\kappa_2$
 such that 
 \begin{equation}
   R_z(\kappa_1+\theta) R_x(\beta)  = \left(R_z(\kappa_2+\theta) R_x(\beta)\right)^{-1}.
 \end{equation}
 %This yields:
 After writing down the left and the right hand side of this equation in the matrix form and comparing them element-wise     (see e.g., \cite{QuantumLMoment}) we conclude  that both rotations about the $z$-direction must be by $\pi$ modulo $2\pi$:
 \begin{equation}
 \kappa_1+\theta=\pi +2\pi p_1, \qquad   \kappa_2+\theta=\pi +2\pi p_2,
\end{equation}
with $p_1,p_2\in\mathds{Z}$, or equivalently:  
\begin{equation}
 \frac{\sigma_1+\sigma'_1}{2j+1}=\frac{1+2p_1-\theta/\pi}{2a}, \qquad   \frac{\sigma_2+\sigma'_2}{2j+1}=\frac{1+2p_2-\theta/\pi}{2a}. \label{eq:apx:dualEvec:solution}
\end{equation}
Since $-j \leq \sigma_i, \sigma'_i\leq j$, the above solutions exist only if  the interval $[-2{a} +\frac{\theta}{\pi}, 2{a} +\frac{\theta}{\pi}]$ contains a point from  $\{1+2k | k\in\mathbb{Z}\}$. This is, in fact, precisely the condition for the existence of 4-dimensional manifolds.  In particular, for  the case $b_z=0$   this condition reduces to $|a|<1/2$.

 \subsubsection*{Single PO manifold.} In what follows we will consider parameters  $a, \theta$  such that \eqref{eq:apx:dualEvec:solution} admits at most  one solution $\sigma_1, \sigma_2 \in [-j, j]$, $p_1= p_2=p$  for each  pair $\sigma'_1, \sigma'_2$. 
 In that case  we can write  $\Uto= (2j+1)\Utp \Utg  $, where  
 \begin{eqnarray}
 \langle \sigma_1 \sigma_2|\Utp |\sigma'_1 \sigma'_2\rangle &=& \delta_{\sigma_1+\sigma'_1, g } \, \delta_{\sigma_2 +\sigma_2, g},\\  &&-j\leq g= \left\lfloor\frac{(2j+1)(1+2p -\theta/\pi)}{2a}\right\rfloor\leq j\end{eqnarray}
 is a truncated permutation  while  $\Utg$ has a band like structure.  The last matrix   has approximately unity  elements on  the diagonal $\langle \sigma_1 \sigma_2|\Utg |\sigma_2 \sigma_1\rangle \approx 1$, while its off-diagonal elements are highly fluctuating  with  absolute values  decaying  as distance from the diagonal grows: 
 \[|\langle \sigma_1 \sigma_2|\Utg |\sigma'_1 \sigma'_2\rangle| \sim \left((\sigma'_1- \sigma_1)^2 +   (\sigma_2 -\sigma'_2)^2\right)^{-1/2}.\]

   To facilitate the study of the spectrum of  $\Ut$ we  make a crude  approximation $\Utg\approx \mathds{1} $ (resp.   $\Uto\approx (2j+1)\Utp $) in the body of the paper.  The above  approximation amounts to picking up the largest  
 element from each row of the matrix $\Uto$. 
 Since $\Uti$ is a diagonal matrix and  $\Utp$ is a  permutation,  the eigenvectors of $\Uti \Utp$ take a simple form: 
 \begin{equation}
  \psi_{(\sigma_1,\sigma_2)}= C_1|\sigma_1 \sigma_2\rangle \pm  C_2|g-\sigma_1\, g- \sigma_2\rangle,
 \end{equation}
 with
 \begin{equation}
 \left({C_1}/{C_2}\right)^2= {\langle \sigma_1 \sigma_2|\Uti |\sigma_1 \sigma_2\rangle }/{\langle g-\sigma_1 \, g-\sigma_2|\Uti |g-\sigma_1 \, g-\sigma_2\rangle}
 \end{equation}
 and the corresponding eigenvalues $\tilde{\Lambda}_{(\sigma_1,\sigma_2)}$ given by 
 \begin{equation}
  \tilde{\Lambda}^2_{(\sigma_1,\sigma_2)}= \langle \sigma_1 \sigma_2|\Uti |\sigma_2 \sigma_1\rangle \langle g-\sigma_1 \, g-\sigma_2|\Uti |g-\sigma_1 \, g-\sigma_2\rangle .
 \end{equation}
 This in turn can be written down in terms of Wigner $d$-functions as 
 \begin{equation}
  \tilde{\Lambda}_{(\sigma_1,\sigma_2)}= e^{i(\theta-\pi) g} d^j_{\sigma_2,\sigma_1}(\beta) \, d^j_{g-\sigma_2,g-\sigma_1} (\beta). 
 \end{equation}

\subsubsection*{Multiple  PO manifolds.} In this case   eq.~\eqref{eq:apx:dualEvec:solution} admits multiple solutions corresponding to  several different combinations of $(p_1,p_2)$.  Each pair $(p_1,p_2)$   determines uniquely the pair of constants 
\[g_1= \left\lfloor\frac{(2j+1)(1+2p_1 -\theta/\pi)}{2a}\right\rfloor, \qquad g_2= \left\lfloor\frac{(2j+1)(1+2p_2 -\theta/\pi)}{2a}\right\rfloor  \]  such that 
$\Uto$ can be thought as an approximate sum of permutations (if only the largest elements in each row are left), i.e.,  $\Uto\approx\Utp$, where
\begin{eqnarray}
&&\Utp = (2j+1)\sum_{(p_1,p_2)}\Utp_{(p_1,p_2)},
\\ && \langle \sigma_1\, \sigma_2|\Utp_{(p_1,p_2)} |\sigma'_1\, \sigma'_2\rangle = \delta_{\sigma_1+\sigma'_1, g_1 } \, \delta_{\sigma_2 +\sigma_2, g_2}.\nonumber 
\label{append:sumperm}
\end{eqnarray}

As opposed to the single manifold case, even within   the above approximation it seems to be impossible  to provide explicit formula for    the spectrum     of $\Uti\Utp$ for generic system parameters. 
However, after the crossover from the  regime of single PO manifold  to one of multiple PO manifolds   there exists a  certain  range  of parameters where permutations $\Utp_{(p_1,p_2)}$
are mutually orthogonal: \[\Utp_{(p'_1,p'_2)}\Utp_{(p_1,p_2)}=0, \qquad \mbox{ for } \quad (p'_1,p'_2)  \neq(p_1,p_2). \]
In this case the total spectrum of 
 $\Uti\Utp$ is composed of subspectra of the operators  of  $\Uti \Utp_{(p_1,p_2)}$ and can be easily evaluated. As in the single manifold case,  the  eigenvectors take a simple  form 
\begin{equation}
  \psi_{(\sigma_1,\sigma_2)}= C_1|\sigma_1\,  \sigma_2\rangle \pm C_2|g_1-\sigma_1\, g_2- \sigma_2\rangle, \label{append:multstates}
 \end{equation}
 with the corresponding eigenvalues given by:
 \begin{equation}
  \tilde{\Lambda}_{(\sigma_1,\sigma_2)}= e^{i(\theta-\pi) (g_1+g_2)/2} d^j_{\sigma_2,\sigma_1}(\beta) \, d^j_{g_2-\sigma_2,g_1-\sigma_1} (\beta). 
 \end{equation}
 Note that the  eigenstates with the largest eigenvalues    must be  localized at the boundary  of the ellipse \eqref{eq:spec:transitReg}. The localization points ($\sigma_1, \sigma_2$)  are, therefore,  determined by the demand  that both points ($g_1-\sigma_1, g_2-\sigma_2$) and ($\sigma_1, \sigma_2$) belong to the ellipse boundary. 
 To see that these are also  eigenstates of $\Uti\Utp$ it is sufficient to notice that action of other permutations $\Utp_{(p'_1,p'_2)}\neq\Utp_{(p_1,p_2)}$ on the  states \eqref{append:multstates}  brings them to zero.  
 
The numerical computation  of the actual spectrum of  the operator  $\Uti\Uto$ shows that the localization points of its  eigenvectors associated with the highest eigenvalues are indeed have the same localization points as the  states  \eqref{append:multstates}, see 
 fig. \ref{fig:spec:T2diagSpec}. Furthermore, as can be seen on the same figure,   the bulk of the spectrum is composed of  a number of cross-like  structures.   Each  such cross    is associated with one of the  pairs  $(p_1,p_2)$ in the sum \eqref{append:sumperm}.

%\nocite{*}
%\printbibliography 
\bibliographystyle{ieeetr}
\bibliography{refs}

\begin{thebibliography}{10}

\bibitem{stoeckmann}
H.-J. St{\"o}ckmann, {\em {Quantum Chaos -- an introduction}}.
\newblock Cambridge University Press, 2006.

\bibitem{haake}
F.~Haake, {\em {Quantum Signatures of Chaos}}.
\newblock {Springer Series in Synergetics}, Springer, 3~ed., 2010.

\bibitem{BGS1984}
O.~Bohigas, M.~J. Giannoni, and C.~Schmit, ``{Characterization of Chaotic
  Quantum Spectra and Universality of Level Fluctuation Laws},'' {\em Phys.
  Rev. Lett.}, vol.~52, pp.~1--4, Jan 1984.

\bibitem{CVGG1980}
G.~Casati, F.~Valz-Gris, and I.~Guarnieri, ``{On the connection between
  quantization of nonintegrable systems and statistical theory of spectra},''
  {\em Lettere al Nuovo Cimento (1971-1985)}, vol.~28, no.~8, pp.~279--282,
  1980.

\bibitem{guhr}
T.~Guhr, A.~M{\"u}ller-Groeling, and H.~A. Weidenm{\"u}ller, ``{Random-matrix
  theories in quantum physics: common concepts},'' {\em Physics Reports},
  vol.~299, no.~4--6, pp.~189--425, 1998.

\bibitem{berry2}
M.~V. Berry, ``{Semiclassical Theory of Spectral Rigidity},'' {\em Proceedings
  of the Royal Society of London A: Mathematical, Physical and Engineering
  Sciences}, vol.~400, no.~1819, pp.~229--251, 1985.

\bibitem{sieberRichter}
M.~Sieber and K.~Richter, ``{Correlations between periodic orbits and their
  r{\^o}le in spectral statistics},'' {\em Physica Scripta}, vol.~2001,
  no.~T90, p.~128, 2001.

\bibitem{haakePRL1}
S.~M{\"u}ller, S.~Heusler, P.~Braun, F.~Haake, and A.~Altland, ``{Semiclassical
  Foundation of Universality in Quantum Chaos},'' {\em Phys. Rev. Lett.},
  vol.~93, p.~014103, Jul 2004.

\bibitem{regen1}
T.~Engl, J.~Dujardin, A.~Arg{\"u}elles, P.~Schlagheck, K.~Richter, and J.~D.
  Urbina, ``{Coherent Backscattering in Fock Space: A Signature of Quantum
  Many-Body Interference in Interacting Bosonic Systems},'' {\em Phys. Rev.
  Lett.}, vol.~112, p.~140403, Apr 2014.

\bibitem{regen2}
T.~Engl, J.~D. Urbina, and K.~Richter, ``{Periodic mean-field solutions and the
  spectra of discrete bosonic fields: Trace formula for Bose-Hubbard models},''
  {\em Phys. Rev. E}, vol.~92, p.~062907, Dec 2015.

\bibitem{regen4}
T.~Engl, P.~Pl{\"o}{\ss}l, J.~D. Urbina, and K.~Richter, ``The semiclassical
  propagator in fermionic fock space,'' {\em Theoretical Chemistry Accounts},
  vol.~133, p.~1563, Sep 2014.

\bibitem{dubert}
R.~Dubertrand and S.~M{\"u}ller, ``{Spectral statistics of chaotic many-body
  systems},'' {\em New Journal of Physics}, vol.~18, no.~3, p.~033009, 2016.

\bibitem{TSimula}
T.~Simula, ``{Collective dynamics of vortices in trapped Bose-Einstein
  condensates},'' {\em Phys. Rev. A}, vol.~87, p.~023630, Feb 2013.

\bibitem{h17}
D.~A. Butts and D.~S. Rokhsar, ``{Predicted signatures of rotating
  Bose-Einstein condensates},'' {\em Nature}, pp.~327--329, 1999.

\bibitem{superParaRev}
M.~Knobel, W.~C. Nunes, L.~M. Socolovsky, E.~{De Biasi}, J.~M. Vargas, and
  J.~C. Denardin, ``{Superparamagnetism and Other Magnetic Features in Granular
  Materials: A Review on Ideal and Real Systems},'' {\em Journal of Nanoscience
  and Nanotechnology}, vol.~8, no.~6, pp.~2836--2857, 2008.

\bibitem{h10}
W.~Nazarewicz, ``{The nuclear collective motion},'' in {\em {An Advanced Course
  in Modern Nuclear Physics}} (J.~Arias and M.~Lozano, eds.), vol.~581 of {\em
  {Lecture Notes in Physics}}, pp.~102--140, Springer Berlin Heidelberg, 2001.

\bibitem{VanDerWoude1987217}
A.~V.~D. Woude, ``{Giant resonances},'' {\em Progress in Particle and Nuclear
  Physics}, vol.~18, no.~0, pp.~217--293, 1987.

\bibitem{f12}
K.~A. Snover, ``{Giant Resonances in Excited Nuclei},'' {\em Annual Review of
  Nuclear and Particle Science}, vol.~36, no.~1, pp.~545--603, 1986.

\bibitem{f14}
F.~E. Bertrand, ``{Giant multipole resonances — perspectives after ten
  years},'' {\em Nuclear Physics A}, vol.~354, no.~1-2, pp.~129--156, 1981.

\bibitem{akilaPRL}
M.~Akila, D.~Waltner, B.~Gutkin, P.~Braun, and T.~Guhr, ``Semiclassical
  identification of periodic orbits in a quantum many-body system,'' {\em Phys.
  Rev. Lett.}, vol.~118, p.~164101, Apr 2017.

\bibitem{cat_berry2}
M.~V. Berry, J.~P. Keating, and H.~Schomerus, ``Universal twinkling exponents
  for spectral fluctuations associated with mixed chaology,'' {\em Proceedings
  of the Royal Society of London A: Mathematical, Physical and Engineering
  Sciences}, vol.~456, no.~1999, pp.~1659--1668, 2000.

\bibitem{cat_schomerus}
H.~Schomerus and M.~Sieber, ``{Bifurcations of periodic orbits and uniform
  approximations},'' {\em Journal of Physics A: Mathematical and General},
  vol.~30, no.~13, p.~4537, 1997.

\bibitem{cat_bestPaper}
A.~M.~O. de~Almeida and J.~H. Hannay, ``Resonant periodic orbits and the
  semiclassical energy spectrum,'' {\em Journal of Physics A: Mathematical and
  General}, vol.~20, no.~17, p.~5873, 1987.

\bibitem{primack}
H.~Primack and U.~Smilansky, ``{On the accuracy of the semiclassical trace
  formula},'' {\em Journal of Physics A: Mathematical and General}, vol.~31,
  no.~29, p.~6253, 1998.

\bibitem{wintgen2}
D.~Wintgen, ``{Connection between long-range correlations in quantum spectra
  and classical periodic orbits},'' {\em Phys. Rev. Lett.}, vol.~58,
  pp.~1589--1592, Apr 1987.

\bibitem{richterHE}
D.~Wintgen, K.~Richter, and G.~Tanner, ``{The semiclassical helium atom},''
  {\em Chaos: An Interdisciplinary Journal of Nonlinear Science}, vol.~2,
  no.~1, pp.~19--33, 1992.

\bibitem{MF1}
A.~Georges, G.~Kotliar, W.~Krauth, and M.~J. Rozenberg, ``Dynamical mean-field
  theory of strongly correlated fermion systems and the limit of infinite
  dimensions,'' {\em Rev. Mod. Phys.}, vol.~68, pp.~13--125, Jan 1996.

\bibitem{MPS1}
F.~Verstraete, V.~Murg, and J.~Cirac, ``Matrix product states, projected
  entangled pair states, and variational renormalization group methods for
  quantum spin systems,'' {\em Advances in Physics}, vol.~57, no.~2,
  pp.~143--224, 2008.

\bibitem{MPS2}
D.~Perez-Garcia, F.~Verstraete, M.~M. Wolf, and J.~I. Cirac, ``Matrix product
  state representations,'' {\em Quantum Info. Comput.}, vol.~7, pp.~401--430,
  July 2007.

\bibitem{gutkin}
B.~Gutkin and V.~Osipov, ``{Classical foundations of many-particle quantum
  chaos},'' {\em Nonlinearity}, vol.~29, no.~2, p.~325, 2016.

\bibitem{akila2}
M.~Akila, D.~Waltner, B.~Gutkin, and T.~Guhr, ``{Particle-time duality in the
  kicked Ising spin chain},'' {\em Journal of Physics A: Mathematical and
  Theoretical}, vol.~49, no.~37, p.~375101, 2016.

\bibitem{prosen2}
T.~c.~v. Prosen, ``{General relation between quantum ergodicity and fidelity of
  quantum dynamics},'' {\em Phys. Rev. E}, vol.~65, p.~036208, Feb 2002.

\bibitem{prosenJt-2}
T.~Prosen, ``{Exact Time-Correlation Functions of Quantum Ising Chain in a
  Kicking Transversal Magnetic FieldSpectral Analysis of the Adjoint Propagator
  in Heisenberg Picture},'' {\em Progress of Theoretical Physics Supplement},
  vol.~139, p.~191, 2000.

\bibitem{prosen2007b}
C.~Pineda and T.~c.~Å. Prosen, ``{Universal and nonuniversal level statistics
  in a chaotic quantum spin chain},'' {\em Phys. Rev. E}, vol.~76, p.~061127,
  Dec 2007.

\bibitem{prosenB3-d}
T.~Prosen, ``{Chaos and complexity of quantum motion},'' {\em Journal of
  Physics A: Mathematical and Theoretical}, vol.~40, no.~28, p.~7881, 2007.

\bibitem{keppeler}
S.~Keppeler, {\em {Spinning particles: Semiclassical quantisation and spectral
  statistics}}.
\newblock PhD thesis, Universit{\"a}t Ulm, 2002.

\bibitem{lieb}
E.~Lieb, T.~Schultz, and D.~Mattis, ``{Two soluble models of an
  antiferromagnetic chain},'' {\em Annals of Physics}, vol.~16, no.~3,
  pp.~407--466, 1961.

\bibitem{atas}
Y.~Y. Atas and E.~Bogomolny, ``{Spectral density of a one-dimensional quantum
  Ising model: Gaussian and multi-Gaussian approximations},'' {\em Journal of
  Physics A: Mathematical and Theoretical}, vol.~47, no.~33, p.~335201, 2014.

\bibitem{waltner3}
D.~Waltner, P.~Braun, M.~Akila, and T.~Guhr, ``{Trace formula for interacting
  spins},'' {\em Journal of Physics A: Mathematical and Theoretical}, vol.~50,
  no.~8, p.~085304, 2017.

\bibitem{wintgen3}
D.~Wintgen, ``Connection between long-range correlations in quantum spectra and
  classical periodic orbits,'' {\em Phys. Rev. Lett.}, vol.~58, pp.~1589--1592,
  Apr 1987.

\bibitem{stockStein}
H.-J. St\"ockmann and J.~Stein, ````quantum'' chaos in billiards studied by
  microwave absorption,'' {\em Phys. Rev. Lett.}, vol.~64, pp.~2215--2218, May
  1990.

\bibitem{welge}
A.~Holle, J.~Main, G.~Wiebusch, H.~Rottke, and K.~H. Welge, ``{Quasi-Landau
  Spectrum of the Chaotic Diamagnetic Hydrogen Atom},'' {\em Phys. Rev. Lett.},
  vol.~61, pp.~161--164, Jul 1988.

\bibitem{kus}
M.~{Ku\ifmmode \acute{s}\else {\'s}\fi{}}, F.~Haake, and D.~Delande,
  ``{Prebifurcation periodic ghost orbits in semiclassical quantization},''
  {\em Phys. Rev. Lett.}, vol.~71, pp.~2167--2171, Oct 1993.

\bibitem{cat_manderfeld}
C.~Manderfeld and H.~Schomerus, ``{Semiclassical singularities from bifurcating
  orbits},'' {\em Phys. Rev. E}, vol.~63, p.~066208, May 2001.

\bibitem{cat_biff}
M.~de~Aguiar, C.~Malta, M.~Baranger, and K.~Davies, ``Bifurcations of periodic
  trajectories in non-integrable hamiltonian systems with two degrees of
  freedom: Numerical and analytical results,'' {\em Annals of Physics},
  vol.~180, no.~2, pp.~167 -- 205, 1987.

\bibitem{braunRot}
P.~A. Braun, P.~Gerwinski, F.~Haake, and H.~Schomerus, ``Semiclassics of
  rotation and torsion,'' {\em Zeitschrift f{\"u}r Physik B Condensed Matter},
  vol.~100, no.~1, pp.~115--127, 1996.

\bibitem{namias}
V.~NAMIAS, ``The fractional order fourier transform and its application to
  quantum mechanics,'' {\em IMA Journal of Applied Mathematics}, vol.~25,
  no.~3, pp.~241--265, 1980.

\bibitem{candan}
C.~Candan, M.~A. Kutay, and H.~M. Ozaktas, ``The discrete fractional fourier
  transform,'' {\em IEEE Transactions on Signal Processing}, vol.~48,
  pp.~1329--1337, May 2000.

\bibitem{braun}
C.~Braun, F.~Li, A.~Garg, and M.~Stone, ``{The semiclassical coherent state
  propagator in the Weyl representation},'' {\em Journal of Mathematical
  Physics}, vol.~56, no.~12, p.~122106, 2015.

\bibitem{QuantumLMoment}
D.~Varshalovich, A.~Moskalev, and V.~Khersonskii, {\em Quantum Theory Of
  Angular Momemtum}.
\newblock World Scientific, 1988.

\end{thebibliography}

\end{document}